\documentclass[11pt,a4paper]{article}
\pdfoutput=1

\usepackage[T1]{fontenc}
\usepackage{lmodern}
\usepackage[utf8]{inputenc}
\usepackage[english]{babel}

\usepackage[centertags]{amsmath}
\usepackage{amssymb}
\usepackage{mathtools}
\usepackage{mathrsfs}
\usepackage{slashed}

\usepackage{jheppub}

\newcommand{\dd}{\mathrm{d}}
\newcommand{\ii}{\mathrm{i}}
\newcommand{\e}{\mathrm{e}}
\newcommand{\half}{\tfrac12}
\newcommand{\Ga}{\Gamma}
\usepackage{xcolor}

\begin{document}

\title{Flat holography for spinor fields}

\author[a,b]{Dmitry S. Ageev}
\author[c]{and Anna S. Bernakevich}

\affiliation[a]{Steklov Mathematical Institute, Russian Academy of Sciences,\\ Gubkin str.~8, 119991 Moscow, Russia}
\affiliation[b]{Institute for Theoretical and Mathematical Physics, Lomonosov Moscow State University, 119991 Moscow, Russia}
\affiliation[c]{Moscow Institute of Physics and Technology, Institutskii per. 9, 141702 Dolgoprudny, Russia}

\emailAdd{ageev@mi-ras.ru}
\emailAdd{bernakevich.as@phystech.edu}

\hypersetup{
pdftitle={Flat holography for spinor fields},
pdfauthor={Dmitry S. Ageev and Anna S. Bernakevich}
}

\abstract{We extend the hyperbolic Milne-slicing construction of flat holography in four-dimensional Minkowski spacetime from scalar fields to massless spin-$\tfrac12$ fields. We solve the massive mode equation and restrict the boundary source--response analysis to the massless sector. Decomposition into harmonics on three-dimensional hyperbolic space, labeled by a continuous principal-series parameter, yields a separated-point nonlocal kernel up to the action normalization and local contact terms. The kernel has the universal form required by two-dimensional conformal covariance for spin-$\tfrac12$ principal-series primaries. Then we construct regular source-normalized conformal-primary wavefunctions in planar and global coordinates on the celestial sphere $S^2$. We show that the planar source--response kernel is naturally identified with the spin-$\tfrac12$ shadow transform, while inverse shadowing recovers the angular delta-function structure of the unshadowed basis. We also analyze radial renormalization by analytic continuation from the principal-series problem to a real-mass AdS$_3$ problem.}


\arxivnumber{2603.06794}

\maketitle
\flushbottom

\section{Introduction}

The holographic principle---motivated historically by black hole entropy \cite{Bekenstein:1973ur,Hawking:1975vc}---states that a quantum theory of gravity in $d{+}1$ dimensions admits an equivalent description in terms of a non-gravitational theory in $d$ dimensions \cite{tHooft:1993dmi,Susskind:1994vu}. In asymptotically anti-de Sitter (AdS) spacetimes, this idea is realized sharply by the AdS/CFT correspondence \cite{Maldacena:1997re,Gubser:1998bc,Witten:1998qj,Aharony:1999ti}, which provides a concrete and computable holographic dictionary between bulk fields and boundary operators.

A standard expression of the AdS/CFT dictionary is that the (renormalized) on-shell bulk action, viewed as a functional of boundary sources, generates correlation functions of the dual operators
\begin{align}
Z_{\rm bulk}[J]
\;\equiv\;
\exp\!\big(\ii S_{\rm bulk}^{\rm ren}[J]\big)
\;=\;
\Big\langle \exp\!\big(\ii \int_{\partial{\rm AdS}} J\,\mathcal O \big)\Big\rangle_{\rm CFT}\,.
\label{eq:AdSCFT_schematic}
\end{align}
A systematic implementation of \eqref{eq:AdSCFT_schematic} requires holographic renormalization, including covariant counterterms and the careful treatment of variational principles. Classic developments include the holographic Weyl anomaly and reconstruction \cite{Henningson:1998gx,deHaro:2000vlm,Skenderis:2002wp}. Holographic entanglement entropy provides another central entry in the AdS/CFT dictionary \cite{Ryu:2006bv,Lewkowycz:2013nqa}.

Despite these successes, the most pressing conceptual target is holography for spacetimes that are not asymptotically AdS. In particular, our universe is well approximated by asymptotically flat (and perhaps asymptotically de Sitter) geometries. The dS/CFT proposal \cite{Strominger:2001pn,Witten:2001kn} offers one route in a cosmological setting. For Minkowski space, one can sometimes extract approximate flat-space scattering data from AdS/CFT in suitable limits \cite{Giddings:1999qu,Polchinski:1999ry}. However, a direct analogue of \eqref{eq:AdSCFT_schematic} for flat space---a nonperturbative dictionary with a well-defined ``boundary'' theory and a controlled renormalization scheme---remains incomplete.

A crucial structural clue in the asymptotically flat setting is the emergence of infinite-dimensional asymptotic symmetries. The Bondi--Metzner--Sachs (BMS) group \cite{Bondi:1962px,Sachs:1962wk} and its extensions and refinements \cite{Barnich:2009se,Barnich:2010eb} tightly constrain infrared physics. In particular, supertranslation Ward identities reproduce Weinberg's soft graviton theorem \cite{Weinberg:1965nx,He:2014laa}, while a Virasoro enhancement governs subleading soft behavior \cite{Strominger:2013jfa,Kapec:2014opa}. For gauge theory, related structures arise as a two-dimensional Kac--Moody symmetry acting on four-dimensional scattering \cite{He:2015zea}. These developments have evolved into an increasingly rich ``celestial'' symmetry picture \cite{Donnay:2020guq,Pate:2019mfs,Guevara:2021abz,Strominger:2021lvk}, alongside alternative Carrollian formulations of flat holography \cite{Donnay:2022aba,Bagchi:2022cyv,Nguyen:2023miw,Nguyen:2023vfz,Nguyen:2025zhg,Iacobacci:2024laa}, including fermionic Carroll limits \cite{Bergshoeff:2023CarrollFermions}.

The celestial holography program proposes that four-dimensional scattering amplitudes can be reorganized as correlation functions of a putative two-dimensional conformal field theory (CCFT) living on the celestial sphere at null infinity. Operationally, the key step is to expand asymptotic states in a conformal basis: plane waves are Mellin transformed into conformal primary wavefunctions, so that energies are traded for generally complex conformal weights \cite{Pasterski:2016qvg,Pasterski:2017kqt}. In this basis, spin is encoded in $SL(2,\mathbb C)$ quantum numbers, and celestial correlators acquire the standard tensorial structures of a two-dimensional CFT \cite{Pasterski:2017ylz,Pasterski:2020pdk}. Pedagogical overviews and lecture notes include \cite{Strominger:2017zoo,Raclariu:2021zjz,Pasterski:2021rjz,McLoughlin:2022uqa}. Complementary approaches develop conformal wave expansions and celestial partial waves directly at the level of the $S$-matrix \cite{Cheung:2016iub,Liu:2021uge} and sharpen both the CCFT state--operator map, including its extension to multiparticle states, and the role of shadow transforms in obtaining standard power-law two-point functions \cite{Crawley:2021pls,Furugori:2023fga,Kulp:2024scx}. Recent work has also clarified the relations among celestial correlators, AdS data, and ``round-trip'' reconstructions \cite{Iacobacci:2022yjo,Iacobacci:2024PartII,Sleight:2023ojm}. Representative results in this amplitude-first approach include the early construction of tree-level gluon celestial amplitudes \cite{Schreiber:2017jsr}; their interpretation in terms of soft limits and the resulting conformal and current-algebra structures \cite{Fan:2019emx,Pate:2019lpp}; their systematic organization in terms of conformal partial waves and related bases \cite{Nandan:2019jas,Law:2020tsg,Atanasov:2021cje}; studies of resonances and unitarity \cite{Garc_a_Sep_lveda_2022}; and the later development of loop-level gluon OPE data and light-ray correlators in celestial holography \cite{Bhardwaj:2022xlt,Hu:2022wsy,Banerjee:2022wcx}.

In parallel with the amplitude-first viewpoint, an AdS/CFT-like route to flat holography emerges from hyperbolic slicing of Minkowski space. Inside the future (or past) light cone, Minkowski admits a Milne foliation by Euclidean AdS$_3$ slices
\begin{align}
ds^2
=
-\dd\tau^2
+
\tau^2\,ds^2_{\mathbb H^3},
\qquad
\tau>0\,,
\label{eq:Milne_metric_intro}
\end{align}
where $ds^2_{\mathbb H^3}$ is the unit metric on hyperbolic space. This observation underlies early holographic reductions of Minkowski space \cite{deBoer:2003vf} and, more recently, several attempts to formulate a flat-space holographic dictionary directly in bulk terms. These include the Milne-slicing construction for scalars and gravitational probes in \cite{Hao:2023wln} and the Hamilton--Jacobi approach to holographic renormalization for scalar fields with scattering boundary conditions in \cite{Ammon:2025avo}. Related efforts to formulate genuinely bulk versions of flat holography have been particularly successful in chiral higher-spin theories, where explicit flat-space holographic structures have been proposed in a controlled setting \cite{Ponomarev:2022ryp,Ponomarev:2022qkx}; see also the earlier light-front construction \cite{Ponomarev:2016lrm}. In this framework, bulk fields are expanded in harmonics on $\mathbb H^3$, naturally organized by principal-series representations of $SO(1,3)$ \cite{Dobrev:1977qvd}. The conformal boundary of $\mathbb H^3$ is the celestial $S^2$. The radial coordinate within each $\mathbb H^3$ slice controls the holographic source--response split, while the Milne time $\tau$ governs the separate four-dimensional evolution. A distinctive feature is the presence of two independent Milne phases $\tau^{\mp\ii\nu}$, which lead to two a priori independent sets of celestial sources and operators \cite{Hao:2023wln}. We therefore use phase labels that refer only to the displayed Milne powers and defer their global scattering interpretation. Related codimension-two constructions, such as wedge holography, provide an alternative embedding of celestial data into higher-dimensional bulk geometries \cite{Akal:2020wfl,Ogawa:2022fhy,Fukada:2023ohh}.

The goal of this paper is to develop the separated-point nonlocal source--response sector of this Milne-slicing construction for \emph{fermions}. Extending flat holography beyond scalars is essential for at least three reasons. First, realistic quantum field theories, especially supersymmetric ones, contain spin-$\tfrac12$ matter; a CCFT description of scattering must therefore accommodate fermionic operators and their selection rules. Second, fermions introduce genuinely new holographic ingredients: the Dirac action is first order, boundary conditions are chiral, and a consistent variational principle requires spinor-specific boundary terms and renormalization \cite{HenningsonSfetsos:1998,Mueck:1998iz,Henneaux:1999vb}. Third, control of spinor harmonic analysis on $\mathbb H^3$ and spinorial bulk-to-boundary kernels is a prerequisite for building spinning conformal blocks and exchange diagrams in a position-space CCFT \cite{Costa:2011mg,Costa:2014kfa,Camporesi:1995fb}. In the celestial-amplitude literature, conformal primary bases for Dirac spinors and their interpretation have been developed in \cite{Iacobacci:2020cqy,Narayanan:2020amh}, providing a natural target for a bulk derivation.

In this work, we solve the massive Dirac mode equation in the Milne wedge and construct a massless source--response map. Retaining the mass parameter in the mode analysis clarifies how the four-dimensional Dirac problem reduces to a family of effective problems on $\mathbb H^3$. The boundary analysis is restricted to massless fields, whose radiative data are encoded at null infinity, whereas massive excitations are naturally associated with timelike infinity. We use Poincar\'e coordinates for the Fourier-space asymptotic analysis and global coordinates for the full celestial sphere. The former make boundary translations and momentum space manifest; the latter resolve the point at infinity and organize the modes into $SO(3)$ harmonics. They provide complementary planar ($\mathbb R^2$) and spherical ($S^2$) presentations of the same source--response map.

The planar and global mode-sum coefficients are compatible with a common normalization after matching the boundary orientation, spin frame, and source conventions, while the short-distance limit of the global kernel reproduces the planar conformal structure. The comparison is limited to the matched coefficients and short-distance behavior; a finite-separation stereographic pullback would additionally require the endpoint Weyl factors and spin-frame rotations.

Specializing to the massless sector, we apply an asymptotic coefficient prescription and obtain a nonlocal fermionic kernel modulo local terms. It has the standard power-law form of a two-dimensional conformal primary \cite{Belavin:1984vu} with principal-series dimension $\Delta_\nu=1+\ii\nu$ and spin magnitude $|\mathfrak s|=\tfrac12$, and it mixes the two Milne-phase sectors off diagonally. We also construct regular source-normalized spinor wavefunctions in the Poincar\'e patch and as a global harmonic expansion. Covariance and uniqueness of the regular boundary-value problem establish their conformal-primary transformation law. The normalization is defined by a delta-functional source in the chosen boundary Weyl frame rather than by a scattering-state inner product.

At a schematic level, the separated-point result is a bilinear in the two phase-labeled source pairs. Retaining only its nonlocal part, we find
\begin{align}
\left.
\mathcal F_{\bar s r}^{\rm asy}
[\mathbf J,\bar{\mathbf J}]
\right.
={}&
\int_0^\infty\frac{\dd\nu}{2\pi}
\int_{S^2}\dd\Omega_2
\int_{S^2}\dd\Omega'_2
\nonumber\\
&\times
\Big[
\bar{\widetilde J}(\Omega,\nu)\,
\mathcal M_\nu(\Omega,\Omega')\,
J(\Omega',\nu)
\nonumber\\
&\hspace{17mm}
+
\bar J(\Omega,\nu)\,
\mathcal M_\nu(\Omega,\Omega')\,
\widetilde J(\Omega',\nu)
\Big].
\label{eq:flatCFT_schematic_intro}
\end{align}
Here $\mathcal F_{\bar s r}^{\rm asy}$ denotes the intrinsic asymptotic coefficient functional: an independent barred coefficient source is paired with the intrinsic unbarred response coefficient. This definition extracts the coefficients before forming the bilinear. It is not the nonlocal part of the complete finite-cutoff four-component contraction. Equation~\eqref{eq:flatCFT_schematic_intro} defines its separated-point nonlocal part, modulo local contact terms and up to the overall action normalization carried by $\kappa_{\rm P,G}$.

It is useful to distinguish this restricted functional from the complete
on-shell functional. Since a symmetrized Dirac action contains both boundary
orderings, the full cutoff problem must be renormalized in terms of the actual
fixed projected fields rather than only the intrinsic coefficients.
Subsection~\ref{subsec:ads3-renormalization} and appendix~\ref{app:counterterm-audit}
derive the complementary barred response and the required complete-source
subtraction. Here equal weight means that the independent barred and unbarred
fixed sources both scale as $\rho^{1-\mu}$; this complexified prescription
is distinct from imposing a Lorentzian adjoint on the reconstructed field.
In the strip $0<\Re\mu<\tfrac12$ the only divergent term of the
complete equal-weight problem is the derivative-zero cross-block contribution.
For a phase $p=\pm$ it is removed by
\begin{equation}
I_{\rho,{\rm ct}}^{(0),p}
=-\kappa\,\sigma_p
\int_{\Sigma_\rho}\dd^2x\,\sqrt{\gamma_\rho}\,
\bar\Phi_{{\rm src},\rho}^{\,-p}\Gamma^{\hat0}
\Phi_{{\rm src},\rho}^{\,p},
\qquad \sigma_-=-1,\quad\sigma_+=+1.
\label{eq:intro-complete-counterterm}
\end{equation}
After this subtraction the strip limit exists. With the same normalization
$\kappa$ used in the intrinsic coefficient pairing, its separated-point
nonlocal part is
\begin{equation}
\left.I_{\rm ren}^{\rm sym}\right|_{\rm nonlocal}
=2\,\mathcal F_{\bar s r}^{\rm asy},
\label{eq:renormalized-asymptotic-completion}
\end{equation}
modulo finite local contact terms; the factor two follows from the exact
full-source algebra and may be absorbed into the overall action normalization
if that normalization has not yet been fixed. The renormalized result is then
continued to $\mu=\ii\nu$. Equation~\eqref{eq:intro-complete-counterterm} is
local on the celestial cutoff at fixed Mellin phase, but after the two phases
are recombined it contains $\operatorname{sgn}(-\ii\partial_{\log\tau})$ and
is therefore nonlocal in Milne time. Accordingly, this establishes a complete
spectral/celestial radial renormalization, not a conventional local
three-dimensional counterterm scheme for the unreduced Milne-boundary field.

The present paper continues to use $\mathcal F_{\bar s r}^{\rm asy}$ to denote
the intrinsic separated-point source--response coefficient because this is the
quantity from which the planar and spherical kernels are displayed. We do not
identify $\exp(\ii\mathcal F_{\bar s r}^{\rm asy})$ with a complete CCFT
generating functional. The unfolding $\lambda=-\sigma_p\nu$ fixes the signed Mellin label.
The remaining scattering problem is the global map from phase-resolved data
$(p,\nu)$ to the independent labels $(\eta,\lambda)$, including continuation
to $\mathcal I^\pm$, the $\ii0$ prescription, scattering normalization, and
the conserved Dirac inner product; the Milne phase is not an energy-sign label.

The remainder of the paper is organized as follows. In section~\ref{sec:milne}, we review the Milne slicing and formulate the Dirac equation on $\mathbb R^+ \times \mathbb H^3$. In section~\ref{sec:solution}, we solve the system in Poincar\'e and global coordinates while retaining the mass parameter in the mode analysis. In section~\ref{sec:correlator}, we specialize to the massless sector and determine the nonlocal kernels. In section~\ref{sec:cpw}, we construct the spinor CPWs in planar and global harmonic forms. We conclude in section~\ref{sec:conclusion} with an outlook and a discussion of open problems.

\section{Hyperbolic slicing of Minkowski spacetime}\label{sec:milne}

In standard Cartesian coordinates, the Minkowski metric is
\begin{equation}
ds^2=\eta_{\mu\nu}\dd X^\mu \dd X^\nu=-(\dd X^0)^2+(\dd X^1)^2+(\dd X^2)^2+(\dd X^3)^2,
\end{equation}
and the light cone defined by $X^2=0$ naturally divides $\mathbb R^{1,3}$ into three regions:
\begin{equation}
\begin{aligned}
&D: \quad && X^2>0,\\
&A_{+}: &&X^2<0, \quad X^0>0,\\
&A_{-}: &&X^2<0, \quad X^0<0.
\end{aligned}
\end{equation}
Each region admits a foliation by constant-curvature surfaces, reflecting the $SO(1,3)$ symmetry. In region $D$, the leaves are three-dimensional de Sitter spacetimes of radius $R$, satisfying $X^2=R^2$. In $A_{\pm}$, they are three-dimensional Euclidean anti-de Sitter spaces of radius $\tau$, satisfying $X^2=-\tau^2$.

We focus on the region $A_{+}$, in which the Minkowski metric takes the Milne form
\begin{equation}
ds^2 = -\dd\tau^2 + \tau^2 ds^2_{\mathbb H^3}.
\end{equation}
Thus, the future Milne wedge $A_+$ is foliated by hyperbolic ($\mathrm{AdS}_3$) slices; the past wedge $A_-$ admits an analogous foliation. The unit metric $ds^2_{\mathbb H^3}$ may be expressed in any convenient coordinate system. The Milne time $\tau$ labels the slices and sets their curvature radius. This foliation provides the geometric basis for reorganizing four-dimensional data in terms of fields on the two-dimensional celestial sphere introduced below.

Two coordinate systems are useful for this reduction. In terms of the Cartesian coordinates $X^\mu$, the Milne coordinates $(\tau,x_0,x_1,x_2)$ use Poincar\'e coordinates on the $\mathbb H^3$ slices:
\begin{align}
&X^0=\tau\;\frac{1+x_0^2+x_1^2+x_2^2}{2 x_0},\\
&X^1=\tau\;\frac{x_1}{x_0},\\
&X^2=\tau\;\frac{x_2}{x_0},\\
&X^3=\tau\;\frac{1-x_0^2-x_1^2-x_2^2}{2 x_0}.
\end{align}
The coordinate ranges are
\begin{equation}
\tau>0,\qquad x_0>0,\qquad (x_1,x_2)\in\mathbb R^2.
\end{equation}
In these coordinates, the Minkowski metric becomes
\begin{equation} \label{Poincare-Milne}
ds^2=-\dd\tau^2+ {\tau^2}\left(\frac{\dd x_0^2+\dd x_1^2+\dd x_2^2}{x_0^2}\right).
\end{equation}
Alternatively, use Milne coordinates $(\tau,y,\theta,\phi)$ with global coordinates $(y,\theta,\phi)$ on the same $\mathbb H^3$ slices:
\begin{align}
&X^0=\tau\;\cosh{y},\\
&X^1=\tau\;\sin{\theta}\;\sin{\phi}\;\sinh{y},\\
&X^2=\tau\;\sin{\theta}\;\cos{\phi}\;\sinh{y},\\
&X^3=\tau\;\cos{\theta}\sinh{y},
\end{align}
with ranges
\begin{equation}
\tau>0,\qquad y\geq0,\qquad 0\leq\theta\leq\pi,\qquad 0\leq\phi<2\pi.
\end{equation}
Our angular embedding uses
\begin{align}
&\widehat n(\theta,\phi)
=
\bigl(\sin\theta\sin\phi,\,\sin\theta\cos\phi,\,\cos\theta\bigr)
=
R_{12}\widehat n_{\rm std}(\theta,\phi),\\
&R_{12}=
\begin{pmatrix}
0&1&0\\
1&0&0\\
0&0&1
\end{pmatrix},
\qquad \det R_{12}=-1,
\end{align}
where $\widehat n_{\rm std}=(\sin\theta\cos\phi,\sin\theta\sin\phi,\cos\theta)$ is the conventional spherical assignment. Thus our ambient Cartesian coordinates interchange the $X^1$ and $X^2$ axes relative to the standard assignment. Since $R_{12}\in O(3)\setminus SO(3)$, we use it only to specify the ambient-coordinate embedding and do not implement it as a $\mathrm{Spin}(3)$ transformation on spinors. The spin orientation used below is defined intrinsically by the ordered orthonormal coframe $(e^{\hat y},e^{\hat\theta},e^{\hat\phi})$; all gamma matrices and spinor harmonics refer to this intrinsic frame. In these global coordinates, the metric takes the form
\begin{equation} \label{Global-Milne}
ds^2=-\dd\tau^2+ \tau^2\left(\dd y^2+\sinh^2 y\,(\dd\theta^2 + \sin^2\theta\,\dd\phi^2)\right).
\end{equation}
\subsection{Conformal boundary and the celestial sphere}
\label{subsec:celestial-sphere}
In the Milne slicing,
\[
ds^2=-\dd\tau^2+\tau^2\,ds^2_{\mathbb H^3},
\]
each constant-$\tau$ slice is a unit $\mathbb H^3$ up to the Weyl factor $\tau^2$. Since a conformal boundary is defined only up to Weyl rescalings, the conformal boundaries of all $\tau$ slices are naturally identified. This common codimension-two boundary is called the \emph{celestial sphere}.
An invariant way to describe it is as the space of future null directions, i.e.\ the projective light cone in $\mathbb R^{1,3}$ \cite{Sleight:2023ojm}:
\begin{equation}
\label{eq:celestial-projective-cone}
\mathcal C \;\equiv\;
\Big\{P^A\neq0 \ \big|\ P^2=0,\ P^0>0\Big\}\Big/\sim,
\qquad
P^A\sim \alpha P^A,\ \ \alpha\in\mathbb R_{>0}.
\end{equation}
Choosing a section of the projective cone (a ``gauge'' for the null ray) identifies $\mathcal C$ with a two-sphere. In the embedding description of $\mathbb H^3$ as $X^2=-1$ in $\mathbb R^{1,3}$, the $\mathbb H^3$ conformal boundary is precisely the set of null rays approached by $X^A$ at infinity.
In global coordinates on $\mathbb H^3$, with $X^A(y,\Omega)=(\cosh y,\sinh y\,\hat n(\Omega))$ and $\Omega\in S^2$, the boundary point $\Omega$ corresponds to the null ray
\begin{equation}
\label{eq:P-global-celestial}
P^A(\Omega)\ \sim\ (1,\hat n(\Omega)),
\qquad \hat n(\Omega)\in S^2\subset\mathbb R^3.
\end{equation}
In the Poincar\'e patch of $\mathbb H^3$, with boundary coordinates $\vec{x}\in\mathbb R^2$, a standard null representative is
\begin{equation}
\label{eq:P-poincare-celestial}
P^A(\vec{x})=\big(1+|\vec{x}|^2,\ 2x_1,\ 2x_2,\ 1-|\vec{x}|^2\big),
\qquad P(\vec{x})^2=0,
\end{equation}
which is related to \eqref{eq:P-global-celestial} by stereographic projection. Thus, the celestial sphere may be viewed either globally as $S^2$, the boundary of global $\mathbb H^3$, or locally as $\mathbb R^2$ together with the point at infinity, the boundary of Poincar\'e $\mathbb H^3$.

\section{Dirac equation in Milne coordinates}
\label{sec:solution}

We work on the Milne wedge of four-dimensional Minkowski space,
\begin{equation}
\label{eq:Milne-general}
ds^2=-\dd\tau^2+\tau^2\,ds^2_{\mathbb H^3},\qquad \tau>0,
\end{equation}
where constant-$\tau$ slices are unit hyperbolic space $\mathbb H^3$.
Two coordinate patches on $ds^2_{\mathbb H^3}$ will be used below: the upper half--space (``Poincar\'e'') patch \eqref{Poincare-Milne} and the global patch \eqref{Global-Milne}. All frame and spin-connection conventions are collected in appendix~\ref{app:vielbein}.

A massive spinor field $\psi$ is described by
\begin{equation}
\label{eq:action}
S=\mathcal N_S\int \dd^4x\,\sqrt{-g}\;\bar\psi\,(\slashed{\nabla}-m)\psi,
\end{equation}
where $\mathcal N_S$ is an overall complex normalization. We absorb its phase, including the conventional Lorentzian factor of $\ii$, into $\mathcal N_S$ and into the constants multiplying the on-shell functionals below. This does not affect the classical Dirac equation; the phase is fixed only when the real-time generating functional and its reality convention are specified. We suppress $\mathcal N_S$ in intermediate variational formulas. The Dirac equation is
\begin{equation}
\label{eq:dirac-eq}
(\slashed{\nabla}-m)\psi=0,
\qquad
\slashed{\nabla}=\Ga^\mu\nabla_\mu
=\Ga^\mu(\partial_\mu+\Omega_\mu).
\end{equation}
We work in Lorentzian signature $(-,+,+,+)$ with $\{\Gamma^a,\Gamma^b\}=2\eta^{ab}$. Thus, $(\Gamma^{\hat0})^2=-1$ and the spatial frame matrices square to $+1$. Equations \eqref{eq:action} and \eqref{eq:dirac-eq} define our Dirac-operator and mass conventions; all subsequent signs and adjoint relations refer to this choice. For a Milne metric of the form \eqref{eq:Milne-general}, the operator separates as
\begin{equation}
\label{eq:dirac-milne-separated}
\slashed{\nabla}
=
\Ga^{\hat 0}\Big(\partial_\tau+\frac{3}{2\tau}\Big)
+\frac{1}{\tau}\,\mathcal D_{\mathbb H^3},
\end{equation}
where $\mathcal D_{\mathbb H^3}$ is the intrinsic Dirac operator on the unit $\mathbb H^3$ slice (see appendix~\ref{app:vielbein} for explicit forms in each patch).
Introducing the standard Milne rescaling
\begin{equation}
\label{eq:milne-rescaling}
\psi(\tau,x)=\tau^{-3/2}\,\phi(\tau,x),
\end{equation}
the equation becomes
\begin{equation}
\label{eq:dirac-rescaled}
\Bigl(\Ga^{\hat 0}\partial_\tau+\frac{1}{\tau}\mathcal D_{\mathbb H^3}-m\Bigr)\phi=0.
\end{equation}
It is also convenient to square the operator. With the curvature convention
$R[S^2]=2$ and our positive spatial Clifford squares, the Lichnerowicz formula is
\begin{equation}
\label{eq:lichnerowicz}
(\Ga^\mu\nabla_\mu)^2=-\Delta_B-\frac{1}{4}R,
\qquad
\Delta_B\psi=
-\,g^{\mu\nu}\Big(\nabla_\mu\nabla_\nu-\Gamma^\rho{}_{\mu\nu}\nabla_\rho\Big)\psi,
\end{equation}
so, using $R=0$ for Minkowski space, one obtains
\begin{equation}
\label{eq:dirac-square}
(\slashed{\nabla}-m)(\slashed{\nabla}+m)\psi=(-\Delta_B-m^2)\psi.
\end{equation}
In terms of $\phi$ this yields
\begin{equation}
\label{eq:squared-phi}
\Bigl(
-\partial_\tau^2+ \frac{1}{\tau^2}\mathcal D_{\mathbb H^3}^2-\frac{1}{\tau^2}\Ga^{\hat0}\mathcal D_{\mathbb H^3}
-m^2
\Bigr)\phi=0.
\end{equation}
The remaining task is the same in any $\mathbb H^3$ chart: expand $\phi$ in eigenmodes of $\mathcal D_{\mathbb H^3}$ and solve the resulting time-dependent ODEs. We now do this in the two coordinate systems.

\subsection{Poincar\'e coordinates on \texorpdfstring{$\mathbb H^3$}{H3}}

We take the Milne metric in the upper half--space chart,
\begin{equation}\label{Poincare-Milne-again}
ds^2=-\dd\tau^2+ {\tau^2}\left(\frac{\dd x_0^2+\dd x_1^2+\dd x_2^2}{x_0^2}\right),
\end{equation}
with
\begin{equation}
\tau>0,\qquad x_0>0,\qquad (x_1,x_2)\in\mathbb R^2.
\end{equation}
The explicit coframe, spin connection, and the corresponding
$\mathcal D_{\mathbb H^3}$ are summarized in appendix~\ref{app:vielbein}.
For the plane-wave representatives written below we take
\begin{equation}
\lambda\in(0,\infty),\qquad k\in\mathbb R^2\setminus\{0\}.
\end{equation}
The zero-momentum modes are obtained directly from the radial equation and are recorded in appendix~\ref{app:poincare-zero-mode}; the isolated point $k=0$ has zero measure in the plane-wave expansion. The threshold $\lambda=0$, at which the $s=\pm1$ labels coincide, is understood as a separate limiting generalized mode and carries no discrete Plancherel contribution.

We choose the Poincar\'e eigenspinors in one irreducible two-component spatial Clifford sector, denoted by the projector $\Pi_{\rm A}$. For each $k\neq0$, that sector contains one polarization for each sign $s=\pm1$:
\begin{equation}
\label{eq:H3-eigs-poincare}
\mathcal D_{\mathbb H^3}\,\chi^{(s)}_{\lambda,k}=\ii\;s\lambda\,\chi^{(s)}_{\lambda,k},
\qquad
s=\pm1.
\end{equation}
The complementary Clifford sector is generated by $\Gamma^{\hat0}\chi^{(s)}_{\lambda,k}$. Explicit representatives are given in appendix~\ref{app:dirac-modes-poincare}; their continuous-spectrum normalization is derived in appendix~\ref{app:plancherel}. Expanding
\begin{equation}
\label{eq:mode-ansatz-uv}
\phi(\tau,x)=u(\tau)\,\chi^{(s)}_{\lambda,k}(x)+v(\tau)\,\Ga^{\hat0}\chi^{(s)}_{\lambda,k}(x),
\end{equation}
and substituting into \eqref{eq:dirac-rescaled} yields
\begin{equation}
\label{eq:uv-system}
\begin{cases}
&u'(\tau) = \Bigl(m+\dfrac{\ii s\lambda}{\tau}\Bigr)\,v(\tau),\\[6pt]
&v'(\tau) = -\Bigl(m-\dfrac{\ii s\lambda}{\tau}\Bigr)\,u(\tau).
\end{cases}
\end{equation}
As in the global analysis below, it is convenient to work with the combinations
\begin{equation}
\label{eq:eplusminus-def}
e^{(s)}_{\pm}(\lambda,k;x)=\chi^{(s)}_{\lambda,k}(x)\pm \Ga^{\hat0}\chi^{(s)}_{\lambda,k}(x),
\qquad
\phi(\tau,x)=f_+(\tau)\,e^{(s)}_{+}+f_-(\tau)\,e^{(s)}_{-}.
\end{equation}
Equivalently $u=f_+ + f_-$ and $v=f_+ - f_-$. In this basis the squared equation \eqref{eq:squared-phi} diagonalizes, and one finds
\begin{equation}
\label{eq:fpm-odes}
\begin{cases}
&f_+''(\tau)+\Bigl(m^2+\dfrac{\lambda^2+\ii s\lambda}{\tau^2}\Bigr)f_+(\tau)=0,\\[6pt]
&f_-''(\tau)+\Bigl(m^2+\dfrac{\lambda^2-\ii s\lambda}{\tau^2}\Bigr)f_-(\tau)=0.
\end{cases}
\end{equation}
Defining the (branch-dependent) Bessel orders
\begin{equation}
\label{eq:nu-orders-unified}
\nu_{+}(s,\lambda)=\half-\ii s\lambda,\qquad
\nu_{-}(s,\lambda)=\half+\ii s\lambda,
\end{equation}
a convenient basis of time-dependent solutions is
\begin{equation}
\label{eq:fpm-sol}
\begin{aligned}
f_+(\tau)&=\sqrt{\tau}\Big(C_1\,J_{\nu_+(s,\lambda)}(m\tau)+C_2\,Y_{\nu_+(s,\lambda)}(m\tau)\Big),\\
f_-(\tau)&=\sqrt{\tau}\Big(D_1\,J_{\nu_-(s,\lambda)}(m\tau)+D_2\,Y_{\nu_-(s,\lambda)}(m\tau)\Big),
\end{aligned}
\end{equation}
where $J_\nu$ and $Y_\nu$ are Bessel functions. Although the solution contains four integration constants, the original first-order Dirac equation imposes two linear relations among them. Equivalently, for $m\neq0$, $f_{-}$ is determined by $f_{+}$ through
\begin{equation}
f_{-}=\frac{1}{m}\left(\frac{\ii s \lambda}{\tau}f_{+}-f_{+}' \right).
\end{equation}
For $m=0$, the branch coefficients $f_\pm$, which diagonalize
$\Gamma^{\hat0}\mathcal D_{\mathbb H^3}$ in the basis
\eqref{eq:eplusminus-def}, decouple:
\[
f_+'=\frac{\ii s\lambda}{\tau}f_+,
\qquad
f_-'=-\frac{\ii s\lambda}{\tau}f_-,
\]
and hence
\[
f_+(\tau)=C_+\,\tau^{\ii s\lambda},
\qquad
f_-(\tau)=C_-\,\tau^{-\ii s\lambda}.
\]
Combining \eqref{eq:milne-rescaling} with \eqref{eq:eplusminus-def} gives the separated solutions
\begin{equation}
\label{eq:psi-separated-poincare}
\psi^{(s)}_{\lambda,k}(\tau,x)
=
\tau^{-3/2}\Big[
f_+(\tau)\,e^{(s)}_{+}(\lambda,k;x)
+
f_-(\tau)\,e^{(s)}_{-}(\lambda,k;x)
\Big].
\end{equation}
The two signs $s=\pm1$ are retained: with the explicit irreducible spatial-sector construction of appendix~\ref{app:dirac-modes}, they supply the two independent Clifford-sector pairs and are not redundant. In the delta-normalized convention of appendix~\ref{app:plancherel}, a complete generalized mode expansion is
\begin{equation}
\label{eq:psi-general-poincare}
\psi(\tau,x)
=
\sum_{s=\pm1}
\int_{0}^{\infty}\!\dd\lambda
\int_{\mathbb R^2}\!\frac{\dd^2k}{(2\pi)^2}\;
\Big[
A_s(\lambda,k)\,\psi^{(s)}_{\lambda,k}(\tau,x)
+
B_s(\lambda,k)\,\widetilde{\psi}{}^{(s)}_{\lambda,k}(\tau,x)
\Big],
\end{equation}
where $\widetilde{\psi}{}^{(s)}_{\lambda,k}$ denotes the second independent time solution. Equivalently, if the unnormalized regular eigenspinors are used, the normalization density displayed in appendix~\ref{app:plancherel} is included in the integration measure. The coefficients are fixed by initial or boundary data.

\subsection{Global coordinates on \texorpdfstring{$\mathbb H^3$}{H3}}

We now take the Milne metric in global $\mathbb H^3$ coordinates,
\begin{equation}\label{Global-Milne-again}
ds^2=-\dd\tau^2+ \tau^2\left(\dd y^2+\sinh^2 y\,(\dd\theta^2 + \sin^2\theta\,\dd\phi^2)\right),
\end{equation}
with
\begin{equation}
\tau>0,\qquad y\geq0,\qquad 0\leq\theta\leq\pi,\qquad 0\leq\phi<2\pi.
\end{equation}
The intrinsic Dirac operator $\mathcal D_{\mathbb H^3}$ and the construction and normalization of a separated eigenbasis are collected in appendices~\ref{app:dirac-modes} and~\ref{app:plancherel}.
We use global $\mathbb H^3$ eigenspinors $\chi^{(s)}_{\lambda l m_j}$, solving
\begin{equation}
\label{eq:H3-eigs-global}
\mathcal D_{\mathbb H^3}\,\chi^{(s)}_{\lambda l m_j}(y,\Omega)
=
\ii\;s\lambda\;\chi^{(s)}_{\lambda l m_j}(y,\Omega),
\qquad
\lambda\in(0,\infty),\quad s=\pm1,\quad l\in\mathbb Z_{\ge0},
\end{equation}
with $\Omega=(\theta,\phi)$ and
\begin{equation}
m_j=-l-\tfrac12,-l+\tfrac12,\ldots,l+\tfrac12.
\end{equation}
The threshold $\lambda=0$ is obtained by a limiting prescription; at that point $\chi^{(+)}_{0 l m_j}=\chi^{(-)}_{0 l m_j}$, so it is not counted twice and gives no separate discrete term in the Plancherel decomposition.
Expanding $\phi$ as in the Poincar\'e case,
\begin{equation}
\label{eq:mode-ansatz-global}
\phi(\tau,y,\Omega)=u(\tau)\,\chi^{(s)}_{\lambda l m_j}(y,\Omega)
+v(\tau)\,\Ga^{\hat0}\chi^{(s)}_{\lambda l m_j}(y,\Omega),
\end{equation}
one obtains the same time-dependent system \eqref{eq:uv-system}, with $\lambda$ and $s$ as above. Define again
\begin{equation}
\label{eq:eplusminus-def-global}
\begin{aligned}
e^{(s)}_{\pm}(\lambda,l,m_j;y,\Omega)
&=
\chi^{(s)}_{\lambda l m_j}(y,\Omega)
\pm \Ga^{\hat0}\chi^{(s)}_{\lambda l m_j}(y,\Omega),
\\
\phi(\tau,y,\Omega)
&=f_+(\tau)e^{(s)}_{+}+f_-(\tau)e^{(s)}_{-}.
\end{aligned}
\end{equation}
The decoupled equations for $f_\pm$ are still \eqref{eq:fpm-odes}, hence the time dependence is again given by \eqref{eq:fpm-sol} with the same unified orders \eqref{eq:nu-orders-unified}.

The separated solutions take the same form as \eqref{eq:psi-separated-poincare}
\begin{equation}
\label{eq:psi-separated-global}
\psi^{(s)}_{\lambda l m_j}(\tau,y,\Omega)
=
\tau^{-3/2}\Big[
f_+(\tau)\,e^{(s)}_{+}(\lambda,l,m_j;y,\Omega)
+
f_-(\tau)\,e^{(s)}_{-}(\lambda,l,m_j;y,\Omega)
\Big].
\end{equation}
Because the explicit harmonics in appendix~\ref{app:dirac-modes-global} are constructed in one irreducible spatial Clifford sector, both $s=+1$ and $s=-1$ families must be retained. Together with their $\Gamma^{\hat0}$ partners in \eqref{eq:mode-ansatz-global}, they span the two independent Clifford-sector pairs. In the delta-normalized convention of appendix~\ref{app:plancherel}, the complete generalized expansion is
\begin{equation}
\label{eq:psi-general-global}
\begin{aligned}
\psi(\tau,y,\Omega)
={}&
\sum_{s=\pm1}
\int_{0}^{\infty}\!\dd\lambda\;
\sum_{l=0}^{\infty}
\sum_{m_j=-l-\frac{1}{2}}^{l+\frac{1}{2}}
\Big[
A_{s;\lambda l m_j}\,\psi^{(s)}_{\lambda l m_j}(\tau,y,\Omega)
\\[-1mm]
&\hspace{44mm}
+
B_{s;\lambda l m_j}\,\widetilde\psi^{(s)}_{\lambda l m_j}(\tau,y,\Omega)
\Big].
\end{aligned}
\end{equation}
Here $\widetilde\psi^{(s)}_{\lambda l m_j}$ denotes the second independent time solution. If the regular-at-origin modes without the factor $c_3(\lambda,l)$ are used instead, the corresponding normalization density in appendix~\ref{app:plancherel} is included in the measure. Explicit hypergeometric expressions for the $\mathbb H^3$ radial profiles are given in appendix~\ref{app:dirac-modes}.

\section{Celestial spinor source--response kernels}
\label{sec:correlator}

\subsection{Milne phases and Mellin pairing}
\label{subsec:milne-mellin}

We now restrict to the massless sector and give the main steps of the construction. In this section, we rename the positive spectral parameter $\lambda$ of section~\ref{sec:solution} as $\nu$, the conventional principal-series label. The body of the paper retains the variational input, the Mellin selection rule, and the source--response extraction that produce the kernels. Appendix~\ref{app:coefficient-kernels} preserves the full finite-cutoff derivation, including the source--response maps evaluated separately in the two irreducible Clifford blocks, subleading terms, and distributional details. The rescaled field contains two Milne phase branches,
\begin{equation}
\phi(\tau,x)
=
\int_0^\infty\frac{\dd\nu}{2\pi}
\left[
\tau^{-\ii\nu}\phi_\nu^{(-)}(x)
+
\tau^{+\ii\nu}\phi_\nu^{(+)}(x)
\right],
\qquad \nu>0.
\label{eq:main-phase-expansion}
\end{equation}
The labels $-$ and $+$ refer only to the displayed powers of $\tau$; they do not by themselves distinguish incoming from outgoing data. Treating the barred and unbarred fields as independent Grassmann variables, we expand both in the same phase basis. The relevant Mellin integrals are
\begin{align}
\int_0^\infty\frac{\dd\tau}{\tau}\,
\tau^{-\ii\nu}\tau^{+\ii\nu'}
&=2\pi\delta(\nu-\nu'),
\label{eq:main-opposite-phase-pairing}
\\
\int_0^\infty\frac{\dd\tau}{\tau}\,
\tau^{\mp\ii\nu}\tau^{\mp\ii\nu'}
&=2\pi\delta(\nu+\nu').
\label{eq:main-equal-phase-pairing}
\end{align}
For $\nu,\nu'>0$, the second distribution has no support in the open spectral domain. Hence only opposite phases pair, and the quadratic functional is off-diagonal in the phase basis. Smooth compactly supported wave packets in $\log\tau$ justify the intermediate integrations; pure phases are recovered distributionally.

\subsection{Boundary polarization and asymptotic source–response prescription}
\label{subsec:boundary-polarization}

Since the Dirac equation is first order, the radial boundary-value problem requires a polarization: only one projected component is fixed independently at the cutoff surface, while the complementary component is determined by the bulk equation and regularity in the interior. We call the fixed component the \emph{source} and the dependent component the \emph{response}. The barred and unbarred fields are treated as independent Grassmann variables during variation.

The same outward-normal polarization takes different coordinate forms in the two descriptions of $\mathbb H^3$. In the Poincar\'e patch the regulated region is $x_0\geq\epsilon$, so the outward normal points toward decreasing $x_0$. With
\begin{equation}
\Pi_\pm=\frac{1}{2}\left(1\pm\Gamma^{\hat1}\right),
\end{equation}
fixing the outward-$+$ component is equivalent to
\begin{equation}
\Phi_{\rm src}=\phi_-\equiv\Pi_-\phi,
\qquad
\Phi_{\rm resp}=\phi_+\equiv\Pi_+\phi.
\end{equation}
In global coordinates the cutoff surface is $y=Y$ and the outward normal points toward increasing $y$. With
\begin{equation}
P_\pm=\frac{1}{2}\left(1\pm\Gamma^{\hat y}\right),
\end{equation}
the corresponding choice is
\begin{equation}
\Phi_{\rm src}=\phi_+\equiv P_+\phi,
\qquad
\Phi_{\rm resp}=\phi_-\equiv P_-\phi.
\end{equation}
The apparent interchange of coordinate projectors is therefore entirely an orientation effect.

To distinguish the complete on-shell action from an individual boundary contraction, we denote the former by $I$ and the latter by $\mathcal B$. The regulated contraction in which an independent barred source multiplies the dependent unbarred response is
\begin{equation}
\mathcal B_{\epsilon;\bar s r}
=
\begin{cases}
\displaystyle
\kappa_{\rm P}
\int_0^\infty\frac{\dd\tau}{\tau}
\int_{\mathbb R^2}\dd^2\vec x\,
\epsilon^{-2}\,
\bar\phi\,\Pi_+\phi,
& \text{Poincar\'e},
\\[4mm]
\displaystyle
\kappa_{\rm G}
\int_0^\infty\frac{\dd\tau}{\tau}
\int_{S^2}\dd\Omega_2\,
\sinh^2Y\,
\bar\phi\,P_-\phi,
& \text{global}.
\end{cases}
\label{eq:main-barsr-pairing}
\end{equation}
The complementary ordering is
\begin{equation}
\mathcal B_{\epsilon;\bar r s}
=
\begin{cases}
\displaystyle
\kappa_{\rm P}
\int_0^\infty\frac{\dd\tau}{\tau}
\int_{\mathbb R^2}\dd^2\vec x\,
\epsilon^{-2}\,
\bar\phi\,\Pi_-\phi,
& \text{Poincar\'e},
\\[4mm]
\displaystyle
\kappa_{\rm G}
\int_0^\infty\frac{\dd\tau}{\tau}
\int_{S^2}\dd\Omega_2\,
\sinh^2Y\,
\bar\phi\,P_+\phi,
& \text{global}.
\end{cases}
\label{eq:main-barrs-pairing}
\end{equation}
The projector assignments for barred spinors and the variational origin of these two contractions are derived in appendix~\ref{app:dirac-boundary}.

The bare symmetrized on-shell functional is
\begin{equation}
I_{\epsilon,{\rm os}}^{\rm sym}
=
\frac{1}{2}\left(
\mathcal B_{\epsilon;\bar s r}
+
\mathcal B_{\epsilon;\bar r s}
\right).
\label{eq:main-sym-action-pairings}
\end{equation}
Before removing the cutoff, we allow a counterterm functional that cancels
radial divergences and define the renormalized cutoff functional by
\begin{equation}
I_{\epsilon,{\rm ren}}^{\rm sym}
=
I_{\epsilon,{\rm os}}^{\rm sym}
+
I_{\epsilon,{\rm ct}}^{\rm div}.
\label{eq:main-ren-action}
\end{equation}
For the intrinsic fixed-Clifford-block map, the divergent part of the
odd-derivative counterterm operator is empty:
\begin{equation}
C_{\gamma,\mathrm{intr}}^{\rm div}=0.
\label{eq:main-no-divergent-ct}
\end{equation}
This does not imply $I_{\epsilon,{\rm ct}}^{\rm div}=0$ for the complete
four-component action. The algebraic cross-block term in the full branch
map is not covered by the intrinsic recursion. Subsection~\ref{subsec:ct-full-pairing}
derives the barred radial branch and shows that in the equal-weight
fixed-Mellin problem the complete divergent operator is instead
$C_{\gamma,p}^{\rm div}=-\sigma_p\Gamma^{\hat0}$.

We compute the intrinsic separated-point coefficient associated with $\mathcal B_{\epsilon;\bar s r}$ explicitly. Regularity determines the unbarred response from the unbarred source. The complementary response obtained from the left-acting barred Dirac equation, including the momentum reversal and matrix ordering, is derived separately in subsection~\ref{subsec:ct-full-pairing}; it is needed to assemble and renormalize the complete symmetrized functional.

The source--response split on the principal-series line requires an analytic definition. For real $\nu$, the two radial behaviors $\rho^{1-\ii\nu}$ and $\rho^{1+\ii\nu}$ have the same real part. We therefore continue
\begin{equation}
\ii\nu\longrightarrow\mu,
\qquad
0<\Re\mu<\frac{1}{2},
\label{eq:main-continuation-strip}
\end{equation}
so that the two powers
\begin{equation}
\rho^{1-\mu},
\qquad
\rho^{1+\mu}
\end{equation}
are strictly ordered by real part. Here $\rho=x_0$ in the Poincar\'e patch and $\rho=2e^{-y}$ in global coordinates. The coefficient of $\rho^{1-\mu}$ defines the source and the coefficient of $\rho^{1+\mu}$ defines the response. Only after these coefficients have been separated do we return to the physical line $\mu=\ii\nu$ from $\Re\mu>0$.

For any quantity with a radial expansion
\begin{equation}
X(\rho)\sim\sum_a\rho^a X_{(a)},
\end{equation}
whose powers are distinct in the continuation strip, we write
\begin{equation}
\bigl[X\bigr]_{\rho^a}\equiv X_{(a)}.
\label{eq:asymptotic-component-notation}
\end{equation}
On the principal-series line this bracket means analytic continuation of the coefficient from the strip; it is not a pointwise radial limit.

For the intrinsic fixed-block problem, the exact regular source--response map contains a finite nonanalytic term together with a local source-branch derivative expansion. At derivative weight $2n+1$, the latter contributes to the regulated quadratic functional with the radial factor
\begin{equation}
\rho^{2n+1-2\mu}.
\label{eq:main-local-radial-scaling}
\end{equation}
Every such term vanishes in the strip \eqref{eq:main-continuation-strip}, and after continuation to $\mu=\ii\nu$ its modulus is $\rho^{2n+1}$. Hence this intrinsic odd-derivative sector has no divergent terms to subtract. The formal local recursion does not terminate in general, but its terms are cutoff corrections rather than required counterterms in the physical principal-series problem. A detailed derivation, including the planar Riccati recursion and its spherical curvature completion, is given in appendix~\ref{app:counterterm-audit}.

Finite local functionals of the asymptotic sources may still be added as contact-term scheme choices. They do not affect separated boundary points. We therefore denote by $\mathcal F_{\bar s r}^{\rm asy}$ the separated-point coefficient extracted from the intrinsic fixed-block map, modulo finite contact terms. The complete four-component problem is renormalized separately in subsection~\ref{subsec:ads3-renormalization}; its full-source counterterm changes the relation between the complete on-shell functional and this intrinsic coefficient. The quantity $\exp(\ii\mathcal F_{\bar s r}^{\rm asy})$ is not identified here with a complete CCFT generating functional.

For Grassmann sources we differentiate from the left with respect to barred sources and from the right with respect to unbarred sources. With this convention, a bilinear $\bar J\mathcal M J$ contributes
\begin{equation}
\mathscr M=\frac{\mathcal M}{\ii}
\end{equation}
to the asymptotic source--response kernel. The indexed convention is recorded in appendix~\ref{app:coefficient-kernels}.

The central element of the spatial Clifford algebra decomposes the four-component spinor into two irreducible blocks, denoted by $A$ and $B$. Their projectors and algebraic properties are collected in appendix~\ref{app:coefficient-kernels}. The radial source--response operator preserves these blocks, so the intrinsic separated-point coefficient is block diagonal. This statement excludes the algebraic cross-block terms of the complete cutoff pairing. For a fixed block $I\in\{{\rm A},{\rm B}\}$ we use
\begin{equation}
\mathfrak s_{\rm A}=+\frac{1}{2},
\qquad
\mathfrak s_{\rm B}=-\frac{1}{2},
\end{equation}
in the adopted boundary orientation. The complete result is
\begin{equation}
\left.\mathcal F_{\bar s r}^{\rm asy}\right|_{\rm nonlocal}
=
\mathcal F_{\bar s r,{\rm A}}^{\rm asy}
+
\mathcal F_{\bar s r,{\rm B}}^{\rm asy},
\label{eq:main-block-sum}
\end{equation}
and both blocks have
\begin{equation}
\Delta_\nu=1+\ii\nu,
\qquad
h_I=\frac{\Delta_\nu+\mathfrak s_I}{2},
\qquad
\bar h_I=\frac{\Delta_\nu-\mathfrak s_I}{2}.
\label{eq:main-weights}
\end{equation}
For one block, let $(J,\bar J)$ and $(\widetilde J,\bar{\widetilde J})$ denote the sources associated with the phases $\tau^{-\ii\nu}$ and $\tau^{+\ii\nu}$, respectively. The labels refer only to the displayed Mellin phases. By the selection rule in subsection~\ref{subsec:milne-mellin}, only opposite phases pair, and the finite separated-point functional is
\begin{align}
\mathcal F_{\bar s r}^{\rm asy}
[\mathbf J,\bar{\mathbf J}]
=
\int_0^\infty\frac{\dd\nu}{2\pi}
\int_{S^2}\dd\Omega_2
\int_{S^2}\dd\Omega_2'
&\Big[
\bar{\widetilde J}(\Omega,\nu)
\mathcal M_\nu(\Omega,\Omega')
J(\Omega',\nu)
\nonumber\\
&\quad+
\bar J(\Omega,\nu)
\mathcal M_\nu(\Omega,\Omega')
\widetilde J(\Omega',\nu)
\Big].
\label{eq:main-coefficient-functional}
\end{align}
Equal-phase terms vanish by Mellin orthogonality, while $A$--$B$ cross terms vanish by block diagonality. Subsections~\ref{subsec:planar-kernel} and~\ref{subsec:spherical-kernel} determine $\mathcal M_\nu$ in planar and global representations.

\subsection{Planar kernel}
\label{subsec:planar-kernel}

In the Poincar\'e patch, let $\Pi_\pm=\tfrac12(1\pm\Gamma^{\hat1})$ and $\phi_\pm=\Pi_\pm\phi$. The regulated region is $x_0\geq\epsilon$, so the outward gamma matrix is $-\Gamma^{\hat1}$; fixing the outward $+$ polarization is therefore equivalent to fixing the coordinate component $\phi_-$. For a regular spatial eigenspinor in a fixed Clifford block, the projected response is
\begin{equation}
\chi^\pm_{\nu,\vec k}=\Pi_\pm\chi_{\nu,\vec k},
\qquad
\chi^+_{\nu,\vec k}(\epsilon)
=
A_\epsilon(\vec k,\nu)\,
\chi^-_{\nu,\vec k}(\epsilon),
\qquad
A_\epsilon(\vec k,\nu)
=
\frac{\ii\slashed k}{k}
\frac{K_{-\frac{1}2+\ii\nu}(k\epsilon)}{K_{\frac{1}2+\ii\nu}(k\epsilon)}.
\label{eq:main-planar-cutoff-map}
\end{equation}
The quotient is used where $K_{\frac{1}{2}+\ii\nu}(k\epsilon)\neq0$; at a denominator zero, the equivalent homogeneous source--response relation remains well defined. Here $\kappa_{\rm P}$ below is the patch-specific version of the boundary-action normalization $\kappa$. The $\Gamma^{\hat0}$ partner in the complete four-dimensional branch exchanges the Clifford blocks. It does not alter the designated $\epsilon^{1-\mu}$ source coefficient or $\epsilon^{1+\mu}$ response coefficient; the exact fixed-block and homogeneous statements are given in appendix~\ref{app:coefficient-kernels}.

The continuation strip makes the coefficient extraction explicit. With $z=k\epsilon$,
\begin{equation}
\frac{K_{\frac{1}2-\mu}(z)}{K_{\frac{1}2+\mu}(z)}
=
\frac{\Gamma(\frac{1}2-\mu)}{\Gamma(\frac{1}2+\mu)}
\left(\frac{z}{2}\right)^{2\mu}
+
\frac{z/2}{\mu-\frac{1}2}
+\cdots.
\label{eq:main-bessel-ratio}
\end{equation}
The first term gives the nonanalytic source--response coefficient. The second term and its momentum-polynomial descendants are local contact terms. Including the two source falloffs supplies the factor $\epsilon^{-2\mu}$, so the first term has a finite limit. After isolating the nonanalytic coefficient and taking the cutoff limit in the continuation strip, we continue $\mu\to\ii\nu$ and obtain
\begin{equation}
\mathcal M^{\rm P}(\vec k,\nu)
=
\kappa_{\rm P}
\frac{\ii\slashed k}{k}
\frac{\Gamma(\frac{1}2-\ii\nu)}{\Gamma(\frac{1}2+\ii\nu)}
\left(\frac{k}{2}\right)^{2\ii\nu},
\qquad
\slashed k=k_1\Gamma^{\hat2}+k_2\Gamma^{\hat3}.
\label{eq:main-planar-momentum-kernel}
\end{equation}
The factor $(k/2)^{2\ii\nu}$ uses the real logarithm for $k>0$. Fourier transformation of $\mathscr M^{\rm P}=\mathcal M^{\rm P}/\ii$, modulo contact terms, gives
\begin{equation}
\mathcal K_\nu^{\rm P}(\vec x-\vec y)
=
-\frac{\kappa_{\rm P}}{\ii}\frac{1+2\ii\nu}{2\pi}
\frac{\Gamma^{\hat\alpha}(x-y)_{\hat\alpha}}{|\vec x-\vec y|^{3+2\ii\nu}},
\qquad \hat\alpha=\hat2,\hat3.
\label{eq:main-planar-position-kernel}
\end{equation}
Thus, the only nonzero asymptotic kernels in the computed $\bar s r$ pairing mix the two Milne phases:
\begin{align}
\left\langle\mathcal O(\vec x,\nu)
\bar{\widetilde{\mathcal O}}(\vec y,\nu')\right\rangle_{\rm asy}
&=
2\pi\delta(\nu-\nu')\,
\mathcal K_\nu^{\rm P}(\vec x-\vec y),
\\
\left\langle\widetilde{\mathcal O}(\vec x,\nu)
\bar{\mathcal O}(\vec y,\nu')\right\rangle_{\rm asy}
&=
2\pi\delta(\nu-\nu')\,
\mathcal K_\nu^{\rm P}(\vec x-\vec y).
\label{eq:main-planar-mixed-kernels}
\end{align}

\subsection{Spherical kernel}
\label{subsec:spherical-kernel}

In global coordinates, the conformal boundary is the full celestial sphere $S^2$. We regulate it at $y=Y$, write $\epsilon=2\e^{-Y}$, and use $P_\pm=\tfrac12(1\pm\Gamma^{\hat y})$. The $P_+$ component is fixed and carries the $\epsilon^{1-\ii\nu}$ source coefficient; the dependent $P_-$ component carries the $\epsilon^{1+\ii\nu}$ response coefficient. For the regular harmonic of angular momentum $l$, their asymptotic ratio is
\begin{equation}
\mathcal R_\epsilon(l,\nu)
=
\frac{\Gamma(\frac{1}2-\ii\nu)}{\Gamma(\frac{1}2+\ii\nu)}
\frac{\Gamma(l+\frac{3}2+\ii\nu)}{\Gamma(l+\frac{3}2-\ii\nu)}
\left(\frac{\epsilon}{2}\right)^{2\ii\nu}.
\label{eq:main-global-response-ratio}
\end{equation}
This ratio follows from the two large-$Y$ coefficients of the regular hypergeometric solution; the exact finite-$Y$ quotient and its homogeneous form at possible denominator zeros are retained in appendix~\ref{app:coefficient-kernels}. The cutoff measure obeys $\epsilon^2\sinh^2Y\to1$. Combining it with the two source falloffs and \eqref{eq:main-global-response-ratio} gives the finite harmonic coefficient
\begin{equation}
\mathcal M^{\rm G}(l,\nu)
=
\kappa_{\rm G}2^{-2\ii\nu}
\frac{\Gamma(\frac{1}2-\ii\nu)}{\Gamma(\frac{1}2+\ii\nu)}
\frac{\Gamma(l+\frac{3}2+\ii\nu)}{\Gamma(l+\frac{3}2-\ii\nu)}.
\label{eq:main-global-harmonic-kernel}
\end{equation}
Here $\kappa_{\rm G}$ is the global-frame boundary-action normalization. The displayed coefficients are compatible with the choice $\kappa_{\rm G}=\kappa_{\rm P}$ after matching boundary orientation, spin frame, and source normalization; an independent finite-separation matching is not supplied here.
Let $u_{l m_j}$ and $v_{l m_j}$ be the normalized source and response harmonics defined in \eqref{eq:global-boundary-harmonics}. The position-space kernel is reconstructed as
\begin{equation}
\mathcal K_\nu^{S^2}(\Omega,\Omega')
=
\sum_{l=0}^\infty\sum_{m_j}
\frac{\mathcal M^{\rm G}(l,\nu)}{\ii}\,
v_{l m_j}(\Omega)u_{l m_j}^\dagger(\Omega').
\label{eq:main-global-harmonic-reconstruction}
\end{equation}
The fixed-$l$ spinor addition theorem followed by Abel summation of the remaining Jacobi series, both derived in appendix~\ref{app:modesum}, yields
\begin{equation}
\mathcal K_\nu^{S^2}(\Omega,\Omega')
=
-\frac{\kappa_{\rm G}}{\ii}\frac{1+2\ii\nu}{2\pi}
\frac{
P_-(\Omega)\Gamma^{\hat\alpha}\Xi_{\hat\alpha}(\Omega,\Omega')
\mathcal P(\Omega,\Omega')P_+(\Omega')
}{[2(1-\cos\gamma)]^{1+\ii\nu}},
\qquad 0<\gamma<\pi.
\label{eq:main-sphere-kernel}
\end{equation}
Here $\gamma$ is the geodesic separation, $\Xi_{\hat\alpha}=\nabla_{\hat\alpha}\gamma$, and $\mathcal P(\Omega,\Omega')$ is the spin parallel propagator from $\Omega'$ to $\Omega$. The formula is pointwise valid for $0<\gamma<\pi$; the Abel-regulated harmonic expansion specifies the global distribution, including contact terms and the antipodal point. For the coefficient comparison we choose the common normalization $\kappa_{\rm G}=\kappa_{\rm P}$ with matched orientation and source conventions. Since $2(1-\cos\gamma)\sim\gamma^2$, the short-distance limit of \eqref{eq:main-sphere-kernel} reproduces \eqref{eq:main-planar-position-kernel}.

\subsection{Radial counterterms and the complete spectral functional}
\label{subsec:ads3-renormalization}

At fixed spectral parameter the intrinsic equation on $\mathbb H^3$ is the
Euclidean AdS$_3$ Dirac equation with
\begin{equation}
m_{\rm AdS}L=\ii\nu.
\end{equation}
For the radial analysis we therefore introduce the ordinary real-mass AdS
variable $\mu=m_{\rm AdS}L$, work first in
\begin{equation}
0<\Re\mu<\frac12,
\label{eq:main-ren-strip}
\end{equation}
and analytically continue $\mu\to\ii\nu$ only after the cutoff limit has
been taken. This order of operations is important: a power that decays in the
open strip can become an oscillatory phase on the principal-series line.

By the \emph{equal-weight prescription} we mean the complexified variational
problem in which the independent barred and unbarred fixed sources both
scale as $\rho^{1-\mu}$ in this strip. This is the same-weight source
prescription used for the intrinsic coefficient functional. The Lorentzian
adjoint instead relates conjugate principal-series weights, as well as
opposite phases and complementary spatial Clifford blocks, after the full
spectral field is reconstructed. The complete-source divergence derived
below refers specifically to the equal-weight continuation.

Consider first the planar intrinsic problem in one irreducible spatial
Clifford block. The exact regular source--response map is
\begin{equation}
A_\epsilon(\vec k,\mu)
=
\frac{\ii\slashed k}{k}
\frac{K_{\frac12-\mu}(k\epsilon)}{K_{\frac12+\mu}(k\epsilon)}.
\label{eq:main-finiteness-exact-map}
\end{equation}
Its near-boundary expansion consists of a nonanalytic response branch,
\begin{equation}
A_\epsilon^{\rm nonloc}(\vec k,\mu)
=
\frac{\ii\slashed k}{k}
2^{-2\mu}
\frac{\Gamma(\frac12-\mu)}{\Gamma(\frac12+\mu)}
(k\epsilon)^{2\mu}+\cdots,
\label{eq:main-finiteness-nonlocal}
\end{equation}
and a local source branch of odd derivative weights,
\begin{equation}
A_\epsilon^{\rm loc}
=
\sum_{n=0}^{\infty}b_n(\mu)D_\epsilon^{2n+1},
\qquad D_\epsilon=\epsilon\slashed\partial.
\label{eq:main-finiteness-local-series}
\end{equation}
The recursion for $b_n$ is given in appendix~\ref{app:counterterm-audit}.
For the rescaled intrinsic sources the local term of weight $2n+1$ enters the
regulated quadratic functional as
\begin{equation}
\epsilon^{2n+1-2\mu}.
\label{eq:main-all-local-cutoff-terms}
\end{equation}
It can diverge only for $2n+1<2\Re\mu$. The set is empty in
\eqref{eq:main-ren-strip}; after continuation to $\mu=\ii\nu$ its modulus is
$\epsilon^{2n+1}$. The same weight counting holds on the round spherical
cutoff, with each curvature counted as two derivatives. Thus the standard
AdS$_3$ intrinsic subtraction is empty in the strip,
\begin{equation}
C_{\gamma,{\rm intr}}^{\rm div}=0,
\qquad 0<\Re\mu<\frac12.
\label{eq:main-ct-zero-result}
\end{equation}
The exact Bessel estimate in appendix~\ref{app:counterterm-audit} controls the
remainder and establishes the intrinsic limit without interchanging an
infinite formal derivative expansion with the cutoff limit.

The complete four-component source is different. Let
$G=\Gamma^{\hat0}$, $G^2=-\mathbf1$, and let
$a=\sigma_p=\pm1$ label the unbarred Milne phase. For the equal-weight
independent barred source, opposite-phase Mellin pairing gives the same
algebraic sign in the barred projection. The complete finite-cutoff fields can
be written
\begin{align}
s&=(1+aGA_\epsilon)J,
& r&=(A_\epsilon+aG)J,\nonumber\\
\bar s&=\bar J(1-aGA_\epsilon),
& \bar r&=\bar J(A_\epsilon+aG),
\label{eq:main-full-source-fields}
\end{align}
where the barred regular response obeys
$\bar T(-\vec k)=\bar J(-\vec k)A_\epsilon(\vec k,\mu)$; its derivation from
the left-acting Dirac equation is given in
subsection~\ref{subsec:ct-full-pairing}. Since
$\{G,A_\epsilon\}=0$, the two boundary orderings are governed by the same
exact full-source response operator
\begin{equation}
\mathbb F_{a,\epsilon}
=(A_\epsilon+aG)(1+aGA_\epsilon)^{-1}
=
\frac{2A_\epsilon+aG(1+A_\epsilon^2)}{1-A_\epsilon^2}.
\label{eq:main-complete-response-operator}
\end{equation}
For sufficiently small cutoff in the strip, $A_\epsilon\to0$, so the first
term in the complete response is the derivative-zero cross-block operator
$aG$. With $s_\epsilon\sim\epsilon^{1-\mu}j$ and
$\bar s_\epsilon\sim\epsilon^{1-\mu}\bar j$, it produces
\begin{equation}
I_{\epsilon}^{(0),p}
=
\kappa\,a\,\epsilon^{-2\mu}
\int\dd^2x\,\bar j\,G\,j+\cdots,
\label{eq:main-full-leading-divergence}
\end{equation}
which is power divergent throughout \eqref{eq:main-ren-strip}. This term is
not part of the odd-derivative AdS$_3$ recursion. It is local in the actual
fixed cutoff sources and is removed by the unique weight-zero subtraction
\begin{equation}
I_{\epsilon,{\rm ct}}^{(0),p}
=-\kappa\,\sigma_p
\int_{\Sigma_\epsilon}\dd^2x\,\sqrt{\gamma_\epsilon}\,
\bar\Phi_{{\rm src},\epsilon}^{\,-p}
\Gamma^{\hat0}
\Phi_{{\rm src},\epsilon}^{\,p}
\,.
\label{eq:main-full-counterterm}
\end{equation}
Equivalently, the complete divergent counterterm operator in this strip is
\begin{equation}
C_{\gamma,p}^{\rm div}=-\sigma_p\Gamma^{\hat0}.
\label{eq:main-full-ct-operator}
\end{equation}
No further divergent term occurs: every intrinsic local descendant carries at
least one positive radial weight, and every term quadratic or higher in
$A_\epsilon$ is suppressed after the subtraction in the open strip. The same
counterterm applies on the spherical cutoff, with the corresponding induced
round metric and source projector.

The subtraction also gives a simple exact check. From
\eqref{eq:main-complete-response-operator},
\begin{equation}
\mathbb F_{a,\epsilon}-aG
=
\frac{2A_\epsilon+2aGA_\epsilon^2}{1-A_\epsilon^2}.
\label{eq:main-subtracted-response-operator}
\end{equation}
Using
\begin{equation}
\epsilon^{-2\mu}A_\epsilon(\vec k,\mu)
\longrightarrow
\mathcal K_\mu(\vec k)
\equiv
\frac{\ii\slashed k}{k}
2^{-2\mu}
\frac{\Gamma(\frac12-\mu)}{\Gamma(\frac12+\mu)}k^{2\mu},
\end{equation}
and $A_\epsilon\to0$ in \eqref{eq:main-ren-strip}, one finds
\begin{equation}
\lim_{\epsilon\to0}
\epsilon^{-2\mu}
\bigl(\mathbb F_{a,\epsilon}-aG\bigr)
=2\mathcal K_\mu.
\label{eq:main-full-ren-limit}
\end{equation}
Thus, with the same $\kappa$ used in the intrinsic coefficient functional,
\begin{equation}
\left.I_{\rm ren}^{\rm sym}\right|_{\rm nonlocal}
=2\,\mathcal F_{\bar s r}^{\rm asy}
\label{eq:main-full-ren-kernel}
\end{equation}
modulo finite local contact terms. The factor two is fixed by the complete
source algebra; because the overall bulk normalization is not fixed in the
present analysis, it does not alter the conformal structure of the kernels
displayed above.

The counterterm \eqref{eq:main-full-counterterm} has an important locality
qualification. At fixed $\nu$ and phase it is a derivative-zero local
functional on the celestial cutoff. If the half-line $\nu>0$ and the two
phases are unfolded using the convention of section~\ref{sec:cpw}, then
\begin{equation}
\tau^{\sigma_p\ii\nu}=\tau^{-\ii\lambda},
\qquad \lambda=-\sigma_p\nu,
\qquad \sigma_p=-\operatorname{sgn}\lambda.
\label{eq:main-counterterm-frequency}
\end{equation}
With $u=\log\tau$, the operator $-\ii\partial_u$ has eigenvalue
$-\lambda=\sigma_p\nu$ on $\e^{-\ii\lambda u}$. Thus
$\operatorname{sgn}(-\ii\partial_u)$ acts by $\sigma_p$, and the
phase-summed subtraction is 
\begin{equation}
I_{\epsilon,{\rm ct}}^{(0)}
=-\kappa\int\dd u\,\dd^2x\,\sqrt{\gamma_\epsilon}\,
\bar\Phi_{\rm src}
\Gamma^{\hat0}
\operatorname{sgn}(-\ii\partial_u)
\Phi_{\rm src}.
\label{eq:main-milne-nonlocal-counterterm}
\end{equation}
It is consequently nonlocal in Milne time. The result above should therefore
be understood as a complete fixed-Mellin (spectral/celestial) radial
renormalization. It is not a conventional local three-dimensional
holographic counterterm scheme for the unreduced Milne-boundary field.
Furthermore, the direct fixed-mode limit at $\mu=\ii\nu$ remains oscillatory;
the prescription used here is to renormalize and take the cutoff limit in the
open strip and only then continue to the principal-series line.

The planar Riccati recursion, the left-acting barred response, the exact
full-source algebra, and the spherical curvature completion are given in
appendix~\ref{app:counterterm-audit}.

\subsection{Mellin localization and a shock-profile example}
\label{subsec:mellin-shock-profile}

The planar and spherical kernels determine the angular response at fixed
principal-series parameter $\nu$. To clarify what the phase label measures
kinematically, we now consider localization on a positive multiplicative
half-line. This example does not derive a null limit; it only shows how
angular localization and localization in logarithmic scale are encoded by
the phase-resolved source data.

Let $\upsilon>0$ be a multiplicative variable, defined only up to the
rescaling $\upsilon\mapsto a\upsilon$ with $a>0$. Until a null-limit
prescription is supplied, $\upsilon$ is an abstract kinematical variable.
If such a prescription identifies it with a positive affine or scale
coordinate, the construction below may be interpreted along a chosen null
generator. Consider the idealized spinor source
\begin{equation}
\label{eq:localized-shock-profile}
\mathcal J_{\rm sh}(\upsilon,\Omega)
=
\delta(\upsilon-\upsilon_0)\,
\delta_{S^2}(\Omega,\Omega_0)\,
\zeta,
\qquad
\upsilon_0>0,
\end{equation}
Here $\delta_{S^2}(\Omega,\Omega_0)$ denotes the invariant delta distribution
on the unit round sphere, normalized with respect to the measure
$\dd\Omega_2$ by
\begin{equation}
\int_{S^2}\dd\Omega_2\,
\delta_{S^2}(\Omega,\Omega_0)f(\Omega)
=
f(\Omega_0),
\label{eq:sphere-delta-normalization}
\end{equation}
for every smooth test function $f$. In local coordinates $x^a$, with
$\dd\Omega_2=\sqrt{\gamma(x)}\,\dd^2x$, it is represented by
$\delta_{S^2}(\Omega,\Omega_0)=
\delta^{(2)}\!\bigl(x-x(\Omega_0)\bigr)/\sqrt{\gamma(x)}$. When it multiplies a spinor
polarization, the identity map on the spinor fiber along the diagonal
$\Omega=\Omega_0$ is implicit.
In the Clifford block under consideration,
$\zeta\in\operatorname{im}\!\bigl(P_+(\Omega_0)\Pi_I\bigr)$ belongs to the
independent source fiber at $\Omega_0$. The profile is localized both at the
celestial direction $\Omega_0$ and at the multiplicative position
$\upsilon_0$. We use the term ``shock profile'' only in this kinematical
sense, and not in the sense of a dynamically generated gravitational shock wave
\cite{Aichelburg:1970dh,Dray:1984ha}.

The $\upsilon$ dependence is resolved by the half-density Mellin identity
\begin{equation}
\label{eq:half-density-delta-mellin}
\delta(\upsilon-\upsilon_0)
=
\frac{1}{2\pi\sqrt{\upsilon\upsilon_0}}
\int_{-\infty}^{+\infty}\dd\nu\,
\left(\frac{\upsilon}{\upsilon_0}\right)^{\ii\nu}.
\end{equation}
Thus $\nu$ is the logarithmic frequency conjugate to $\log\upsilon$.
With the convention used in this paper, in which $\nu>0$ and the two
Mellin phases are displayed separately,
\eqref{eq:half-density-delta-mellin} becomes
\begin{equation}
\label{eq:half-density-delta-positive-nu}
\delta(\upsilon-\upsilon_0)
=
\frac{1}{2\pi\sqrt{\upsilon\upsilon_0}}
\int_0^\infty\dd\nu\,
\left[
\left(\frac{\upsilon}{\upsilon_0}\right)^{\ii\nu}
+
\left(\frac{\upsilon}{\upsilon_0}\right)^{-\ii\nu}
\right].
\end{equation}

To compare this half-density identity with the pure phase basis
$\upsilon^{\mp\ii\nu}$ used in
\eqref{eq:main-phase-expansion}, introduce the auxiliary phase-expanded
profile
\begin{align}
\widehat{\mathcal J}_{\rm sh}(\upsilon,\Omega)
&\equiv
\sqrt{\upsilon}\,\mathcal J_{\rm sh}(\upsilon,\Omega)=
\int_0^\infty\frac{\dd\nu}{2\pi}
\Big[
\upsilon^{-\ii\nu}J_-^{\rm sh}(\Omega,\nu)
+
\upsilon^{+\ii\nu}J_+^{\rm sh}(\Omega,\nu)
\Big],
\label{eq:shock-source-phase-expansion}
\\
J_-^{\rm sh}(\Omega,\nu)
&=
\upsilon_0^{-\frac{1}{2}+\ii\nu}
\delta_{S^2}(\Omega,\Omega_0)\zeta,
\nonumber\\
J_+^{\rm sh}(\Omega,\nu)
&=
\upsilon_0^{-\frac{1}{2}-\ii\nu}
\delta_{S^2}(\Omega,\Omega_0)\zeta.
\label{eq:shock-source-phase-coefficients}
\end{align}
Within this auxiliary convention, $J_-^{\rm sh}$ and $J_+^{\rm sh}$ provide a
kinematical realization of the source coefficients denoted by $J$ and
$\widetilde J$, respectively, in $\mathcal F_{\bar s r}^{\rm asy}$. A
profile sharply localized at $\upsilon_0$
requires a coherent superposition over all $\nu>0$ and both phase branches.
A single value of $\nu$, or a single phase branch, does not describe a
localized profile.

Indeed, retaining only the $+\ii\nu$ Mellin phase gives
\begin{equation}
\label{eq:one-sided-mellin-profile}
\frac{1}{2\pi\sqrt{\upsilon\upsilon_0}}
\int_0^\infty\dd\nu\,
\left(\frac{\upsilon}{\upsilon_0}\right)^{\ii\nu}
=
\frac{1}{2}\delta(\upsilon-\upsilon_0)
+
\frac{\ii}{2\pi\sqrt{\upsilon\upsilon_0}}
\operatorname{PV}
\frac{1}{\log(\upsilon/\upsilon_0)}.
\end{equation}
The one-sided branch is therefore an analytic-signal distribution: besides
one half of the localized source, it contains a principal-value tail in
$\log\upsilon$. Multiplication by $\sqrt{\upsilon}$ gives the corresponding
decomposition of the rescaled source
\eqref{eq:shock-source-phase-expansion} in the pure phase basis. The two Milne
phases should therefore be regarded as Mellin-frequency components of the
source data rather than as individually localized profiles.

A pure-phase source is nevertheless a valid element of the complexified
source space. In the following one-point notation, $\langle\mathcal O_p\rangle_{\rm asy}$ denotes the coefficient response $\delta_L\mathcal F_{\bar s r}^{\rm asy}/\delta\bar J_p$; the factor $1/\ii$ remains part of the two-point convention \eqref{eq:grassmann-derivative-convention}. For a negative-phase unbarred source, $J_-=J$ and
$\widetilde J=0$, the quadratic functional gives
\begin{equation}
\left\langle\widetilde{\mathcal O}\right\rangle_{\rm asy}
=
\mathcal M J,
\qquad
\left\langle\mathcal O\right\rangle_{\rm asy}
=
0.
\label{eq:pure-phase-coefficient-response}
\end{equation}
For example, if
\begin{equation}
J(\Omega,\nu)
=
j(\nu)\,
\delta_{S^2}(\Omega,\Omega_0)\,
\zeta,
\end{equation}
then the spherical response is
\begin{equation}
\left\langle
\widetilde{\mathcal O}(\Omega,\nu)
\right\rangle_{\rm asy}
=
j(\nu)\,
\mathcal M_\nu(\Omega,\Omega_0)\,\zeta
=
\ii\,j(\nu)\,
\mathcal K_\nu^{S^2}(\Omega,\Omega_0)\,\zeta.
\label{eq:localized-angular-response}
\end{equation}
For the negative-phase coefficient of
\eqref{eq:localized-shock-profile}, one sets
$j(\nu)=\upsilon_0^{-1/2+\ii\nu}$; the assignments and the sign of the
exponent are reversed for the positive-phase source. Equation
\eqref{eq:localized-angular-response} shows how the spherical kernel maps a
phase-resolved source localized at one celestial direction, whereas
\eqref{eq:half-density-delta-positive-nu} and
\eqref{eq:shock-source-phase-coefficients} specify the superposition required
to localize the profile in $\upsilon$ as well.

If the coordinate represented by $\upsilon$ takes both signs, its positive
and negative multiplicative half-lines must be treated as separate Mellin
sectors. The source \eqref{eq:localized-shock-profile} is only a
kinematical model of localized celestial data. A genuine distributional
spinor solution must also satisfy the Dirac equation, the appropriate null
gamma projection, and the boundary conditions selected by the null-limit
prescription.

\subsection{Relation to the energy-Mellin and shadow bases}
\label{subsec:shadow-basis}

The smooth kernel obtained above should not be identified directly with the distributional pairing of two unshadowed energy-Mellin scattering states. The relation is instead naturally expressed through the spin-$\frac{1}{2}$ shadow transform \cite{Iacobacci:2020cqy}.

In two boundary dimensions, the spinor shadow intertwiner maps a primary of dimension $1-\ii\nu$ to one of dimension $1+\ii\nu$. Comparison with \eqref{eq:main-planar-position-kernel} identifies the nonlocal planar source--response kernel with the normalized shadow kernel defined in subsection~\ref{subsec:cpw-shadow-normalization}:
\begin{equation}
\mathcal K_\nu^{\rm P}(\vec x-\vec y)
=
\kappa_{\rm P}\,\mathsf S_{1-\ii\nu}(\vec x,\vec y).
\label{eq:milne-kernel-shadow-form}
\end{equation}
In that normalization, inverse shadowing recovers the angular delta distribution in the Fourier-defined contact completion. This relates the nonlocal angular structure of the Milne kernel to the distributional pairing of unshadowed states, without fixing their scattering normalization or energy-sign pairing. The normalization and operator inverse are derived in subsection~\ref{subsec:cpw-shadow-normalization}.

The spectral delta function requires a separate comment. In the standard massless spinor conformal-primary basis, the unshadowed Dirac inner product may be written with a delta function in the principal-series parameter together with a delta function in celestial position \cite{Iacobacci:2020cqy}. Other celestial two-point conventions, in which the principal-series weights are paired without the same conjugation convention, are often written schematically with a factor $\delta(\lambda_1+\lambda_2)$. The present Milne calculation instead restricts to $\nu>0$, keeps the two Milne phase branches explicitly, and obtains $\delta(\nu-\nu')$ from the opposite-phase Mellin pairing.

The phase homogeneity provides the kinematical unfolding
$\lambda=-\sigma_p\nu$, as implemented in section~\ref{sec:cpw}. For
opposite phases $q=-p$,
\begin{equation}
\lambda_p=-\sigma_p\nu,
\qquad \lambda_q=-\sigma_q\nu'=\sigma_p\nu',
\qquad \delta(\nu-\nu')=\delta(\lambda_p+\lambda_q).
\label{eq:half-full-line-delta}
\end{equation}
The last equality uses $|\sigma_p|=1$ and holds on the corresponding
opposite-sign half-lines. This resolves the difference between the two
bilinear spectral delta conventions. It does not identify $p$, or the sign
of $\lambda$, with the independent scattering energy-sign label $\eta$,
and it does not fix the global $\ii0$ prescription, helicity normalization,
or conserved four-dimensional Dirac inner product. Moreover, unfolding
the phases does not by itself identify the intrinsic source bases:
section~\ref{sec:cpw} separately normalizes the $-\ii\nu$ eigenspinor
used for $\lambda=-\nu$.

\subsection{Scope of the result}
\label{subsec:scope-result}

Equations
\eqref{eq:main-coefficient-functional}--\eqref{eq:main-sphere-kernel}
define a separated-point source--response map at fixed $\nu$, modulo local
contact terms and the common action normalization. The shock-profile example
provides a conditional kinematical model for these data: the two source
coefficients encode angular position, spinor polarization, and logarithmic
frequency, while localization on the positive multiplicative half-line
requires integration over $\nu$ and both Mellin phases.

This construction does not identify $\upsilon$ with a particular Bondi
coordinate or affine parameter. Its interpretation at $\mathcal I^\pm$
requires a global null-limit and scattering prescription in which the
energy-sign label is independent of the Milne phase. Conditional on that additional
input, $\Omega_0$ may be interpreted as a null direction and $\upsilon_0$
as a position or scale along the associated generator. Without it,
\eqref{eq:localized-shock-profile} is a test source on an abstract
multiplicative half-line, not a dynamically derived shockwave or a complete
distributional solution of the Dirac equation.

Promoting the asymptotic source--response map to a complete CCFT generating functional still requires the global map from phase-resolved data $(p,\nu)$ to scattering data $(\eta,\lambda)$, continuation to $\mathcal I^\pm$, the $\ii0$ prescription, scattering normalization, the conserved Dirac inner product, and a compatible Hermiticity prescription. The kinematical relation $\lambda=-\sigma_p\nu$ is already fixed. The barred response and the complete fixed-Mellin radial subtraction are derived in subsection~\ref{subsec:ads3-renormalization} and appendix~\ref{app:counterterm-audit}. Their counterterm is local on the celestial cutoff at fixed Mellin phase but becomes nonlocal in Milne time after the phases are recombined, so this radial result is not by itself a conventional local three-dimensional CCFT generating functional. Finite local contact terms remain conventional. Appendices~\ref{app:counterterm-audit} and~\ref{app:coefficient-kernels} contain the detailed finite-cutoff source--response maps underlying the kernels displayed in this section.

\section{Spinor conformal primary wavefunctions}
\label{sec:cpw}

The source--response problem of section~\ref{sec:correlator} was organized
using a half-line spectral parameter $\nu>0$ together with two Milne phases.
That convention is convenient for the variational problem, where both phases
can be represented using a common $+\ii\nu$ intrinsic $\mathbb H^3$
eigenspinor. For comparison with the ordinary energy--Mellin
conformal-primary basis it is more natural to unfold the two phases to a
single signed principal-series parameter
\begin{equation}
\lambda\in\mathbb R,
\qquad
\lambda=-\sigma_p\nu,
\qquad
\sigma_-=-1,
\qquad
\sigma_+=+1.
\label{eq:lambda-unfolding-cpw}
\end{equation}
Thus
\begin{equation}
p=-\quad\Longleftrightarrow\quad\lambda=+\nu,
\qquad
p=+\quad\Longleftrightarrow\quad\lambda=-\nu,
\end{equation}
and the two Milne homogeneities become
\begin{equation}
\tau^{-3/2+\sigma_p\ii\nu}
=
\tau^{-3/2-\ii\lambda}.
\end{equation}
We define
\begin{equation}
\Delta_\lambda=1+\ii\lambda,
\qquad
\delta_\lambda\equiv2-\Delta_\lambda=1-\ii\lambda.
\label{eq:Delta-lambda-cpw}
\end{equation}
The four-dimensional homogeneity is therefore
\begin{equation}
\tau^{-3/2-\ii\lambda}
=
\tau^{-\Delta_\lambda-\frac12},
\label{eq:cpw-milne-homogeneity}
\end{equation}
which is the homogeneity of a massless spin-$\frac12$
energy--Mellin conformal-primary wavefunction of dimension
$\Delta_\lambda$ \cite{Iacobacci:2020cqy}.

The scattering energy-sign label will be denoted by $\eta=\pm$. It is
independent of the sign of $\lambda$. The hyperbolic boundary-value problem
constructed below fixes the conformal-primary wavefunction inside the future
Milne wedge, while $\eta$ is fixed only after specifying a global $\ii0$
prescription and continuation to the scattering regions.

The essential difference from the half-line bookkeeping used in
section~\ref{sec:correlator} is that we now choose a separately
source-normalized $\mathbb H^3$ eigenspinor for every
$\lambda\in\mathbb R$:
\begin{equation}
\mathcal D_{\mathbb H^3}K_\lambda(x_0,\vec x;\vec x',\zeta)
=
\ii\lambda K_\lambda(x_0,\vec x;\vec x',\zeta).
\label{eq:cpw-H3-eigenvalue}
\end{equation}
For $\lambda=+\nu$ this is the same intrinsic representative used for the
negative Milne phase in section~\ref{sec:correlator}. For
$\lambda=-\nu$ we instead use a regular $-\ii\nu$ eigenspinor and normalize
its own independent boundary coefficient to a delta function. This
distinction removes the shadow-basis mismatch that occurs if both Milne
phases are represented using the same $+\ii\nu$ source-normalized
eigenspinor.

\subsection{Poincar\'e coordinates on \texorpdfstring{$\mathbb H^3$}{H3}}
\label{subsec:spinor-cpw-poincare}

We work in upper half--space coordinates $(x_0,\vec x)$, with
$\vec x=(x_1,x_2)$ and $x_0>0$. In the planar irreducible Clifford sector
$A$, appendix~\ref{app:dirac-modes-poincare} gives a regular eigenspinor
for either sign of the spectral parameter,
\begin{equation}
\label{eq:chi-mode}
\widehat\chi_{\lambda,\vec k}(x_0,\vec x)
=
\e^{\ii\vec k\cdot\vec x}\,x_0^{3/2}
\left[
a^+(k)K_{-\frac12+\ii\lambda}(|k|x_0)
+
a^-(k)K_{\frac12+\ii\lambda}(|k|x_0)
\right],
\end{equation}
with
\begin{equation}
\label{eq:spinor-constraint}
a^-(k)
=
-\frac{\ii}{|k|}\slashed k\,a^+(k),
\qquad
\slashed k
=
k_1\Gamma^{\hat2}+k_2\Gamma^{\hat3},
\end{equation}
and
\begin{equation}
\mathcal D_{\mathbb H^3}
\widehat\chi_{\lambda,\vec k}
=
\ii\lambda
\widehat\chi_{\lambda,\vec k}.
\end{equation}
The continuum Plancherel normalization is not included here because the
normalization of the conformal-primary wavefunction will instead be fixed
directly by its boundary source.

For a boundary insertion at $\vec x'$ and a constant polarization spinor
$\zeta$ satisfying
\begin{equation}
\Pi_-\zeta=\zeta,
\qquad
\Pi_{\rm A}\zeta=\zeta,
\end{equation}
we impose
\begin{equation}
\label{eq:spinor-bc}
\left[
\Pi_-K_\lambda^{\rm P}(x_0,\vec x;\vec x',\zeta)
\right]_{x_0^{\,1-\ii\lambda}}
=
\delta^{(2)}(\vec x-\vec x')\zeta.
\end{equation}
Equivalently,
\begin{equation}
\left[
\Pi_-K_{\Delta_\lambda}^{\rm P}
\right]_{x_0^{\,2-\Delta_\lambda}}
=
\delta^{(2)}(\vec x-\vec x')\zeta.
\end{equation}
The complementary radial polarization is not prescribed independently; it
is determined by the intrinsic Dirac equation together with regularity as
$x_0\to\infty$.

Using
\begin{equation}
K_\alpha(z)
\sim
2^{\alpha-1}\Gamma(\alpha)z^{-\alpha},
\qquad
z\to0,
\qquad
\Re\alpha>0,
\label{eq:K-small}
\end{equation}
the source term is obtained from the $a^-$ component. We choose
\begin{equation}
\label{eq:a-choice}
a^-(k)
=
C_\lambda
|k|^{\frac12+\ii\lambda}
\e^{-\ii\vec k\cdot\vec x'}\zeta,
\qquad
a^+(k)
=
\ii\frac{\slashed k}{|k|}a^-(k),
\end{equation}
where $C_\lambda
=
\frac{
2^{\frac12-\ii\lambda}
}{
\Gamma(\frac12+\ii\lambda)
}.$ This gives
\begin{equation}
\left[
\Pi_-K_\lambda^{\rm P}
\right]_{x_0^{1-\ii\lambda}}
=
\delta^{(2)}(\vec x-\vec x')\zeta
\end{equation}
for either sign of $\lambda$. The momentum-space kernel is therefore
\begin{align}
\label{eq:spinor-cpw-momentum}
K_\lambda^{\rm P}
={}&
C_\lambda
\int\frac{\dd^2k}{(2\pi)^2}
\e^{\ii\vec k\cdot(\vec x-\vec x')}
x_0^{3/2}|k|^{\frac12+\ii\lambda}
\left[
\ii\frac{\slashed k}{|k|}
K_{-\frac12+\ii\lambda}(|k|x_0)
+
K_{\frac12+\ii\lambda}(|k|x_0)
\right]\zeta.
\end{align}
The Fourier--Bessel integrals give the exactly source-normalized spinor
Poisson kernel
\begin{equation}
\label{eq:spinor-poisson-coordinate}
K_{\Delta_\lambda}^{\rm P}
(x_0,\vec x;\vec x',\zeta)
=
-\frac{\Delta_\lambda-\frac12}{\pi}
\frac{
x_0^{\Delta_\lambda}
\bigl(
x_0\Gamma^{\hat1}+\slashed r
\bigr)
}{
\bigl(
x_0^2+|\vec r|^2
\bigr)^{\Delta_\lambda+\frac12}
}\zeta,
\qquad
\vec r=\vec x-\vec x'.
\end{equation}
It satisfies
\begin{equation}
\mathcal D_{\mathbb H^3}
K_{\Delta_\lambda}^{\rm P}
=
\ii\lambda
K_{\Delta_\lambda}^{\rm P},
\end{equation}
and
\begin{equation}
\left[
\Pi_-K_{\Delta_\lambda}^{\rm P}
\right]_{x_0^{2-\Delta_\lambda}}
=
\delta^{(2)}(\vec x-\vec x')\zeta.
\label{eq:spinor-poisson-properties}
\end{equation}
Within the class of Fourier modes regular as $x_0\to\infty$ and with no
separate $k=0$ Plancherel contribution, this boundary-value problem is
unique. For $k>0$ the regular solution is the $K$-Bessel branch and its source coefficient is nonzero for every real $\lambda$. Hence a regular
solution with vanishing source coefficient vanishes identically.

\subsection{Exact source--response map and normalized shadow transform}
\label{subsec:cpw-shadow-normalization}

We now fix the normalization of the spinor shadow operator used throughout
this section. Let an input boundary spinor have conformal dimension
$\delta$. We define
\begin{equation}
\bigl(\widehat{\mathsf S}_\delta f\bigr)(\vec x)
=
\int_{\mathbb R^2}\dd^2y\,
\mathsf S_\delta(\vec x,\vec y)f(\vec y),
\end{equation}
with
\begin{equation}
\label{eq:normalized-spinor-shadow}
\mathsf S_\delta(\vec x,\vec y)
=
\frac{\ii(3-2\delta)}{2\pi}
\frac{
\Gamma^{\hat\alpha}(x-y)_{\hat\alpha}
}{
|\vec x-\vec y|^{\,2(2-\delta)+1}
},
\qquad
\hat\alpha=\hat2,\hat3.
\end{equation}
The singular kernel is understood by analytic continuation as a
distribution, with its contact completion fixed by the Fourier-space
definition below.

We use the Fourier conventions
\begin{equation}
f(\vec k)
=
\int\dd^2x\,
\e^{-\ii\vec k\cdot\vec x}f(\vec x),
\qquad
f(\vec x)
=
\int\frac{\dd^2k}{(2\pi)^2}
\e^{\ii\vec k\cdot\vec x}f(\vec k).
\end{equation}
Then \eqref{eq:normalized-spinor-shadow} is exactly equivalent to
\begin{equation}
\label{eq:spinor-shadow-momentum-exact}
\mathsf S_\delta(\vec k)
=
\frac{\slashed k}{|k|}
\frac{
\Gamma(\delta-\frac12)
}{
\Gamma(\frac32-\delta)
}
\left(
\frac{|k|}{2}
\right)^{2-2\delta}.
\end{equation}
Since $\left(
\frac{\slashed k}{|k|}
\right)^2
=
\mathbf1,$ this normalization obeys
\begin{equation}
\label{eq:spinor-shadow-involution}
\mathsf S_{2-\delta}(\vec k)
\mathsf S_\delta(\vec k)
=
\mathbf1.
\end{equation}
Equivalently,
\begin{equation}
\int\dd^2z\,
\mathsf S_{2-\delta}(\vec x,\vec z)
\mathsf S_\delta(\vec z,\vec y)
=
\delta^{(2)}(\vec x-\vec y)\mathbf1
\label{eq:spinor-shadow-involution-position}
\end{equation}
as a distributional identity.

For the CPW boundary-value problem the independent source has dimension
\begin{equation}
\delta_\lambda
=
2-\Delta_\lambda
=
1-\ii\lambda.
\end{equation}
Thus
\begin{equation}
\label{eq:spinor-shadow-principal-exact}
\mathsf S_{1-\ii\lambda}(\vec x,\vec y)
=
\frac{\ii(1+2\ii\lambda)}{2\pi}
\frac{
\Gamma^{\hat\alpha}(x-y)_{\hat\alpha}
}{
|\vec x-\vec y|^{3+2\ii\lambda}
},
\end{equation}
and
\begin{equation}
\label{eq:spinor-shadow-principal-momentum}
\mathsf S_{1-\ii\lambda}(\vec k)
=
\frac{\slashed k}{|k|}
\frac{
\Gamma(\frac12-\ii\lambda)
}{
\Gamma(\frac12+\ii\lambda)
}
\left(
\frac{|k|}{2}
\right)^{2\ii\lambda}.
\end{equation}
Let $\widehat{\mathsf S}_{1\pm\ii\lambda}$
denote the integral operators with kernels
$\mathsf S_{1\pm\ii\lambda}$, defined by
\begin{equation}
(\widehat{\mathsf S}_{1\pm\ii\lambda}f)
(\vec x)
=
\int_{\mathbb R^2}\dd^2\vec y\,
\mathsf S_{1\pm\ii\lambda}
(\vec x,\vec y)f(\vec y).
\end{equation}
In particular, their Fourier-defined
contact completion satisfies
\begin{equation}
\label{eq:spinor-shadow-principal-inverse}
\widehat{\mathsf S}_{1+\ii\lambda}
\widehat{\mathsf S}_{1-\ii\lambda}
=
\operatorname{Id}_{\rm src}.
\end{equation}
Writing $\mathcal K_\lambda^{\rm P}$ for the
signed extension of the source--response kernel
in this separately normalized basis, define
the corresponding integral operator by
\begin{equation}
(\widehat{\mathcal K}_\lambda^{\rm P}f)
(\vec x)
=
\int_{\mathbb R^2}\dd^2\vec y\,
\mathcal K_\lambda^{\rm P}
(\vec x,\vec y)f(\vec y).
\end{equation}
The kernel and operator relations are
\begin{equation}
\label{eq:cpw-inverse-shadow-pairing}
\mathcal K_\lambda^{\rm P}
=
\kappa_{\rm P}\mathsf S_{1-\ii\lambda},
\qquad
\frac{1}{\kappa_{\rm P}}\,
\widehat{\mathsf S}_{1+\ii\lambda}
\widehat{\mathcal K}_\lambda^{\rm P}
=
\operatorname{Id}_{\rm src}.
\end{equation}
This identity uses the Fourier-defined contact
completion above and recovers the angular delta
distribution as the kernel of
$\operatorname{Id}_{\rm src}$. It fixes no
scattering-state norm or energy-sign pairing.

The complementary coefficient of the regular
Poisson kernel \eqref{eq:spinor-poisson-coordinate} is then fixed exactly
\begin{equation}
\label{eq:cpw-response-exact-shadow}
\left[
\Pi_+K_{\Delta_\lambda}^{\rm P}
(x_0,\vec x;\vec x',\zeta)
\right]_{x_0^{1+\ii\lambda}}
=
\ii\,\mathsf S_{1-\ii\lambda}
(\vec x,\vec x')\zeta.
\end{equation}
Using \eqref{eq:spinor-shadow-principal-exact},
\begin{equation}
\label{eq:cpw-response-exact-coordinate}
\left[
\Pi_+K_{\Delta_\lambda}^{\rm P}
\right]_{x_0^{1+\ii\lambda}}
=
-\frac{1+2\ii\lambda}{2\pi}
\frac{\Gamma^{\hat\alpha}
(x-x')_{\hat\alpha}}
{|\vec x-\vec x'|^{3+2\ii\lambda}}
\zeta.
\end{equation}
Therefore, for an arbitrary source $J_\lambda$,
\begin{equation}
\label{eq:exact-source-response-shadow}
R_\lambda
=
\ii\,\widehat{\mathsf S}_{1-\ii\lambda}
J_\lambda.
\end{equation}
This is the precise sense in which the
regularity-induced boundary response is a
shadow transform. It does not imply that
the complete four-dimensional bulk wavefunction
is a shadow-transformed conformal-primary
wavefunction. The source--response map and the
scattering-state basis are different objects.

The four-dimensional source-normalized conformal-primary wavefunction in
the future Milne wedge is
\begin{equation}
\label{eq:spinor-cpw-coordinate}
\Psi_{\lambda}^{\rm src}
(\tau,x_0,\vec x;\vec x',\zeta)
=
\tau^{-3/2-\ii\lambda}
(1-\Gamma^{\hat0})
K_{\Delta_\lambda}^{\rm P}
(x_0,\vec x;\vec x',\zeta),
\qquad
\lambda\in\mathbb R.
\end{equation}
Using
\begin{equation}
\{
\Gamma^{\hat0},
\mathcal D_{\mathbb H^3}
\}=0,
\qquad
\mathcal D_{\mathbb H^3}K_{\Delta_\lambda}^{\rm P}
=
\ii\lambda K_{\Delta_\lambda}^{\rm P},
\end{equation}
one verifies directly that \eqref{eq:spinor-cpw-coordinate} solves the
four-dimensional massless Dirac equation.

The two half-line representatives are now
\begin{align}
\label{eq:cpw-negative-phase-new}
\Psi_{+\nu}^{\rm src}
&=
\tau^{-3/2-\ii\nu}
(1-\Gamma^{\hat0})
K_{1+\ii\nu}^{\rm P},
\\
\label{eq:cpw-positive-phase-new}
\Psi_{-\nu}^{\rm src}
&=
\tau^{-3/2+\ii\nu}
(1-\Gamma^{\hat0})
K_{1-\ii\nu}^{\rm P}.
\end{align}
The second line is the important change relative to the common-$+\nu$
basis used in the source--response bookkeeping. The positive Milne phase
is represented by a separately delta-source-normalized $-\ii\nu$
eigenspinor rather than by the same $K_{1+\ii\nu}^{\rm P}$ used in the
negative phase.

The conformal-primary transformation law follows from covariance and
uniqueness. If $g\in SO^+(1,3)$ acts on $\mathbb H^3$ with bulk spin lift
$S_g$ and boundary spin rotation $s_g$, and if
$f_g^*\gamma=\Omega_g^2\gamma$, then
\begin{equation}
S_g(X)
K_{\Delta_\lambda}^{\rm P}
(g^{-1}X;x',\zeta)
=
K_{\Delta_\lambda}^{\rm P}
\left(
X;
gx',
\Omega_g(x')^{\Delta_\lambda}s_g(x')\zeta
\right).
\label{eq:cpw-covariance}
\end{equation}
Hence \eqref{eq:spinor-cpw-coordinate} transforms as a spin-$\frac12$
conformal primary of dimension $\Delta_\lambda$.

\subsection{Global source-normalized harmonic expansion}
\label{subsec:spinor-cpw-global}

We now give the corresponding construction on the full celestial sphere.
Use the global block-$B$ Pauli realization of appendix~\ref{app:modesum},
and extend the radial profiles in \eqref{eq:phi-psi-radial-basis} to signed
$\lambda\in\mathbb R$. The function $\phi_{\lambda l}$ is even in
$\lambda$, while $\psi_{\lambda l}$ is odd. Thus the regular, unnormalized
intrinsic harmonic
\begin{equation}
\widehat\chi^{\rm B}_{\lambda l m_j}
=\phi_{\lambda l}\hat\chi^{(-)}_{l m_j}
+\ii\psi_{\lambda l}\hat\chi^{(+)}_{l m_j},
\qquad
\mathcal D_{\mathbb H^3}\widehat\chi^{\rm B}_{\lambda l m_j}
=\ii\lambda\widehat\chi^{\rm B}_{\lambda l m_j}
\label{eq:cpw-global-signed-harmonic}
\end{equation}
works for either sign. No factor $c_3$ is included: the normalization here
is fixed by the boundary source, not by the bulk Plancherel measure.

Let $\rho=2\e^{-y}$ and use the orthonormal source and response harmonics
$u_{l m_j}$ and $v_{l m_j}$ of \eqref{eq:global-boundary-harmonics}.
The hypergeometric connection coefficients simplify by the duplication
formula to
\begin{equation}
\mathcal A_\pm(l,\lambda)
=\frac{4}{\sqrt\pi}
\frac{\Gamma(l+\frac32)\Gamma(\frac12\pm\ii\lambda)}
{\Gamma(l+\frac32\pm\ii\lambda)}.
\label{eq:cpw-global-connection-coefficients}
\end{equation}
In particular,
\begin{align}
\bigl[P_+\widehat\chi^{\rm B}_{\lambda l m_j}\bigr]_{\rho^{1-\ii\lambda}}
&=\frac{2^{-1+\ii\lambda}}{\sqrt2}\,
\mathcal A_+(l,\lambda)u_{l m_j},\nonumber\\
\bigl[P_-\widehat\chi^{\rm B}_{\lambda l m_j}\bigr]_{\rho^{1+\ii\lambda}}
&=\frac{2^{-1-\ii\lambda}}{\sqrt2}\,
\mathcal A_-(l,\lambda)v_{l m_j}.
\label{eq:cpw-global-source-coefficient}
\end{align}
The brackets use the same analytic-continuation prescription as in
\eqref{eq:asymptotic-component-notation}, now with $\mu\to\ii\lambda$.
The formula has a continuous $\lambda=0$ limit; this endpoint is counted
only once and introduces no discrete spectral contribution.

For $\zeta\in\operatorname{im}(P_+(\Omega')\Pi_{\rm B})$, define
\begin{equation}
K_{\lambda}^{\rm G,B}(y,\Omega;\Omega',\zeta)
=\sum_{l=0}^{\infty}\sum_{m_j}
\frac{\sqrt2\,2^{1-\ii\lambda}}{\mathcal A_+(l,\lambda)}
\widehat\chi^{\rm B}_{\lambda l m_j}(y,\Omega)
u_{l m_j}^{\dagger}(\Omega')\zeta.
\label{eq:cpw-global-source-sum}
\end{equation}
For boundary distributions the sum is defined by inserting $q^l$ and
taking $q\to1^-$ after pairing with smooth test sections. Every mode is
regular at the origin. The source coefficient is exactly
\begin{align}
\bigl[P_+K_{\lambda}^{\rm G,B}\bigr]_{\rho^{1-\ii\lambda}}
&=\sum_{l,m_j}u_{l m_j}(\Omega)u_{l m_j}^{\dagger}(\Omega')\zeta=\delta_{S^2}(\Omega,\Omega')\zeta,
\label{eq:cpw-global-delta-source}
\end{align}
by completeness in the $P_+\Pi_{\rm B}$ source bundle. The delta
distribution includes the identity on that fiber and is normalized with
respect to $\dd\Omega_2$. The response coefficient is
\begin{equation}
\bigl[P_-K_{\lambda}^{\rm G,B}\bigr]_{\rho^{1+\ii\lambda}}
=\sum_{l,m_j}2^{-2\ii\lambda}
\frac{\mathcal A_-(l,\lambda)}{\mathcal A_+(l,\lambda)}
v_{l m_j}(\Omega)u_{l m_j}^{\dagger}(\Omega')\zeta,
\label{eq:cpw-global-response-sum}
\end{equation}
which reproduces \eqref{eq:main-global-harmonic-kernel} after multiplication
by $\kappa_{\rm G}$ and restriction to $\lambda=\nu>0$.
For real $\lambda$ none of the gamma functions in
\eqref{eq:cpw-global-connection-coefficients} has a zero or pole.
Thus the regular radial solution for each angular harmonic has a nonzero
source coefficient, and the regular problem with zero source is trivial.
This proves uniqueness in the harmonic class just specified.

The complementary block is explicit as well. Put
\begin{equation}
\widehat\chi^{\rm A}_{\lambda l m_j}
=\Gamma^{\hat0}\widehat\chi^{\rm B}_{-\lambda,l m_j},
\qquad
u^{\rm A}_{l m_j}=\Gamma^{\hat0}v_{l m_j},
\qquad
v^{\rm A}_{l m_j}=\Gamma^{\hat0}u_{l m_j}.
\label{eq:cpw-global-complementary-block}
\end{equation}
Anticommutation with the intrinsic Dirac operator gives eigenvalue
$\ii\lambda$, and $\Gamma^{\hat0}$ exchanges both the radial projectors
and Clifford blocks. Since $\mathcal A_-(l,-\lambda)=\mathcal A_+(l,\lambda)$,
the same normalization in \eqref{eq:cpw-global-source-sum}, with
$(\widehat\chi^{\rm B},u,v)$ replaced by
$(\widehat\chi^{\rm A},u^{\rm A},v^{\rm A})$, defines
$K_{\lambda}^{\rm G,A}$ and gives the same delta-source check in block $A$.

Finally, for either $I\in\{{\rm A},{\rm B}\}$,
\begin{equation}
\Psi_{\lambda,I}^{\rm src,G}
=\tau^{-3/2-\ii\lambda}(1-\Gamma^{\hat0})K_{\lambda}^{\rm G,I}.
\label{eq:cpw-global-four-dimensional}
\end{equation}
The cancellation used for \eqref{eq:spinor-cpw-coordinate} proves the
massless four-dimensional Dirac equation. Covariance and the harmonic
uniqueness argument give the same conformal-primary transformation law in
the round boundary Weyl frame. These are intrinsic delta-source
normalizations; the complete four-component projected field still has the
additional terms discussed in subsection~\ref{subsec:ct-full-pairing}.

\subsection{Relation to the ordinary conformal-primary basis}
\label{subsec:cpw-standard-comparison}

The full-line parametrization makes the relation to the usual
energy--Mellin conformal-primary basis immediate. The two Milne phases
used in section~\ref{sec:correlator} are unfolded according to
\begin{equation}
\lambda=-\sigma_p\nu,
\qquad
\Delta_\lambda=1+\ii\lambda,
\end{equation}
so that their common time dependence is
\begin{equation}
\tau^{-3/2-\ii\lambda}
=
\tau^{-\Delta_\lambda-\frac12}.
\end{equation}
This is precisely the Milne homogeneity of a massless spin-$\frac12$
energy--Mellin conformal-primary wavefunction
\cite{Iacobacci:2020cqy}. In particular, the branch
$\tau^{-3/2+\ii\nu}$ corresponds to $\lambda=-\nu$ and therefore to
$\Delta=1-\ii\nu$. It must consequently be represented by a separately
source-normalized $-\ii\nu$ eigenspinor, rather than by reusing the
$+\ii\nu$ kernel.

With this phase-matched choice, the regular kernel
$K_{\Delta_\lambda}^{\rm P}$ has the same Milne homogeneity,
$\mathbb H^3$ Dirac eigenvalue, and delta-functional boundary
normalization as the hyperbolic restriction of a spin-$\frac12$
conformal-primary wavefunction of dimension
$\Delta_\lambda=1+\ii\lambda$.
Thus it provides the source-normalized conformal-primary basis in the
future Milne wedge. Matching this normalization to the conventional
scattering normalization additionally requires fixing the Minkowski
spin frame, the energy-sign/$\ii0$ prescription, and the conserved
four-dimensional Dirac inner product.

This is directly analogous to the scalar hyperbolic construction of
Ogawa, Takayanagi, Tsuda and Waki \cite{Ogawa:2022fhy}: a regular
hyperbolic solution with a delta-functional boundary source gives the
ordinary conformal-primary wavefunction restricted to the Milne wedge.
The nonlocal complementary coefficient does not make the bulk mode a
shadow conformal primary. In the present spinor problem regularity instead
implies the exact boundary relation
\begin{equation}
R_\lambda
=
\ii\,\widehat{\mathsf S}_{1-\ii\lambda}
J_\lambda
\end{equation}
so the shadow transform characterizes the source--response map, not the
bulk conformal-primary wavefunction itself.

Subsection~\ref{subsec:spinor-cpw-global} supplies the explicit global
harmonic sum, its normalization, and its delta-source check. The Poincar\'e
and global descriptions solve a common boundary-value problem with
conformal-primary covariance in their respective boundary Weyl frames and
the common signed label $\lambda\in\mathbb R$. Their finite-separation
coordinate comparison requires the endpoint Weyl factors and spin-frame
rotations. The phase variables $(p,\nu)$ are related kinematically by
$\lambda=-\sigma_p\nu$; the independent energy sign $\eta$ still belongs
to the global scattering prescription.

\section{Discussion and conclusions}
\label{sec:conclusion}

We developed a Milne-slicing source--response construction for a free spinor field in four-dimensional Minkowski spacetime. Constant-$\tau$ hypersurfaces are copies of $\mathbb H^3$, so the four-dimensional Dirac problem separates into Milne-time evolution and hyperbolic spinor harmonics labeled by a continuous principal-series parameter. The massive mode equation was solved in this basis, while the celestial source--response analysis was restricted to the massless sector.

\subsection{Summary of results}

The first main result is the explicit phase-resolved boundary-value problem. After the Milne rescaling $\psi=\tau^{-3/2}\phi$, the two massless time dependences are $\tau^{\mp\ii\nu}$. They are retained as independent phase labels rather than being identified a priori with incoming and outgoing scattering states. Mellin orthogonality pairs only opposite phases. Together with the decomposition into the two irreducible spatial Clifford blocks, this gives the off-diagonal source structure in \eqref{eq:main-coefficient-functional}.

The second main result is the nonlocal source--response kernel. In the Poincar\'e patch it is
\begin{equation}
\mathcal K_\nu^{\rm P}(\vec x-\vec y)
\propto
\frac{\Gamma^{\hat\alpha}(x-y)_{\hat\alpha}}{|\vec x-\vec y|^{3+2\ii\nu}},
\end{equation}
while the global harmonic sum gives its covariant counterpart on $S^2$. Both have principal-series dimension
\begin{equation}
\Delta_\nu=1+\ii\nu,
\end{equation}
and signed spins $\mathfrak s_{\rm A}=+\tfrac12$ and $\mathfrak s_{\rm B}=-\tfrac12$ in the adopted orientation. The planar kernel has the functional form of the spin-$\tfrac12$ shadow intertwiner. The global and planar coefficients are compatible with a common normalization after matching the boundary orientation, spin frame, and source conventions, and the short-distance limit of the spherical kernel reproduces the planar conformal structure.

The third result is the radial counterterm analysis. Every intrinsic local
source-branch monomial of odd derivative weight $2n+1$ contributes as
$\rho^{2n+1-2\mu}$ and therefore requires no subtraction in
$0<\Re\mu<\tfrac12$. The complete four-component equal-weight source problem
contains one additional derivative-zero cross-block term. At fixed Mellin
phase $p$ it is removed by
$C_{\gamma,p}^{\rm div}=-\sigma_p\Gamma^{\hat0}$, or equivalently by the
cutoff-source counterterm \eqref{eq:main-full-counterterm}. After this
subtraction the strip limit is finite and its separated-point nonlocal part
is twice the intrinsic coefficient functional in the normalization used here.
The result is then analytically continued to $\mu=\ii\nu$. Since the
phase-summed counterterm contains
$\operatorname{sgn}(-\ii\partial_{\log\tau})$, this is a complete
fixed-Mellin radial renormalization but not a local three-dimensional
counterterm scheme before Mellin decomposition.

Finally, we constructed regular source-normalized spinor conformal primary wavefunctions. The Poincar\'e solution is a spinor Poisson kernel multiplied by the universal Milne phases, and the global construction gives the corresponding harmonic expansion on the full celestial sphere. Their normalization is fixed by a delta-functional boundary source rather than by a scattering-state inner product, while covariance and uniqueness establish the conformal-primary transformation law.

\subsection{Open issues and future directions}

The complete fixed-Mellin radial functional is renormalized by the cross-block subtraction derived above, but its locality is spectral rather than three-dimensional: after inverse Mellin transformation the required sign operator is nonlocal in Milne time. Determining whether there is an alternative formulation with a fully local covariant three-dimensional subtraction remains open. The signed Mellin relation $\lambda=-\sigma_p\nu$ and the normalized shadow inverse are established here. A complete CCFT interpretation still requires the global map between phase-resolved data $(p,\nu)$ and the independent scattering labels $(\eta,\lambda)$, continuation to $\mathcal I^\pm$, the energy-sign/$\ii0$ prescription, scattering normalization, the conserved Dirac inner product, and Hermiticity. An explicit finite-separation stereographic comparison also requires the endpoint Weyl factors and spin-frame rotations.

The extension to massive fields is a separate open problem. Massive excitations are naturally associated with timelike infinity, so a flat/CFT dictionary must explain how hyperbolic data there are related to celestial observables and how source--response boundary conditions are to be imposed on the Bessel-type Milne modes \cite{Laddha:2022nmj}. The massive solutions derived in section~\ref{sec:solution} provide the required mode basis, but the corresponding boundary interpretation, scattering normalization, and operator dictionary require additional input. Interactions and higher-spin fields are further natural extensions and would provide access to higher-point celestial correlators and OPE data.

\acknowledgments

We thank Mihailo \v{C}ubrovi\'{c} for useful Wolfram Language code. We also thank Vasilii Pushkarev and Aleksandr Belokon for their collaboration during the initial stages of this project.

\appendix

\section{Vielbein, spin connection, and intrinsic Dirac operators}
\label{app:vielbein}

\subsection{Poincar\'e coordinates on \texorpdfstring{$\mathbb H^3$}{H3}}
\label{app:vielbein-poincare}

We choose the diagonal orthonormal coframe (vielbein one-forms)
\begin{equation}
\label{eq:vielbein}
e^{\hat 0}=\dd\tau,\qquad
e^{\hat 1}=\frac{\tau}{x_0}\dd x_0,\qquad
e^{\hat 2}=\frac{\tau}{x_0}\dd x_1,\qquad
e^{\hat 3}=\frac{\tau}{x_0}\dd x_2,
\end{equation}
so that $g=\eta_{ab}\,e^{a}\otimes e^{b}$ with $\eta=\mathrm{diag}(-1,1,1,1)$.
In components,
\[
e^a{}_\mu=\mathrm{diag}\!\left(1,\frac{\tau}{x_0},\frac{\tau}{x_0},\frac{\tau}{x_0}\right),
\qquad
e_a{}^\mu=\mathrm{diag}\!\left(1,\frac{x_0}{\tau},\frac{x_0}{\tau},\frac{x_0}{\tau}\right).
\]
The curved gamma matrices are
\begin{equation}
\label{eq:curved-gammas}
\Ga^\mu=e_a{}^\mu \Ga^a,
\qquad
\{\Ga^a,\Ga^b\}=2\eta^{ab}\mathbf 1_4,
\qquad
\{\Ga^\mu,\Ga^\nu\}=2g^{\mu\nu}\mathbf 1_4.
\end{equation}
With torsion-free spin connection $\omega_{\mu}{}^{ab}$, the spinor covariant derivative is
\begin{equation}
\label{eq:spin-cov-der}
\nabla_\mu=\partial_\mu+\Omega_\mu,
\qquad
\Omega_\mu=\frac{1}{4}\,\omega_{\mu ab}\,\Ga^a\Ga^b.
\end{equation}
For the frame \eqref{eq:vielbein}, a convenient summary is
\begin{equation}
\label{eq:Omega-mu}
\begin{aligned}
\Omega_\tau&=0,
&\Omega_{x_0}&=-\frac{1}{2x_0}\Ga^{\hat0}\Ga^{\hat1},\\
\Omega_{x_1}&=\frac{1}{2x_0}
\bigl(-\Ga^{\hat0}\Ga^{\hat2}+\Ga^{\hat1}\Ga^{\hat2}\bigr),
&\Omega_{x_2}&=\frac{1}{2x_0}
\bigl(-\Ga^{\hat0}\Ga^{\hat3}+\Ga^{\hat1}\Ga^{\hat3}\bigr).
\end{aligned}
\end{equation}
With these conventions, the intrinsic Dirac operator on the unit $\mathbb H^3$ slice is
\begin{equation}
\label{eq:DH3}
\mathcal D_{\mathbb H^3}
=
x_0\Big[
\Ga^{\hat 1}\Big(\partial_{x_0}-\frac{1}{x_0}\Big)
+\Ga^{\hat 2}\partial_{x_1}
+\Ga^{\hat 3}\partial_{x_2}
\Big].
\end{equation}

\subsection{Global coordinates on \texorpdfstring{$\mathbb H^3$}{H3}}
\label{app:vielbein-global}

In global coordinates $(y,\theta,\phi)$ on $\mathbb H^3$,
\[
ds^2_{\mathbb H^3}=\dd y^2+\sinh^2 y\,(\dd\theta^2+\sin^2\theta\,\dd\phi^2),
\]
we use the orthonormal coframe adapted to the Milne metric,
\begin{equation}
\label{eq:milne-global-tetrad}
e^{\hat0}=\dd\tau,\qquad
e^{\hat y}=\tau\,\dd y,\qquad
e^{\hat\theta}=\tau\sinh y\,\dd\theta,\qquad
e^{\hat\phi}=\tau\sinh y\sin\theta\,\dd\phi.
\end{equation}
With this choice, the Milne Dirac operator takes the separated form
\begin{equation}
\label{eq:milne-dirac-separated-app}
\slashed{\nabla}
=
\Ga^{\hat0}\Bigl(\partial_\tau+\frac{3}{2\tau}\Bigr)
+\frac{1}{\tau}\,\mathcal D_{\mathbb H^3},
\end{equation}
where the intrinsic $\mathbb H^3$ Dirac operator in global coordinates is
\begin{equation}
\label{eq:DH3-global}
\mathcal D_{\mathbb H^3}
=
\Ga^{\hat y}\Bigl(\partial_y+\coth y\Bigr)
+\frac{1}{\sinh y}\,\mathcal D_{S^2},
\end{equation}
and the tangential-frame $S^2$ Dirac operator is
\begin{equation}
\label{eq:DS2-frame}
\mathcal D_{S^2}
=
\Ga^{\hat\theta}\Bigl(\partial_\theta+\tfrac12\cot\theta\Bigr)
+\Ga^{\hat\phi}\csc\theta\,\partial_\phi.
\end{equation}
Introducing $r=\sinh y$, one may equivalently write
\begin{equation}
\label{eq:DH3-global-areal-radius}
\mathcal D_{\mathbb H^3}
=
\sqrt{1+r^2}\,\Ga^{\hat r}\Bigl(\partial_r+\frac{1}{r}\Bigr)
+\frac{1}{r}\,\mathcal D_{S^2},
\end{equation}
where $e^{\hat r}=\tau\,\dd r/\sqrt{1+r^2}$.

\section{\texorpdfstring{$\mathbb H^3$ Dirac eigenmodes}{H3 Dirac eigenmodes}}
\label{app:dirac-modes}

\subsection{Poincar\'e eigenmodes}
\label{app:dirac-modes-poincare}

Let $k=(k_1,k_2)$, $|k|=\sqrt{k_1^2+k_2^2}$, and $x_\perp=(x_1,x_2)$. Let $\Pi_{\rm A}$ project onto the planar irreducible spatial Clifford sector. The explicit global Pauli realization below instead belongs to block $B$. We choose
\begin{equation}
\Pi_{\rm A}^2=\Pi_{\rm A},
\qquad
[\Pi_{\rm A},\Gamma^{\hat i}]=0\quad(\hat i=\hat1,\hat2,\hat3),
\qquad
\Gamma^{\hat0}\Pi_{\rm A}(\Gamma^{\hat0})^{-1}=\Pi_{\rm B}=\mathbf1_4-\Pi_{\rm A}.
\end{equation}
For $k\neq0$, choose a normalized vector $a^+(k)$ in the one-dimensional subspace
\begin{equation}
\Pi_{\rm A}a^+=a^+,
\qquad
\Gamma^{\hat1}a^+=+a^+,
\qquad
(a^+)^\dagger a^+=1,
\end{equation}
and define
\begin{equation}
\label{eq:a-minus}
a^-(k)= -\,\frac{\ii}{|k|}\,(k_1\Gamma^{\hat2}+k_2\Gamma^{\hat3})\,a^+(k).
\end{equation}
Then $\Pi_{\rm A}a^-=a^-$, $\Gamma^{\hat1}a^-=-a^-$, and $(a^-)^\dagger a^-=1$. The regular but not yet delta-normalized eigenspinor
\begin{equation}
\label{eq:chi-mode-app}
\widehat\chi_{\lambda,k}(x_0,x_\perp)
=
\e^{\ii k\cdot x_\perp}\,x_0^{3/2}
\left[
a^+(k)\,K_{-\half+\ii\lambda}(|k|x_0)
+
a^-(k)\,K_{\half+\ii\lambda}(|k|x_0)
\right]
\end{equation}
satisfies
\begin{equation}
\label{eq:H3-eig-poincare-app}
\mathcal D_{\mathbb H^3}\,\widehat\chi_{\lambda,k}
=
\ii \lambda\,\widehat\chi_{\lambda,k},
\qquad \lambda\in\mathbb R.
\end{equation}
The delta-normalized mode used in the generalized expansion is
\begin{equation}
\label{eq:chi-mode-app-normalized}
\chi_{\lambda,k}
=
\mathscr N_{\rm P}(\lambda,k)\,\widehat\chi_{\lambda,k},
\qquad
\mathscr N_{\rm P}(\lambda,k)
=
\frac{\sqrt{|k|\cosh(\pi\lambda)}}{\pi},
\end{equation}
with the normalization derived in appendix~\ref{app:plancherel-poincare}. Moreover, since $\{\Gamma^{\hat0},\mathcal D_{\mathbb H^3}\}=0$,
\begin{equation}
\label{eq:DH-on-G0chi}
\mathcal D_{\mathbb H^3}(\Gamma^{\hat0}\chi_{\lambda,k})
=-\ii\lambda\,\Gamma^{\hat0}\chi_{\lambda,k}.
\end{equation}
For uniformity with the global decomposition, restrict to $\lambda>0$ and introduce $s=\pm1$ by
\begin{equation}
\label{eq:poincare-branch-def}
\chi^{(s)}_{\lambda,k}=\chi_{s\lambda,k},
\qquad
\lambda\in(0,\infty).
\end{equation}
Then $\mathcal D_{\mathbb H^3}\chi^{(s)}_{\lambda,k}=\ii s\lambda\,\chi^{(s)}_{\lambda,k}$ as in \eqref{eq:H3-eigs-poincare}. At $\lambda=0$ the two definitions coincide, so the endpoint is included only through a limiting prescription and is not counted twice.

\subsubsection{The $k=0$ modes}
\label{app:poincare-zero-mode}
The definition \eqref{eq:a-minus} is singular at $k=0$, but the zero-momentum eigenproblem is elementary. Choose normalized vectors $a_q$ in the same irreducible sector,
\begin{equation}
\Pi_{\rm A}a_q=a_q,
\qquad
\Gamma^{\hat1}a_q=q\,a_q,
\qquad q=\pm1.
\end{equation}
For
\begin{equation}
\chi^{(s,q)}_{\lambda,0}(x_0)
=
x_0^{\beta}\,a_q,
\end{equation}
one obtains directly from \eqref{eq:DH3}
\begin{align}
\mathcal D_{\mathbb H^3}\chi^{(s,q)}_{\lambda,0}
&=
x_0\Gamma^{\hat1}\left(\partial_{x_0}-\frac{1}{x_0}\right)
x_0^\beta a_q
\\
&=
q(\beta-1)\chi^{(s,q)}_{\lambda,0}.
\end{align}
Thus the eigenvalue equation $\mathcal D_{\mathbb H^3}\chi=\ii s\lambda\chi$ fixes
\begin{equation}
\beta=1+\ii s q\lambda,
\qquad
\chi^{(s,q)}_{\lambda,0}(x_0)
=
x_0^{1+\ii s q\lambda}a_q.
\end{equation}
These generalized power-law modes are the $k\to0$ endpoint of the Poincar\'e spectral family after a basis-dependent recombination. Since $k=0$ is a set of zero measure in $\dd^2k$, it does not add a separate term to the Plancherel expansion, but it must be treated using the direct formula above whenever a strict zero-momentum source is imposed.

\subsection{Global eigenmodes}
\label{app:dirac-modes-global}

Separation of variables on $\mathbb H^3$ in global coordinates is most naturally organized using spinor harmonics on $S^2$.
In a standard Pauli basis one may define the Hermitian operator
\begin{equation}
\label{eq:DS2-hermitian}
D_{S^2}= \ii\Big[\sigma_1\big(\partial_\theta+\tfrac12\cot\theta\big)
+ \sigma_2\csc\theta\,\partial_\phi\Big],
\end{equation}
whose spectrum is real. The tangential-frame operator \eqref{eq:DS2-frame} is anti-Hermitian and has purely imaginary spectrum. In the Pauli frame used in appendix~\ref{app:modesum}, the two operators are related simply by
\begin{equation}
\label{eq:DS2-rotation-relation}
\mathcal D_{S^2}=-\ii D_{S^2}.
\end{equation}
Any other constant tangential spin frame is related to this one by a unitary rotation.

\subsubsection{Spinor harmonics on $S^2$}
The tangential operator \eqref{eq:DS2-frame} is anti-Hermitian.
It is convenient to introduce the Hermitian angular operator
\begin{equation}
\label{eq:K-operator}
\mathcal K
=
\Gamma^{\hat y}\mathcal D_{S^2}.
\end{equation}
Its normalized eigenspinors are denoted by
$\hat\chi^{(\pm)}_{l m_j}$:
\begin{equation}
\label{eq:K-eigs}
\mathcal K\hat\chi^{(\pm)}_{l m_j}
=
\pm(l+1)\hat\chi^{(\pm)}_{l m_j},
\qquad
l\in\mathbb Z_{\ge0},
\qquad
m_j=-l-\tfrac12,-l+\tfrac12,\ldots,l+\tfrac12.
\end{equation}
The phases are chosen so that
\begin{equation}
\label{eq:hatted-basis}
\hat\chi^{(+)}_{l m_j}
=
\Gamma^{\hat y}\hat\chi^{(-)}_{l m_j}.
\end{equation}
Equivalently,
\begin{equation}
\label{eq:mixed-hatted}
\mathcal D_{S^2}\hat\chi^{(\pm)}_{l m_j}
=
\pm(l+1)\Gamma^{\hat y}\hat\chi^{(\pm)}_{l m_j}.
\end{equation}
An explicit realization in terms of normalized spin-weighted spherical
harmonics, together with the fixed-$l$ addition theorem, is given in
appendix~\ref{app:modesum}.

\subsubsection{Global $\mathbb H^3$ eigenspinors}
We define global $\mathbb H^3$ spinor harmonics $\chi^{(s)}_{\lambda l m_j}$ by
\begin{equation}
\label{eq:H3-Dirac-eigenproblem}
\mathcal D_{\mathbb H^3}\,\chi^{(s)}_{\lambda l m_j}(y,\Omega)
=
\ii\,s\lambda\;\chi^{(s)}_{\lambda l m_j}(y,\Omega),
\qquad
\lambda\in(0,\infty),\;\; s=\pm1.
\end{equation}
A separated basis regular at $y=0$ may be written as
\begin{equation}
\label{eq:H3-spinor-mode-basis}
\chi^{(s)}_{\lambda l m_j}(y,\Omega)
=
c_{3}(\lambda,l)\,\Big[
\phi_{\lambda l}(y)\,\hat\chi^{(-)}_{l m_j}(\Omega)
+s\,\ii\,\psi_{\lambda l}(y)\,\hat\chi^{(+)}_{l m_j}(\Omega)
\Big],
\end{equation}
where we choose the delta-normalizing phase convention
\begin{equation}
\label{eq:c3-normalization}
c_3(\lambda,l)
=
\frac{1}{2}\,\frac{\Gamma(l+\frac{3}2+\ii\lambda)}{\Gamma(l+\frac{3}2)\Gamma(\frac{1}2+\ii\lambda)}.
\end{equation}
Its modulus and the associated Plancherel formula are derived in appendix~\ref{app:plancherel-global}. The radial profiles may be chosen as
\begin{equation}
\label{eq:phi-psi-radial-basis}
\begin{aligned}
\phi_{\lambda l}(y)
&=
\Bigl(\cosh\frac{y}{2}\Bigr)^{l+1}
\Bigl(\sinh\frac{y}{2}\Bigr)^{l}
\\[-1mm]
&\quad\times
{}_2F_1\!\Bigl(\frac{3}{2}+l+\ii\lambda,\;\frac{3}{2}+l-\ii\lambda;
\;\frac{3}{2}+l;\;-\sinh^2\frac{y}{2}\Bigr),
\\[5pt]
\psi_{\lambda l}(y)
&=
\frac{2\lambda}{3+2l}
\Bigl(\cosh\frac{y}{2}\Bigr)^{l}
\Bigl(\sinh\frac{y}{2}\Bigr)^{l+1}
\\[-1mm]
&\quad\times
{}_2F_1\!\Bigl(\frac{3}{2}+l+\ii\lambda,\;\frac{3}{2}+l-\ii\lambda;
\;\frac{5}{2}+l;\;-\sinh^2\frac{y}{2}\Bigr).
\end{aligned}
\end{equation}
Upon restriction to the spatial Clifford algebra, a four-component Dirac spinor decomposes into two irreducible two-component spinor sectors. We take $\chi^{(s)}_{\lambda l m_j}$ to lie in one such sector; $\Gamma^{\hat0}\chi^{(s)}_{\lambda l m_j}$ supplies the complementary sector and reverses the $\mathbb H^3$ Dirac eigenvalue, as in \eqref{eq:DH-on-G0chi}. Since the first sector contains independent $+\ii\lambda$ and $-\ii\lambda$ eigenspinors, both $s=+1$ and $s=-1$ families are retained. The pairs $\{\chi^{(+)},\Gamma^{\hat0}\chi^{(+)}\}$ and $\{\chi^{(-)},\Gamma^{\hat0}\chi^{(-)}\}$ then cover the two Clifford-sector pairs entering the four-dimensional solution. At the threshold $\lambda=0$ the two regular families coincide and are included only by a limiting prescription, with no separate discrete contribution. No explicit block representation is needed in the subsequent calculation.

\section{Spinor Plancherel normalization}
\label{app:plancherel}

This appendix fixes the generalized-eigenfunction conventions used in the mode expansions. The intrinsic spinor inner product on a unit hyperbolic slice is
\begin{equation}
\label{eq:H3-spinor-inner-product}
(\chi_1,\chi_2)_{\mathbb H^3}
=
\int_{\mathbb H^3}\dd^3x\,\sqrt{g_{\mathbb H^3}}\,
\chi_1^\dagger(x)\chi_2(x).
\end{equation}
For this positive-definite spatial spectral problem we use the ordinary adjoint conventions
\begin{equation}
\label{eq:spatial-gamma-adjoints}
(\Gamma^{\hat0})^\dagger=-\Gamma^{\hat0},
\qquad
(\Gamma^{\hat i})^\dagger=\Gamma^{\hat i}
\quad (\hat i=\hat1,\hat2,\hat3),
\qquad
\Pi_{\rm A}^\dagger=\Pi_{\rm A},\qquad\Pi_{\rm B}^\dagger=\Pi_{\rm B}.
\end{equation}
They imply that $\ii\mathcal D_{\mathbb H^3}$ is self-adjoint on its standard domain and that
\begin{equation}
(\Gamma^{\hat0})^\dagger\Gamma^{\hat0}=\mathbf1_4,
\end{equation}
so $\Gamma^{\hat0}$ maps the two spatial Clifford sectors unitarily.

We invoke the standard spectral resolution of the intrinsic Dirac operator on the complete space $\mathbb H^3$ \cite{Camporesi:1995fb}. On the standard self-adjoint domain, the spectrum in each irreducible spatial Clifford sector is purely continuous and is exhausted by the generalized eigenspinors displayed below; there is no discrete $L^2$ contribution, and the threshold $\lambda=0$ carries no separate spectral atom. We therefore use $\lambda\in(0,\infty)$ and retain the two signs $s=\pm1$, with the endpoint obtained only as a limiting generalized mode and not counted twice.

\subsection{Global normalization}
\label{app:plancherel-global}

The normalized angular harmonics obey
\begin{equation}
\int_{S^2}\dd\Omega\,
\hat\chi^{(\sigma)\dagger}_{l m_j}(\Omega)
\hat\chi^{(\sigma')}_{l' m_j'}(\Omega)
=
\delta_{\sigma\sigma'}\delta_{ll'}\delta_{m_jm_j'}.
\end{equation}
For the radial functions in \eqref{eq:phi-psi-radial-basis}, hypergeometric continuation gives, as $y\to\infty$,
\begin{align}
\phi_{\lambda l}(y)
&\sim
c_l(\lambda)\,\e^{(-1+\ii\lambda)y}
+c_l(\lambda)^*\,\e^{(-1-\ii\lambda)y},
\\
\psi_{\lambda l}(y)
&\sim
-\ii\left[
c_l(\lambda)\,\e^{(-1+\ii\lambda)y}
-c_l(\lambda)^*\,\e^{(-1-\ii\lambda)y}
\right],
\end{align}
where
\begin{equation}
\label{eq:global-clambda}
c_l(\lambda)
=
\frac{2}{\sqrt\pi}
\frac{\Gamma(l+\frac{3}2)\Gamma(\frac{1}2+\ii\lambda)}{\Gamma(l+\frac{3}2+\ii\lambda)}.
\end{equation}
The generalized orthogonality is most directly obtained from the finite-cutoff Green identity. For $R>0$, the first-order radial equations imply
\begin{align}
&(\lambda'-\lambda)
\int_0^R\dd y\,\sinh^2y\,
\left[
\phi_{\lambda l}(y)\phi_{\lambda' l}(y)
+
\psi_{\lambda l}(y)\psi_{\lambda' l}(y)
\right]
\nonumber\\
&\hspace{15mm}=
\left.
\sinh^2y\,
\left[
\phi_{\lambda l}(y)\psi_{\lambda' l}(y)
-
\psi_{\lambda l}(y)\phi_{\lambda' l}(y)
\right]
\right|_{y=0}^{y=R}.
\label{eq:global-radial-green-identity}
\end{align}
The regular lower endpoint vanishes. Writing
$c_l(\lambda)=|c_l(\lambda)|\e^{\ii\delta_l(\lambda)}$, the asymptotic forms above give
\begin{align}
&\sinh^2R\,
\left[
\phi_{\lambda l}(R)\psi_{\lambda' l}(R)
-
\psi_{\lambda l}(R)\phi_{\lambda' l}(R)
\right]
\nonumber\\
&\hspace{12mm}\sim
|c_l(\lambda)c_l(\lambda')|
\sin\!\left[(\lambda'-\lambda)R
+\delta_l(\lambda')-\delta_l(\lambda)\right].
\end{align}
Since $\delta_l(\lambda')-\delta_l(\lambda)=O(\lambda'-\lambda)$, the distributional $R\to\infty$ limit of \eqref{eq:global-radial-green-identity} is
\begin{equation}
\int_0^\infty\dd y\,\sinh^2y\,
\left[
\phi_{\lambda l}(y)\phi_{\lambda' l}(y)
+
\psi_{\lambda l}(y)\psi_{\lambda' l}(y)
\right]
=
\pi |c_l(\lambda)|^2\delta(\lambda-\lambda').
\end{equation}
Applying the half-line Fourier identity term by term can leave apparent principal-value contributions depending on the phases. The cutoff Green identity proves that no additional regular off-diagonal contribution remains. For opposite spectral signs, the corresponding delta distribution is proportional to $\delta(\lambda+\lambda')$, which has no support for $\lambda,\lambda'>0$.
Choosing
\begin{equation}
|c_3(\lambda,l)|^2
=
\frac{1}{\pi|c_l(\lambda)|^2}
=
\frac{1}{4}
\frac{|\Gamma(l+\frac{3}2+\ii\lambda)|^2}{\Gamma(l+\frac{3}2)^2|\Gamma(\frac{1}2+\ii\lambda)|^2}
\end{equation}
gives the continuous-spectrum normalization
\begin{equation}
\label{eq:global-delta-normalization}
\left(
\chi^{(s)}_{\lambda l m_j},
\chi^{(s')}_{\lambda' l' m_j'}
\right)_{\mathbb H^3}
=
\delta_{ss'}\delta(\lambda-\lambda')
\delta_{ll'}\delta_{m_jm_j'}.
\end{equation}
The phase choice in \eqref{eq:c3-normalization} has precisely this modulus. These formulas are the $N=3$ specialization of the normalized hyperbolic spinor modes in \cite{Camporesi:1995fb}.

Let $\Pi_{\rm B}$ denote the projector onto the irreducible two-component spatial Clifford sector in which the harmonics \eqref{eq:H3-spinor-mode-basis} are written. Invoking the self-adjoint spectral resolution and the absence of discrete or threshold atoms stated above, the generalized normalization \eqref{eq:global-delta-normalization} gives the completeness relation
\begin{equation}
\label{eq:global-completeness-sector}
\sum_{s=\pm1}
\int_0^\infty\dd\lambda
\sum_{l=0}^\infty
\sum_{m_j=-l-\frac{1}{2}}^{l+\frac{1}{2}}
\chi^{(s)}_{\lambda l m_j}(x)
\chi^{(s)\dagger}_{\lambda l m_j}(x')
=
\delta_{\mathbb H^3}(x,x')\Pi_{\rm B}.
\end{equation}
The complementary sector is generated by $\Gamma^{\hat0}$. Thus, the full four-component spatial resolution of the identity is
\begin{align}
\label{eq:global-completeness-full}
&\sum_{s=\pm1}
\int_0^\infty\dd\lambda
\sum_{l=0}^\infty
\sum_{m_j=-l-\frac{1}{2}}^{l+\frac{1}{2}}
\Big[
\chi^{(s)}_{\lambda l m_j}(x)
\chi^{(s)\dagger}_{\lambda l m_j}(x')
\nonumber\\[-2pt]
&\hspace{45mm}
+
\Gamma^{\hat0}\chi^{(s)}_{\lambda l m_j}(x)
\bigl(\Gamma^{\hat0}\chi^{(s)}_{\lambda l m_j}(x')\bigr)^\dagger
\Big]
=
\delta_{\mathbb H^3}(x,x')\mathbf1_4.
\end{align}
This explicitly shows why both $s$ families are required when the regular harmonics are initially constructed in one irreducible spatial sector.

For comparison with the invariant spherical transform, the spinor Harish--Chandra function on $\mathbb H^3$ may be chosen as
\begin{equation}
\mathfrak c_{1/2}(\lambda)
=
\frac{\Gamma(\frac{1}2+\ii\lambda)}{\Gamma(\frac{3}2+\ii\lambda)}
=
\frac{1}{\frac{1}2+\ii\lambda}.
\end{equation}
The Plancherel measure for each of the two principal-series signs is therefore
\begin{equation}
\label{eq:spinor-plancherel-measure}
\dd\mu_s(\lambda)
=
\frac{1}{2\pi^2}
|\mathfrak c_{1/2}(\lambda)|^{-2}\dd\lambda
=
\frac{\lambda^2+\frac{1}4}{2\pi^2}\,\dd\lambda,
\qquad s=\pm1.
\end{equation}
Equation \eqref{eq:spinor-plancherel-measure} is the invariant representation-space convention. After decomposition into angular $K$-types, each regular radial partial wave carries an additional $l$-dependent normalization. We choose $c_3(\lambda,l)$ so that the resulting partial-wave modes are delta-normalized as in \eqref{eq:global-delta-normalization}, and the expansion then uses the flat measure $\dd\lambda$. If the regular but unnormalized modes $c_3(\lambda,l)^{-1}\chi$ are used instead, their partial-wave measure is $|c_3(\lambda,l)|^2\dd\lambda$. This $l$-dependent partial-wave density should not be identified directly with the $l$-independent invariant Harish--Chandra density in \eqref{eq:spinor-plancherel-measure}.

\subsection{Poincar\'e normalization}
\label{app:plancherel-poincare}

For the upper-half-space metric,
\begin{equation}
\dd V_{\mathbb H^3}
=
\frac{\dd x_0\,\dd^2x_\perp}{x_0^3}.
\end{equation}
The factor $x_0^{3/2}$ in \eqref{eq:chi-mode-app} cancels this volume weight in the radial overlap. At $x_0\to0$, the leading terms of the unnormalized mode are
\begin{align}
\widehat\chi_{s\lambda,k}
\sim
\e^{\ii k\cdot x_\perp}
\Big[
&A_+(s\lambda,k)x_0^{1+\ii s\lambda}a^+(k)
+
A_-(s\lambda,k)x_0^{1-\ii s\lambda}a^-(k)
\Big],
\end{align}
where
\begin{align}
A_+(s\lambda,k)
&=
\frac{1}{2}\Gamma\!\left(\frac{1}{2}-\ii s\lambda\right)
\left(\frac{|k|}{2}\right)^{-\frac{1}{2}+\ii s\lambda},
\\
A_-(s\lambda,k)
&=
\frac{1}{2}\Gamma\!\left(\frac{1}{2}+\ii s\lambda\right)
\left(\frac{|k|}{2}\right)^{-\frac{1}{2}-\ii s\lambda}.
\end{align}
Using
\begin{equation}
|\Gamma(\tfrac12+\ii\lambda)|^2
=
\frac{\pi}{\cosh(\pi\lambda)},
\end{equation}
we derive the radial delta normalization from a cutoff Green identity. Set $\mu=s\lambda$ and $\mu'=s'\lambda'$, and let the transverse integral first impose $k=k'$. For
\begin{equation}
I_\epsilon(\mu,\mu';k)
=
\int_\epsilon^\infty\frac{\dd x_0}{x_0^3}\,
\widehat\chi_{\mu,k}^\dagger(x_0)
\widehat\chi_{\mu',k}(x_0),
\end{equation}
integration by parts gives
\begin{equation}
\ii(\mu'-\mu)I_\epsilon(\mu,\mu';k)
=
\left[
x_0^{-2}\widehat\chi_{\mu,k}^\dagger(x_0)
\Gamma^{\hat1}
\widehat\chi_{\mu',k}(x_0)
\right]_{x_0=\epsilon}^{x_0=\infty}.
\label{eq:poincare-radial-green-identity}
\end{equation}
The upper endpoint vanishes by the exponential decay of the $K$-Bessel functions. For real $\mu$, write
\begin{equation}
A_+(\mu,k)=\rho_\mu\e^{\ii\theta_\mu},
\qquad
A_-(\mu,k)=\rho_\mu\e^{-\ii\theta_\mu},
\qquad
\rho_\mu^2=
\frac{\pi}{2|k|\cosh(\pi\mu)}.
\end{equation}
The lower endpoint of \eqref{eq:poincare-radial-green-identity} then yields
\begin{equation}
I_\epsilon(\mu,\mu';k)
\sim
2\rho_\mu\rho_{\mu'}
\frac{
\sin\!\left[(\mu'-\mu)(-\log\epsilon)
-(\theta_{\mu'}-\theta_\mu)\right]
}{\mu'-\mu}.
\end{equation}
Because $\theta_{\mu'}-\theta_\mu=O(\mu'-\mu)$, its distributional limit is
\begin{equation}
I(\mu,\mu';k)
=
2\pi\rho_\mu^2\delta(\mu-\mu')
=
\frac{\pi^2}{|k|\cosh(\pi\mu)}
\delta(\mu-\mu').
\end{equation}
A separate termwise use of the half-line Fourier identity can leave an apparently phase-dependent regular term and is not sufficient by itself. For $\lambda,\lambda'>0$, $\delta(s\lambda-s'\lambda')$ has no support when $s\neq s'$. Including the transverse plane-wave integral therefore gives
\begin{align}
\left(
\widehat\chi_{s\lambda,k},
\widehat\chi_{s'\lambda',k'}
\right)_{\mathbb H^3}=
(2\pi)^2\delta^{(2)}(k-k')
\delta_{ss'}
\frac{\pi^2}{|k|\cosh(\pi\lambda)}
\delta(\lambda-\lambda').
\end{align}
It follows that the factor in \eqref{eq:chi-mode-app-normalized},
\begin{equation}
\mathscr N_{\rm P}(\lambda,k)
=
\frac{\sqrt{|k|\cosh(\pi\lambda)}}{\pi},
\end{equation}
produces
\begin{equation}
\label{eq:poincare-delta-normalization}
\left(
\chi^{(s)}_{\lambda,k},
\chi^{(s')}_{\lambda',k'}
\right)_{\mathbb H^3}
=
(2\pi)^2\delta^{(2)}(k-k')
\delta(\lambda-\lambda')
\delta_{ss'}.
\end{equation}
Invoking the same self-adjoint spectral resolution, exhaustion of the continuous spectrum, absence of discrete modes, and absence of a separate threshold atom, the normalized Poincar\'e modes give the resolution of the identity in the chosen irreducible sector,
\begin{equation}
\label{eq:poincare-completeness-sector}
\sum_{s=\pm1}
\int_0^\infty\dd\lambda
\int_{\mathbb R^2}\frac{\dd^2k}{(2\pi)^2}
\chi^{(s)}_{\lambda,k}(x)
\chi^{(s)\dagger}_{\lambda,k}(x')
=
\delta_{\mathbb H^3}(x,x')\Pi_{\rm A}.
\end{equation}
Adding the $\Gamma^{\hat0}$-transformed complementary sector gives
\begin{align}
\label{eq:poincare-completeness}
&\sum_{s=\pm1}
\int_0^\infty\dd\lambda
\int_{\mathbb R^2}\frac{\dd^2k}{(2\pi)^2}
\Big[
\chi^{(s)}_{\lambda,k}(x)
\chi^{(s)\dagger}_{\lambda,k}(x')
\nonumber\\[-2pt]
&\hspace{42mm}+
\Gamma^{\hat0}\chi^{(s)}_{\lambda,k}(x)
\bigl(\Gamma^{\hat0}\chi^{(s)}_{\lambda,k}(x')\bigr)^\dagger
\Big]
=
\delta_{\mathbb H^3}(x,x')\mathbf1_4.
\end{align}
The point $k=0$ is absent from this integral only as a measure-zero endpoint; its direct generalized modes are given in appendix~\ref{app:poincare-zero-mode}.

\section{Dirac boundary term and radial polarization}
\label{app:dirac-boundary}

For a first-order Dirac system one fixes only one radial polarization. We
first work with smooth wave packets compactly supported in $u=\log\tau$, so
that the temporal surface terms at $\tau=0$ and $\tau=\infty$ vanish. The
boundary terms displayed below therefore refer to the radial cutoff surface;
pure Mellin modes are recovered only after the calculation, in the
distributional sense.

Throughout this appendix, $I$ denotes an action or complete on-shell
functional, while $\mathcal B$ denotes one bare boundary pairing. In the
pairing labels, $s$ means source, $r$ means response, and the bar identifies
the barred factor: $\mathcal B_{\bar s r}$ is ``barred source times unbarred
response,'' whereas $\mathcal B_{\bar r s}$ is ``barred response times
unbarred source.'' This role-based notation avoids assigning a direction to
either pairing.

All functionals are written in units of the common factor $\mathcal N_S$;
equivalently, every displayed appendix functional denotes the corresponding
full expression divided by $\mathcal N_S$. We start from the symmetrized
action
\begin{equation}
\label{eq:app-Dirac-sym-action}
I_{\rm bulk}^{\rm sym}
=
\frac{1}{2}
\int_M \dd^4x\,\sqrt{-g}\,
\left[
\bar\psi\Gamma^\mu\nabla_\mu\psi
-
(\nabla_\mu\bar\psi)\Gamma^\mu\psi
-
2m\bar\psi\psi
\right].
\end{equation}
Its on-shell variation is
\begin{equation}
\label{eq:app-Dirac-sym-var}
\delta I_{\rm bulk}^{\rm sym}\big|_{\rm os}
=
\frac{1}{2}
\int_{\partial M_\epsilon}\dd^3x\,\sqrt{|h|}
\left[
\bar\psi\Gamma^n\delta\psi
-
\delta\bar\psi\Gamma^n\psi
\right].
\end{equation}
Define
\begin{equation}
\label{eq:app-projectors}
P_\pm=\frac{1}{2}(1\pm\Gamma^n),
\qquad
\psi_\pm=P_\pm\psi,
\qquad
\bar\psi_\pm=\bar\psi P_\mp.
\end{equation}
If $\psi_-$ and $\bar\psi_-$ are fixed, add
\begin{equation}
I_{\rm pol}^{(-)}
=
-\frac{1}{2}
\int_{\partial M_\epsilon}\dd^3x\,\sqrt{|h|}\,
\bar\psi\psi.
\end{equation}
Then
\begin{equation}
\delta
\left(
I_{\rm bulk}^{\rm sym}+I_{\rm pol}^{(-)}
\right)_{\rm os}
=
-
\int_{\partial M_\epsilon}\dd^3x\,\sqrt{|h|}
\left[
\bar\psi_+\delta\psi_-
+
\delta\bar\psi_-\psi_+
\right].
\end{equation}
The full on-shell value is
\begin{equation}
I_{\rm os}^{(-)}
=
-\frac{1}{2}
\int_{\partial M_\epsilon}\dd^3x\,\sqrt{|h|}
\left(
\bar\psi_+\psi_-
+
\bar\psi_-\psi_+
\right)
\end{equation}
For this polarization, the two bare boundary pairings are
\begin{align}
\mathcal B_{\bar s r}^{(-)}
&=
-
\int_{\partial M_\epsilon}
\dd^3x\,\sqrt{|h|}\,
\bar\psi_-\psi_+,
\label{eq:app-pairing-barsr-minus}
\\
\mathcal B_{\bar r s}^{(-)}
&=
-
\int_{\partial M_\epsilon}
\dd^3x\,\sqrt{|h|}\,
\bar\psi_+\psi_-.
\label{eq:app-pairing-barrs-minus}
\end{align}
The exact decomposition is therefore
\begin{equation}
I_{\rm os}^{(-)}
=
\frac{1}{2}
\left(
\mathcal B_{\bar s r}^{(-)}
+
\mathcal B_{\bar r s}^{(-)}
\right)
\end{equation}
These are bare finite-cutoff identities. No counterterms or finite contact terms are included in these variational formulas; their absence here is not a proof that the full cutoff limit exists.

If instead $\psi_+$ and $\bar\psi_+$ are fixed, add
\begin{equation}
I_{\rm pol}^{(+)}
=
+\frac{1}{2}
\int_{\partial M_\epsilon}\dd^3x\,\sqrt{|h|}\,
\bar\psi\psi.
\end{equation}
The variation becomes
\begin{equation}
\delta
\left(
I_{\rm bulk}^{\rm sym}+I_{\rm pol}^{(+)}
\right)_{\rm os}
=
+
\int_{\partial M_\epsilon}\dd^3x\,\sqrt{|h|}
\left[
\bar\psi_-\delta\psi_+
+
\delta\bar\psi_+\psi_-
\right].
\end{equation}
The full on-shell value is
\begin{equation}
I_{\rm os}^{(+)}
=
+\frac{1}{2}
\int_{\partial M_\epsilon}\dd^3x\,\sqrt{|h|}
\left(
\bar\psi_-\psi_+
+
\bar\psi_+\psi_-
\right)
\end{equation}
The corresponding bare boundary pairings are
\begin{align}
\mathcal B_{\bar s r}^{(+)}
&=
+
\int_{\partial M_\epsilon}
\dd^3x\,\sqrt{|h|}\,
\bar\psi_+\psi_-,
\label{eq:app-pairing-barsr-plus}
\\
\mathcal B_{\bar r s}^{(+)}
&=
+
\int_{\partial M_\epsilon}
\dd^3x\,\sqrt{|h|}\,
\bar\psi_-\psi_+.
\label{eq:app-pairing-barrs-plus}
\end{align}
Thus
\begin{equation}
I_{\rm os}^{(+)}
=
\frac{1}{2}
\left(
\mathcal B_{\bar s r}^{(+)}
+
\mathcal B_{\bar r s}^{(+)}
\right)
\end{equation}
Again, this is a bare finite-cutoff identity. The main text evaluates the intrinsic separated-point coefficient associated with $\mathcal B_{\bar s r}$ only. The complementary pairing $\mathcal B_{\bar r s}$ requires the barred response determined by the left-acting Dirac equation and is not inferred from the unbarred source--response map.

For the Milne metric,
\begin{equation}
\dd s^2
=
-\dd\tau^2
+
\tau^2\dd s_{\mathbb H^3}^2,
\end{equation}
the induced measure on
$\partial M_\epsilon\simeq\mathbb R_\tau^+\times
\partial\mathbb H^3_\epsilon$ is
\begin{equation}
\dd^3x\,\sqrt{|h|}
=
\dd\tau\,\tau^2\,\dd^2\Sigma_\epsilon.
\end{equation}
Using $\psi=\tau^{-3/2}\phi$,
\begin{equation}
\dd^3x\,\sqrt{|h|}\,
\bar\psi_\pm\psi_\mp
=
\frac{\dd\tau}{\tau}\,
\dd^2\Sigma_\epsilon\,
\bar\phi_\pm\phi_\mp.
\end{equation}
Equations \eqref{eq:app-pairing-barsr-minus} and
\eqref{eq:app-pairing-barsr-plus} therefore give the $\bar s r$ Milne boundary
functionals used in section~\ref{sec:correlator}.

\section{Spinor addition theorem and global mode sum}
\label{app:modesum}

This appendix derives the fixed-$l$ projector for the angular harmonics of
appendix~\ref{app:dirac-modes-global} and evaluates the sum used in
\eqref{eq:sphere-spinor-kernel}. The final formulas use the gamma matrices
and radial projectors of the main text.

For the calculation, we choose the Pauli realization
\begin{equation}
\Gamma^{\hat y}=\sigma_3,
\qquad
\Gamma^{\hat\theta}=\sigma_1,
\qquad
\Gamma^{\hat\phi}=\sigma_2.
\end{equation}
In this oriented global frame,
$\ii\sigma_3\sigma_1\sigma_2=-\mathbf1_2$.
Thus the displayed two-component realization has $\mathcal C_3=-1$
and belongs to block $B$ under \eqref{eq:explicit-clifford-block-projectors}.

Then
\begin{equation}
\mathcal D_{S^2}
=
\sigma_1\left(\partial_\theta+\frac{1}{2}\cot\theta\right)
+
\sigma_2\csc\theta\,\partial_\phi,
\end{equation}
and the Hermitian operator defined in appendix~\ref{app:dirac-modes-global} is
\begin{equation}
\label{eq:appD-K-explicit}
\mathcal K
=
\Gamma^{\hat y}\mathcal D_{S^2}
=
\begin{pmatrix}
0&
\partial_\theta+\frac{1}{2}\cot\theta
-\ii\csc\theta\,\partial_\phi
\\[2pt]
-\partial_\theta-\frac{1}{2}\cot\theta
-\ii\csc\theta\,\partial_\phi
&0
\end{pmatrix}.
\end{equation}
Let ${{}_sY}_{j m_j}$ be normalized spin-weighted harmonics with
\begin{align}
\eth\,{}_sY_{j m_j}
&=
\sqrt{(j-s)(j+s+1)}\,{}_{s+1}Y_{j m_j},
\\
\bar\eth\,{}_sY_{j m_j}
&=
-\sqrt{(j+s)(j-s+1)}\,{}_{s-1}Y_{j m_j}.
\end{align}
For $j=l+\frac{1}{2}$, define
\begin{equation}
\label{eq:appD-angular-spinors}
\hat\chi^{(+)}_{l m_j}
=
\frac{1}{\sqrt2}
\begin{pmatrix}
{}_{-\frac{1}{2}}Y_{j m_j}\\[2pt]
{}_{+\frac{1}{2}}Y_{j m_j}
\end{pmatrix},
\qquad
\hat\chi^{(-)}_{l m_j}
=
\frac{1}{\sqrt2}
\begin{pmatrix}
{}_{-\frac{1}{2}}Y_{j m_j}\\[2pt]
-{}_{+\frac{1}{2}}Y_{j m_j}
\end{pmatrix}.
\end{equation}
The eth identities give
\begin{equation}
\label{eq:appD-angular-eigenvalues}
\mathcal K\hat\chi^{(\pm)}_{l m_j}
=
\pm(l+1)\hat\chi^{(\pm)}_{l m_j},
\qquad
\hat\chi^{(+)}_{l m_j}
=
\Gamma^{\hat y}\hat\chi^{(-)}_{l m_j}.
\end{equation}
The harmonics are normalized as
\begin{equation}
\int_{S^2}\dd\Omega\,
\hat\chi^{(\sigma)\dagger}_{l m_j}
\hat\chi^{(\sigma')}_{l'm_j'}
=
\delta_{\sigma\sigma'}\delta_{ll'}\delta_{m_jm_j'}.
\end{equation}
We now evaluate the fixed-$l$ spectral projector. Define
\begin{equation}
\label{eq:appD-projector-def}
\Pi_l^{(-)}(\Omega,\Omega')
=
\sum_{m_j}
\hat\chi^{(-)}_{l m_j}(\Omega)
\hat\chi^{(-)\dagger}_{l m_j}(\Omega').
\end{equation}
Let
\begin{equation}
\cos\gamma
=
\hat n(\Omega)\cdot\hat n(\Omega'),
\qquad
\Xi_{\hat\alpha}(\Omega,\Omega')
=
\nabla_{\hat\alpha}\gamma,
\end{equation}
and let $\mathcal P(\Omega,\Omega')$ denote spin parallel transport from
$\Omega'$ to $\Omega$ along the shorter geodesic.

To evaluate \eqref{eq:appD-projector-def}, place $\Omega'$ at the north pole
and $\Omega=(\gamma,0)$, and use the spin frame parallel-transported along
that geodesic. In this frame $\mathcal P=\mathbf1$ and
$\Gamma^{\hat\alpha}\Xi_{\hat\alpha}=\Gamma^{\hat\theta}$. The
spin-weighted harmonics are Wigner matrix elements, so the sum over $m_j$ is
the corresponding $2\times2$ block of the relative rotation. With
$j=l+\frac{1}{2}$,
\begin{align}
d^{\,l+\frac{1}{2}}_{\frac{1}{2},\frac{1}{2}}(\gamma)
&=
d^{\,l+\frac{1}{2}}_{-\frac{1}{2},-\frac{1}{2}}(\gamma)
=
\cos\frac{\gamma}{2}\,
P_l^{(0,1)}(\cos\gamma),
\\
d^{\,l+\frac{1}{2}}_{-\frac{1}{2},\frac{1}{2}}(\gamma)
&=
\sin\frac{\gamma}{2}\,
P_l^{(1,0)}(\cos\gamma),
\\
d^{\,l+\frac{1}{2}}_{\frac{1}{2},-\frac{1}{2}}(\gamma)
&=
-\sin\frac{\gamma}{2}\,
P_l^{(1,0)}(\cos\gamma).
\end{align}
Set
\begin{equation}
\label{eq:appD-varphi-vartheta}
\varphi_l(\gamma)
=
\cos\frac{\gamma}{2}\,
P_l^{(0,1)}(\cos\gamma),
\qquad
\vartheta_l(\gamma)
=
\sin\frac{\gamma}{2}\,
P_l^{(1,0)}(\cos\gamma).
\end{equation}
The projector in the geodesic frame is
\begin{equation}
\Pi_l^{(-)}(\gamma)
=
\frac{l+1}{4\pi}
\begin{pmatrix}
\varphi_l&\vartheta_l\\
-\vartheta_l&\varphi_l
\end{pmatrix}
=
\frac{l+1}{4\pi}
\left[
\varphi_l\mathbf1
+
\vartheta_l\Gamma^{\hat y}\Gamma^{\hat\theta}
\right].
\end{equation}
Restoring covariance gives
\begin{equation}
\label{eq:appD-projector}
\Pi_l^{(-)}(\Omega,\Omega')
=
\frac{l+1}{4\pi}
\left[
\varphi_l(\gamma)\mathbf1
+
\vartheta_l(\gamma)
\Gamma^{\hat y}
\Gamma^{\hat\alpha}\Xi_{\hat\alpha}(\Omega,\Omega')
\right]
\mathcal P(\Omega,\Omega').
\end{equation}
The sign of the second term follows from
$\Xi_{\hat\alpha}=\nabla_{\hat\alpha}\gamma$.

At coincidence,
\begin{equation}
\Pi_l^{(-)}(\Omega,\Omega)
=
\frac{l+1}{4\pi}\mathbf1,
\end{equation}
so
\begin{equation}
\int_{S^2}\dd\Omega\,
\operatorname{tr}\Pi_l^{(-)}(\Omega,\Omega)
=
2(l+1),
\end{equation}
the correct degeneracy. This also shows why the identity-type term in
\eqref{eq:appD-projector} cannot be omitted. The global source and response harmonics are \eqref{eq:global-boundary-harmonics}. Their addition theorem follows by projection
\begin{align}
\label{eq:appD-projected-addition}
\sum_{m_j}
v_{l m_j}(\Omega)u_{l m_j}^\dagger(\Omega')
&=
2P_-(\Omega)
\Pi_l^{(-)}(\Omega,\Omega')
P_+(\Omega')
\nonumber\\
&=
\frac{l+1}{2\pi}
\vartheta_l(\gamma)\,
P_-(\Omega)\Gamma^{\hat y}
\Gamma^{\hat\alpha}\Xi_{\hat\alpha}(\Omega,\Omega')
\mathcal P(\Omega,\Omega')
P_+(\Omega').
\end{align}
The $\varphi_l$ structure vanishes only after the source and response
projections have been applied.

We now evaluate the $l$ sum. Define
\begin{equation}
r_l(\nu)
=
\frac{\Gamma(l+\frac{3}2+\ii\nu)}{\Gamma(l+\frac{3}2-\ii\nu)}.
\end{equation}
The series needed in the global kernel is
\begin{equation}
\label{eq:appD-Abel-sum}
\mathscr S_\nu(\gamma)
=
\lim_{q\to1^-}
\sum_{l=0}^\infty
q^l(l+1)r_l(\nu)\vartheta_l(\gamma),
\qquad
0<\gamma<\pi.
\end{equation}
Set $x=\cos\gamma$. The Jacobi orthogonality relation
\begin{equation}
\int_{-1}^1\dd x\,
(1-x)
P_l^{(1,0)}(x)
P_{l'}^{(1,0)}(x)
=
\frac{2}{l+1}\delta_{ll'}
\end{equation}
gives the coefficient of
$F_\nu(x)=(1-x)^{-3/2-\ii\nu}$
\begin{align}
a_l(\nu)
&=
\frac{l+1}{2}
\int_{-1}^1\dd x\,
(1-x)^{-\frac{1}{2}-\ii\nu}P_l^{(1,0)}(x)
\nonumber\\
&=
2^{-\frac{1}{2}-\ii\nu}(l+1)
\frac{\Gamma(\frac{1}2-\ii\nu)}{\Gamma(\frac{3}2+\ii\nu)}
r_l(\nu).
\end{align}
For completeness, the last equality follows by setting $t=(1-x)/2$, using
\begin{equation}
P_l^{(1,0)}(1-2t)
=
(l+1)\,
{}_2F_1(-l,l+2;2;t),
\end{equation}
and applying the terminating Pfaff--Saalsch\"utz identity
\begin{equation}
{}_3F_2\!\left(
\begin{matrix}
-l,\ l+2,\ a\\
2,\ a+1
\end{matrix};1
\right)
=
\frac{(2-a)_l}{(l+1)(a+1)_l},
\qquad
a=\frac{1}{2}-\ii\nu.
\end{equation}
Therefore,
\begin{equation}
\label{eq:appD-spinor-sum}
\mathscr S_\nu(\gamma)
=
2^{1+2\ii\nu}
\frac{\Gamma(\frac{3}2+\ii\nu)}{\Gamma(\frac{1}2-\ii\nu)}
\bigl[2(1-\cos\gamma)\bigr]^{-1-\ii\nu}.
\end{equation}
The equality is understood by Abel continuation away from coincidence. Terms supported at $\gamma=0$ are local contact terms.

Combining \eqref{eq:Mhat-harm}, \eqref{eq:appD-projected-addition}, and \eqref{eq:appD-spinor-sum} gives
\begin{align}
\mathcal K_\nu^{S^2}(\Omega,\Omega')
&=\frac{\kappa_{\rm G}}{\ii}2^{-2\ii\nu}
\frac{\Gamma(\frac{1}2-\ii\nu)}{\Gamma(\frac{1}2+\ii\nu)}
\sum_{l=0}^\infty r_l(\nu)\sum_{m_j} v_{l m_j}(\Omega)u_{l m_j}^\dagger(\Omega')
\nonumber\\
&=-\frac{\kappa_{\rm G}}{\ii}\frac{1+2\ii\nu}{2\pi}
\frac{P_-(\Omega)\Gamma^{\hat\alpha}\Xi_{\hat\alpha}(\Omega,\Omega')
\mathcal P(\Omega,\Omega')P_+(\Omega')}{[2(1-\cos\gamma)]^{1+\ii\nu}}.
\end{align}
This is \eqref{eq:sphere-spinor-kernel} with \eqref{eq:cnu-sphere} for $0<\gamma<\pi$; the Abel-regulated harmonic sum supplies the global distributional prescription. Its short-distance behavior is $\mathcal K_\nu^{S^2}\sim\slashed\Xi\,\gamma^{-2-2\ii\nu}$, which agrees with the planar kernel.

\section{Radial recursion and complete-source counterterms}
\label{app:counterterm-audit}

This appendix gives the detailed radial analysis underlying subsection~\ref{subsec:ads3-renormalization}. Its purpose is to separate the exact regular intrinsic source--response map, its formal local source-branch recursion, and the divergent terms of the complete four-component cutoff problem. The intrinsic odd-derivative divergent subset is empty in $0<\Re\mu<\tfrac12$. The complete equal-weight problem nevertheless has one derivative-zero cross-block divergence, whose full-source counterterm and renormalized limit are derived below.

\subsection{Planar radial recursion}

Temporarily replace $\ii\nu$ by a complex parameter $\mu$ and work first in the strip
\begin{equation}
0<\Re\mu<\frac{1}{2}.
\label{eq:audit-strip}
\end{equation}
Set $\rho=x_0$, $s=\Pi_-\chi$, $r=\Pi_+\chi$, and
\begin{equation}
D_\rho=\slashed D_{\gamma_\rho}=\rho\slashed\partial,
\qquad
D_\rho^2=\rho^2\partial^2.
\end{equation}
Projecting the intrinsic Dirac equation \eqref{eq:DH3} gives
\begin{equation}
(\rho\partial_\rho-1-\mu)r=-D_\rho s,
\qquad
(\rho\partial_\rho-1+\mu)s=D_\rho r.
\label{eq:ct-projected-radial}
\end{equation}
For a source--response map $r=A_\rho s$, these equations imply the Riccati equation
\begin{equation}
\rho\partial_\rho A_\rho
=2\mu A_\rho-D_\rho-A_\rho D_\rho A_\rho.
\label{eq:ct-riccati}
\end{equation}
The source branch of the local asymptotic solution may be written as
\begin{equation}
A_\rho^{\rm loc}
=\sum_{n=0}^{\infty}b_n(\mu)D_\rho^{2n+1},
\qquad
d_n\equiv2\mu-(2n+1).
\label{eq:audit-local-series}
\end{equation}
Since $\rho\partial_\rho D_\rho^{2n+1}=(2n+1)D_\rho^{2n+1}$, matching powers in \eqref{eq:ct-riccati} gives
\begin{equation}
b_0=\frac{1}{d_0},
\qquad
b_n=\frac{1}{d_n}\sum_{a=0}^{n-1}b_a b_{n-1-a},
\qquad n\geq1.
\label{eq:ct-b-recursion}
\end{equation}
The first terms are
\begin{equation}
A_\rho^{\rm loc}
=
\frac{D_\rho}{d_0}
+
\frac{D_\rho^3}{d_0^2d_1}
+
\frac{2D_\rho^5}{d_0^3d_1d_2}
+
\frac{(10\mu-17)D_\rho^7}{d_0^4d_1^2d_2d_3}
+\cdots.
\label{eq:audit-first-local-terms}
\end{equation}
The recursion does not terminate for generic $\mu$. This fact is sometimes obscured if the recursion is discussed directly as a ``counterterm expansion.'' Equation~\eqref{eq:ct-b-recursion} constructs the complete formal local source branch of the cutoff response. Holographic renormalization, by contrast, uses only those terms whose insertion in the regulated action diverges as the cutoff is removed.

\subsection{Independent Bessel check}

For an interior-regular mode the exact planar map is
\begin{equation}
A_\epsilon(\vec k,\mu)
=
\frac{\ii\slashed k}{k}
\frac{K_{\frac{1}2-\mu}(z)}{K_{\frac{1}2+\mu}(z)},
\qquad
z=k\epsilon.
\label{eq:ads3-response-map-renormalization}
\end{equation}
For nonresonant $\mu$, define
\begin{equation}
F_a(z)=\sum_{j=0}^{\infty}\frac{(z^2/4)^j}{j!\,(a)_j},
\qquad
c_\mu=2^{-2\mu}\frac{\Gamma(\frac{1}2-\mu)}{\Gamma(\frac{1}2+\mu)},
\qquad
q_\mu=-\frac{c_\mu}{1+2\mu}.
\end{equation}
The two Frobenius branches of the Bessel functions give
\begin{equation}
\frac{K_{\frac{1}2-\mu}(z)}{K_{\frac{1}2+\mu}(z)}
=
\frac{c_\mu z^{2\mu}F_{\frac{1}2+\mu}(z)
+(z/d_0)F_{\frac{3}2-\mu}(z)}{F_{\frac{1}2-\mu}(z)
+q_\mu z^{1+2\mu}F_{\frac{3}2+\mu}(z)}.
\label{eq:ct-exact-bessel-split}
\end{equation}
Keeping only the source branch in numerator and denominator yields
\begin{align}
A_\epsilon^{\rm loc}(\vec k,\mu)
&=
\frac{\ii\slashed k}{k}
\frac{z}{d_0}
\frac{F_{\frac{3}2-\mu}(z)}{F_{\frac{1}2-\mu}(z)}
\nonumber\\
&=
\frac{\ii\slashed k}{k}
\left[
\frac{z}{d_0}
-
\frac{z^3}{d_0^2d_1}
+
\frac{2z^5}{d_0^3d_1d_2}
-
\frac{(10\mu-17)z^7}{d_0^4d_1^2d_2d_3}
+O(z^9)
\right].
\label{eq:ct-bessel-local-check}
\end{align}
Because the Fourier symbol of $D_\epsilon$ is $\ii\epsilon\slashed k$, this agrees term by term with \eqref{eq:audit-first-local-terms}. Interior regularity fixes the separate nonanalytic coefficient proportional to $c_\mu z^{2\mu}$; it is not determined by the local recursion.
The exact expression also controls the remainder; decay of each formal
local monomial alone would not establish convergence of the full map.
For fixed $k>0$ and $0\leq\Re\mu<\tfrac12$, define
\begin{equation}
\mathcal K_\mu(\vec k)=
\frac{\ii\slashed k}{k}\,c_\mu k^{2\mu}.
\end{equation}
Since $F_a(k\rho)=1+O(\rho^2)$, the denominator in
\eqref{eq:ct-exact-bessel-split} is
$1+O(\rho^2)+O(\rho^{1+2\Re\mu})$ after its leading Frobenius
factor has been removed. It is therefore bounded away from zero for
sufficiently small $\rho$. Consequently, in any fixed matrix norm,
\begin{equation}
\left\|\rho^{-2\mu}A_\rho(\vec k,\mu)-\mathcal K_\mu(\vec k)\right\|
\leq C\left(
\rho^{1-2\Re\mu}+\rho^2+\rho^{1+2\Re\mu}\right).
\label{eq:audit-intrinsic-remainder-bound}
\end{equation}
The constant and the sufficiently small cutoff can be chosen uniformly
on compact subsets of $0\leq\Re\mu<\tfrac12$ and compact momentum
annuli $0<k_{\min}\leq k\leq k_{\max}<\infty$.
In particular, at fixed $\mu=\ii\nu$ the error is $O(\rho)$.
This proves the direct intrinsic cutoff limit without interchanging
an infinite formal series with a limit. It also permits integration
against test sources supported in these compact parameter sets.
It is not a bound uniform in arbitrarily large momentum or angular
momentum. The position-space and spherical kernels retain their
separate distributional and Abel prescriptions. The Milne endpoint
$\nu=0$ is excluded from opposite-phase pairing here, irrespective of
whether a particular intrinsic formula has a continuous endpoint limit.

\subsection{Potentially divergent intrinsic local terms}

Define rescaled cutoff sources
\begin{equation}
j_\epsilon=\epsilon^{\mu-1}s_\epsilon,
\qquad
\bar j_\epsilon=\epsilon^{\mu-1}\bar s_\epsilon.
\end{equation}
For one opposite-phase intrinsic coefficient pairing the regulated kernel is
\begin{equation}
\mathcal B_{\epsilon;\bar s r}^{\mathrm{intr}}
=
\kappa_{\rm P}
\int\frac{\dd^2k}{(2\pi)^2}
\bar j_\epsilon(-\vec k)
\mathscr K_\epsilon^{\rm reg}(\vec k,\mu)
j_\epsilon(\vec k),
\qquad
\mathscr K_\epsilon^{\rm reg}=\epsilon^{-2\mu}A_\epsilon.
\label{eq:regulated-coefficient-kernel}
\end{equation}
The finite nonlocal contribution is
\begin{equation}
\mathscr K_{\epsilon}^{\rm nonloc}
\longrightarrow
\frac{\ii\slashed k}{k}
2^{-2\mu}
\frac{\Gamma(\frac{1}2-\mu)}{\Gamma(\frac{1}2+\mu)}
k^{2\mu}.
\label{eq:audit-finite-nonlocal}
\end{equation}
Every local term produced by \eqref{eq:ct-b-recursion} is contained in
\begin{equation}
\mathscr K_{\epsilon}^{\rm loc}
=
\sum_{n=0}^{\infty}
b_n(\mu)
\bigl(\ii\slashed k\bigr)(-k^2)^n
\epsilon^{2n+1-2\mu}.
\label{eq:audit-all-local-terms}
\end{equation}
Equation~\eqref{eq:audit-all-local-terms} exhausts the potentially divergent local source-branch terms of the intrinsic operator, not of the full four-component pairing. For a nonresonant $\mu$, the power-divergent subset is therefore
\begin{equation}
\left\{
n\in\mathbb Z_{\geq0}
\;\middle|\;
2n+1<2\Re\mu
\right\}.
\label{eq:audit-divergent-set}
\end{equation}
For any fixed $\mu$ this set is finite. In the strip \eqref{eq:audit-strip} it is empty. On the principal-series line $\mu=\ii\nu$ it remains empty and, more strongly,
\begin{equation}
\left|\epsilon^{2n+1-2\ii\nu}\right|
=\epsilon^{2n+1}.
\end{equation}
Thus each term of the intrinsic local series vanishes as $\epsilon\to0$. The first source descendants in $j_\epsilon$ begin at relative order $\epsilon^2$, while the first correction induced by the response branch begins at a positive real power in the strip. Multiplying by these corrections can only increase the real cutoff exponent and cannot produce a divergence.

If one analytically continues to $\Re\mu>\tfrac12$, a minimal planar counterterm operator can be written as the finite truncation
\begin{equation}
C_\gamma^{\rm div}(\mu)
=-
\sum_{\substack{n\geq0\\2n+1<2\Re\mu}}
b_n(\mu)D_\gamma^{2n+1}.
\label{eq:audit-planar-div-ct}
\end{equation}
At a positive half-integer resonance $\mu=N+\tfrac12$, the colliding Frobenius powers require the usual logarithmic replacement at the resonant order. Neither situation occurs in the strip used here or on the principal-series line. Consequently
\begin{equation}
C_{\gamma,\mathrm{intr}}^{\rm div}=0
\end{equation}
for the physical problem.

\subsection{Round spherical cutoff and curvature completion}

The same conclusion can be checked covariantly on the spherical cutoff. Use
\begin{equation}
\rho=2e^{-y},
\qquad
a(\rho)=\rho^{-1}-\frac{\rho}{4},
\qquad
\gamma_\rho=a(\rho)^2g_{S^2},
\end{equation}
and define
\begin{equation}
D_\rho=a^{-1}\mathcal D_{S^2},
\qquad
w=a^{-2}=\frac{R[\gamma_\rho]}{2},
\qquad
t=D_\rho^2.
\end{equation}
The inward radial gamma matrix is $-\Gamma^{\hat y}$, so the source is $s=P_+\chi$ and the response is $r=P_-\chi$. Their projected equations again imply \eqref{eq:ct-riccati}. Now
\begin{equation}
\rho\partial_\rho D_\rho=hD_\rho,
\qquad
\rho\partial_\rho w=2hw,
\qquad
h=\sqrt{1+w}.
\end{equation}
Write the local solution as
\begin{equation}
A_\rho^{\rm loc}
=D_\rho\sum_{n=0}^{\infty}P_n(t,w),
\end{equation}
where $P_n$ is homogeneous of degree $n$ if $t$ and $w$ are assigned degree one. Expanding
\begin{equation}
h=\sum_{\ell=0}^{\infty}h_\ell w^\ell,
\qquad
h_\ell=\binom{1/2}{\ell},
\end{equation}
gives
\begin{equation}
P_0=\frac{1}{d_0},
\qquad
P_n=\frac{1}{d_n}\left[
t\sum_{a=0}^{n-1}P_aP_{n-1-a}
+\sum_{\ell=1}^{n}h_\ell w^\ell
\bigl(2(n-\ell)+1\bigr)P_{n-\ell}
\right].
\label{eq:ct-curvature-recursion}
\end{equation}
For example,
\begin{align}
P_1(t,w)
&=
\frac{t}{d_0^2d_1}
+
\frac{w}{2d_0d_1},
\\
P_2(t,w)
&=
\frac{2t^2}{d_0^3d_1d_2}
+
\frac{5tw}{2d_0^2d_1d_2}
+
\frac{(9-2\mu)w^2}{8d_0d_1d_2}.
\label{eq:ct-curvature-explicit}
\end{align}
Since $D_\rho=O(\rho)$ and $w=O(\rho^2)$, every term in $D_\rho P_n$ has radial weight $2n+1$. After including the two source factors and the round cutoff measure, its contribution again scales as
\begin{equation}
\rho^{2n+1-2\mu}.
\label{eq:audit-spherical-scaling}
\end{equation}
Thus the intrinsic curved-cutoff recursion produces no new divergence in the stated strip. If one were to continue to a region where counterterms are necessary, the covariant divergent operator on the round sphere would be the finite truncation
\begin{equation}
C_\gamma^{\rm div}(\mu)
=-D_\gamma
\sum_{\substack{n\geq0\\2n+1<2\Re\mu}}
P_n\!\left(D_\gamma^2,\frac{R[\gamma]}{2}\right),
\label{eq:audit-spherical-div-ct}
\end{equation}
with logarithmic replacement at resonances. For $0<\Re\mu<\tfrac12$ and for $\mu=\ii\nu$ this sum is empty.

\subsection{Independent barred sources and the complete boundary pairing}
\label{subsec:ct-full-pairing}

The intrinsic recursion does not exhaust the complete four-component
pairing. We analyze the additional terms directly, using the equal-sign
boundary orderings derived in appendix~\ref{app:dirac-boundary}.
Write $G=\Gamma^{\hat0}$, so $G^2=-\mathbf1$ and
$G\Pi_\pm=\Pi_\mp G$. All products below retain barred variables on the
left; no Grassmann interchange is made. We work at fixed $\nu>0$ and
set $\sigma_-=-1$ and $\sigma_+=+1$. Opposite-phase Mellin pairing is
imposed on the physical line before continuing the spatial radial
problem from $\ii\nu$ to $\mu$. The continued spatial formulas are used
for coefficient extraction; with uncontinued time phases they are not
asserted to solve the four-dimensional equation away from that line.

\medskip\noindent\textit{The barred radial branch.}\
The rescaled massless left equation is
\begin{equation}
(\partial_\tau\bar\phi)G+
\tau^{-1}\bar\phi\overleftarrow{\mathcal D}_{\mathbb H^3}=0.
\label{eq:audit-left-dirac}
\end{equation}
If $\bar\chi\overleftarrow{\mathcal D}_{\mathbb H^3}
=s_{\rm L}\ii\nu\bar\chi$, with $s_{\rm L}=\pm1$, its phase-$q$ solution is
\begin{equation}
\bar\phi^q=\tau^{\sigma_q\ii\nu}
\bar\chi(1+s_{\rm L}\sigma_qG).
\label{eq:audit-left-branch}
\end{equation}
Indeed, anticommutation of the spatial operator with $G$ reduces
\eqref{eq:audit-left-dirac} to
$\ii\nu\bar\chi[\sigma_qG-s_{\rm L}+s_{\rm L}-\sigma_qG]=0$.

For a chosen intrinsic representative, the sign $s_{\rm L}$ is fixed by
which radial coefficient is prescribed as its source. In the
Poincar\'e polarization the intrinsic barred source is
$\bar J=\bar\chi\Pi_+$. At zero tangential momentum, its indicial equation is
\begin{equation}
\rho(\partial_\rho-\rho^{-1})\bar J=\Lambda_{\rm L}\bar J,
\qquad \bar J\sim\rho^{1+\Lambda_{\rm L}},
\qquad \bar\chi\overleftarrow{\mathcal D}_{\mathbb H^3}
=\Lambda_{\rm L}\bar\chi.
\label{eq:audit-barred-indicial}
\end{equation}
Consequently independent sources of the equal analytic weights
$J\sim\rho^{1-\mu}j$ and $\bar J\sim\rho^{1-\mu}\bar j$
require $\Lambda_{\rm L}=-\mu$, or $s_{\rm L}=-1$. The alternative $s_{\rm L}=+1$
has barred source weight $1+\mu$. On the physical principal-series
line that is the weight conjugate to the unbarred source weight;
it is not the equal-weight intrinsic coefficient prescription.
Complex conjugation is not an operation to be continued holomorphically
by simply replacing $\ii\nu$ with $\mu$.

This distinction concerns intrinsic parametrizations, not two disjoint
spaces of four-component solutions. If $\bar\chi_+$ has left eigenvalue
$+\ii\nu$, then $\bar\chi_+G$ has eigenvalue $-\ii\nu$, and
\begin{equation}
\bar\chi_+(1+\sigma_qG)
=\sigma_q(\bar\chi_+G)(1-\sigma_qG).
\label{eq:audit-left-reparametrization}
\end{equation}
Thus the same full branch can be represented using either eigenvalue
sign, provided its intrinsic data are transformed as well. In
particular, for $\bar\chi_- =\sigma_q\bar\chi_+G$,
\begin{equation}
\bar J^{[-1]} =\sigma_q\bar T^{[+1]}G,
\qquad \bar T^{[-1]} =\sigma_q\bar J^{[+1]}G,
\label{eq:audit-left-data-exchange}
\end{equation}
where the bracketed superscripts label the intrinsic left
eigenvalue sign, not the Milne phase. This transformation exchanges
source and response projections and the spatial Clifford block.
It cannot be made while leaving the prescribed intrinsic coefficient
sources unchanged. In particular, the stated $1+\mu$ weight for
$s_{\rm L}=+1$ refers to $\bar J$, not to the complete projection $\bar s$,
which also contains a term proportional to $\bar T G$.

\medskip\noindent\textit{Exact pairing identity.}\
Suppressing phase, spectral, and block indices, define
\begin{equation}
J=\Pi_-\chi,\quad T=\Pi_+\chi,\qquad
\bar J=\bar\chi\Pi_+,\quad \bar T=\bar\chi\Pi_-.
\end{equation}
For the equal-weight branch, the projected left radial equations are
\begin{equation}
\begin{aligned}
(\rho\partial_\rho-1+\mu)\bar J
&=-\bar T\overleftarrow D_\rho,\\
(\rho\partial_\rho-1-\mu)\bar T
&=\bar J\overleftarrow D_\rho,
\qquad
\bar f\overleftarrow D_\rho
=\rho(\partial_{\hat\alpha}\bar f)\Gamma^{\hat\alpha}.
\end{aligned}
\label{eq:audit-left-projected-equations}
\end{equation}
For a paired Fourier mode write
$\bar T(-\vec k)=\bar J(-\vec k)\bar A_\rho(\vec k)$. Since
$\bar f(-\vec k)\overleftarrow D_\rho=-\bar f(-\vec k)D_\rho(\vec k)$,
substitution in \eqref{eq:audit-left-projected-equations} gives
\begin{equation}
\rho\partial_\rho\bar A_\rho
=2\mu\bar A_\rho-D_\rho-\bar A_\rho D_\rho\bar A_\rho.
\label{eq:audit-barred-riccati}
\end{equation}
This is the same Riccati equation as \eqref{eq:ct-riccati}. The regular
interior branch and the source normalization are also the same, hence
\begin{equation}
\bar T(-\vec k)=\bar J(-\vec k)A_\rho(\vec k,\mu).
\label{eq:audit-barred-response-map}
\end{equation}
This equality is an operator statement at the paired momenta; it does not
identify barred and unbarred Grassmann sources. In particular, the local
barred response starts at one tangential derivative and the independent
response branch has weight $1+\mu$.

Set $a=\sigma_p$ and $b=s_{\rm L}\sigma_q$.
After removing the time factors, the full branch projections are
\begin{align}
s&=J+aGT,& r&=T+aGJ,\\
\bar s&=\bar J+b\bar TG,& \bar r&=\bar T+b\bar JG.
\label{eq:audit-full-projections}
\end{align}
Direct multiplication gives
\begin{equation}
\frac12(\bar s r+\bar r s)
=\frac{1-ab}{2}(\bar JT+\bar TJ)
+\frac{a+b}{2}(\bar JGJ+\bar TGT).
\label{eq:audit-exact-full-pairing}
\end{equation}
Mellin pairing imposes $\sigma_q=-\sigma_p$. For the equal-weight
independent source branch $s_{\rm L}=-1$, it follows that $b=a$, and hence
\begin{equation}
\frac12(\bar s r+\bar r s)=a(\bar JGJ+\bar TGT).
\label{eq:audit-independent-full-pairing}
\end{equation}
The derivative-zero terms therefore add rather than cancel.
Moreover, the terms linear in the intrinsic responses cancel in this
particular symmetrized pairing. The intrinsic coefficient extracted
from $\bar JT$ is thus not by itself the kernel of the complete action.
For $s_{\rm L}=+1$ one instead has $b=-a$, and
\eqref{eq:audit-exact-full-pairing} reduces to $\bar JT+\bar TJ$.
That identity cannot be used with the equal-weight barred asymptotics.

\medskip\noindent\textit{Full-source response operator and the divergent term.}\
For the equal-weight independent source branch, $b=a$. Using
\eqref{eq:audit-barred-response-map} and $\{G,A_\rho\}=0$, the complete
cutoff fields may therefore be written
\begin{align}
s&=(1+aGA_\rho)J,
& r&=(A_\rho+aG)J,\nonumber\\
\bar s&=\bar J(1-aGA_\rho),
& \bar r&=\bar J(A_\rho+aG).
\label{eq:audit-full-source-factorization}
\end{align}
For sufficiently small $\rho$ in the strip, $A_\rho\to0$, and the inverses
below exist. Since
\begin{equation}
(1+aGA_\rho)^{-1}
=\frac{1-aGA_\rho}{1-A_\rho^2},
\end{equation}
the unbarred response is $r=\mathbb F_{a,\rho}s$ with
\begin{equation}
\mathbb F_{a,\rho}
=(A_\rho+aG)(1+aGA_\rho)^{-1}
=
\frac{2A_\rho+aG(1+A_\rho^2)}{1-A_\rho^2}.
\label{eq:audit-full-source-response-operator}
\end{equation}
The barred relation gives the same operator from the left,
\begin{equation}
\bar r=\bar s\,\mathbb F_{a,\rho},
\label{eq:audit-barred-full-source-response}
\end{equation}
because
$(1-aGA_\rho)^{-1}(A_\rho+aG)$ reduces to the same expression in
\eqref{eq:audit-full-source-response-operator}. Thus the two orderings of
the symmetrized action agree when expressed in the actual complete cutoff
sources. This equality does not follow from exchanging Grassmann variables;
it follows from the two independently solved radial maps.

With the planar cutoff measure $\sqrt{\gamma_\rho}=\rho^{-2}$, the leading
term of \eqref{eq:audit-full-source-response-operator} is $aG$. For
$s_\rho=\rho^{1-\mu}j+\cdots$ and
$\bar s_\rho=\rho^{1-\mu}\bar j+\cdots$, it produces
\begin{equation}
I_\rho^{(0),p}
=\kappa_{\rm P}a\rho^{-2\mu}
\int\dd^2x\,\bar jGj+\cdots.
\label{eq:audit-cross-block-leading}
\end{equation}
Since $G\Pi_I=\Pi_{\bar I}G$, this is a derivative-zero cross-block term. It
is power divergent for $0<\Re\mu<\tfrac12$. It is not contained in the
intrinsic odd-derivative series \eqref{eq:audit-all-local-terms}.

\medskip\noindent\textit{Counterterm and renormalized strip limit.}\
The divergent term is already a local functional of the actual fixed cutoff
sources. Restoring the opposite barred phase explicitly, the required
counterterm for one phase $p$ is
\begin{equation}
I_{\rho,{\rm ct}}^{(0),p}
=-\kappa_{\rm P}\sigma_p
\int\dd^2x\,\sqrt{\gamma_\rho}\,
\bar s_\rho^{-p}\,G\,s_\rho^p
\,.
\label{eq:audit-full-source-counterterm}
\end{equation}
Equivalently,
\begin{equation}
C_{\gamma,p}^{\rm div}=-\sigma_pG
=-\sigma_p\Gamma^{\hat0},
\qquad 0<\Re\mu<\frac12.
\label{eq:audit-complete-div-operator}
\end{equation}
This is an admissible fixed-Mellin Dirichlet counterterm: it is expressed in
the boundary variables that are actually held fixed, and its variation
vanishes under the allowed variational problem $\delta s=\delta\bar s=0$.
The intrinsic derivative terms need no additional subtraction in this strip,
and terms quadratic or higher in $A_\rho$ carry positive residual radial
weight after \eqref{eq:audit-full-source-counterterm} is removed.

The exact subtracted response operator is
\begin{equation}
\mathbb F_{a,\rho}-aG
=
\frac{2A_\rho+2aGA_\rho^2}{1-A_\rho^2}.
\label{eq:audit-subtracted-full-operator}
\end{equation}
The Bessel estimate \eqref{eq:audit-intrinsic-remainder-bound} implies, for
fixed momentum in the stated compact sets,
\begin{equation}
\rho^{-2\mu}A_\rho\longrightarrow\mathcal K_\mu,
\qquad
A_\rho\longrightarrow0,
\qquad
0<\Re\mu<\frac12.
\end{equation}
Therefore
\begin{equation}
\lim_{\rho\to0}
\rho^{-2\mu}
\bigl(\mathbb F_{a,\rho}-aG\bigr)
=2\mathcal K_\mu
\,.
\label{eq:audit-complete-ren-limit}
\end{equation}
No termwise interchange of an infinite local series with the cutoff limit is
used here. In terms of the intrinsic coefficients, the same result follows
from the exact identities
\begin{equation}
r-aGs=2T,
\qquad
\bar r-a\bar sG=2\bar T.
\label{eq:audit-subtracted-response-identities}
\end{equation}
Consequently, with the same normalization $\kappa_{\rm P}$ used in
$\mathcal F_{\bar s r}^{\rm asy}$,
\begin{equation}
\left.I_{\rm ren}^{\rm sym}\right|_{\rm nonlocal}
=2\,\mathcal F_{\bar s r}^{\rm asy}
\label{eq:audit-complete-functional-relation}
\end{equation}
modulo finite local contact terms. The factor two comes from the complete
source algebra and can be absorbed into the overall action normalization if
that normalization has not independently been fixed.

The round-sphere calculation is unchanged in its power counting. The
intrinsic covariant recursion starts at radial weight one, whereas the same
full-source algebra produces the unique weight-zero term
$\sigma_p\Gamma^{\hat0}$. Hence the spherical counterterm is again
\eqref{eq:audit-full-source-counterterm}, with $\dd^2x\sqrt{\gamma_\rho}$
replaced by the induced round measure and with the global source projector.
Curvature terms begin at weight two and cannot generate another divergence in
the strip.

\medskip\noindent\textit{Principal-series continuation and locality.}\
At $\mu=\ii\nu$ the factor $\rho^{-2\mu}$ is a phase rather than a decaying
or growing power, and $A_\rho$ itself need not tend to zero. Thus
\eqref{eq:audit-complete-ren-limit} is not a direct fixed-mode
principal-series cutoff limit. The prescription is
\begin{equation}
0<\Re\mu<\frac12:
\quad \text{subtract and take }\rho\to0,
\qquad\text{then continue }\mu\to\ii\nu.
\label{eq:audit-order-of-renormalization}
\end{equation}
This is the same analytic-continuation prescription used to define the
source and response coefficients.

There is also a locality qualification. At fixed $\nu$ and phase,
\eqref{eq:audit-full-source-counterterm} is derivative-zero and local on the
celestial cutoff. Unfold the two phases to a single frequency
$\lambda\in\mathbb R$ using $\lambda=-\sigma_p\nu$, so that
$\tau^{\sigma_p\ii\nu}=\e^{-\ii\lambda u}$ with $u=\log\tau$.
Then $\sigma_p=-\operatorname{sgn}\lambda$, while $-\ii\partial_u$
has eigenvalue $-\lambda=\sigma_p\nu$. Consequently the phase-summed
counterterm has the form
\begin{equation}
I_{\rho,{\rm ct}}^{(0)}
=-\kappa_{\rm P}
\int\dd u\,\dd^2x\,\sqrt{\gamma_\rho}\,
\bar s\,G\,
\operatorname{sgn}(-\ii\partial_u)s.
\label{eq:audit-time-nonlocal-counterterm}
\end{equation}
The sign operator is pseudodifferential, so this subtraction is nonlocal in
Milne time. We therefore obtain a complete fixed-Mellin, or spectral/celestial,
radial renormalization of the free quadratic problem. We do not claim that
\eqref{eq:audit-time-nonlocal-counterterm} is a conventional local covariant
three-dimensional holographic counterterm for the unreduced Milne-boundary
field. This distinction is independent of finite local contact-term choices
for the separated-point celestial kernel.

\section{Technical derivation of the celestial source--response kernels}
\label{app:coefficient-kernels}

This appendix contains the finite-cutoff source--response maps, analytic continuation, local-term analysis, harmonic reconstruction, and Mellin identities underlying section~\ref{sec:correlator}. To keep the main argument readable, the explicit Clifford projectors, indexed Grassmann-derivative convention, complete cutoff pairings, subleading asymptotic terms, and homogeneous relations at possible denominator zeros are presented only here. The main text retains the transitional equations that connect phase selection to the planar and spherical kernels.

We now specialize to the massless sector and extract the nonlocal asymptotic
source--response kernel associated with a fixed boundary pairing from the
regulated Dirac action. We use the Milne rescaling
\eqref{eq:milne-rescaling} and the solutions of section~\ref{sec:solution}.
No new Clifford-algebra conventions are introduced.

For the variational problem, it is essential to treat $\phi$ and $\bar\phi$ as independent complex Grassmann fields. Independent variations yield the right- and left-acting Dirac equations and provide the independent Grassmann boundary sources used in the asymptotic functional. Choose the adjoint matrix $B$ so that, in the mostly-plus convention used here,
\begin{equation}
\label{eq:B-adjoint-convention}
B^{-1}(\Gamma^{\hat a})^\dagger B=-\Gamma^{\hat a},
\qquad \hat a=\hat0,\hat1,\hat2,\hat3.
\end{equation}
Consequently, for a spacelike radial projector $P_\pm=\frac{1}{2}(1\pm\Gamma^{\hat r})$,
\begin{equation}
\overline{P_\pm\phi}=\bar\phi P_\mp,
\end{equation}
which fixes the projector assignment in the boundary variational problem. The Lorentzian condition
\begin{equation}
\label{eq:physical-adjoint-slice}
\bar\phi=\phi^\dagger B
\end{equation}
may be imposed after variation and functional differentiation. Since complex conjugation reverses the Milne phase, this condition exchanges the phase labels introduced below.

\subsection{Milne-time phases and two celestial sources}
\label{subsec:milne-branches}

The complete mode expansion retains both $s=\pm1$ Dirac-eigenvalue families in the chosen irreducible spatial Clifford sector, as in section~\ref{sec:solution}. These two values of $s$ do not themselves label the two four-component Clifford blocks: multiplication by $\Gamma^{\hat0}$ generates the complementary spatial sector. To make the block convention explicit, define the central element of the spatial Clifford algebra
\begin{equation}
\label{eq:spatial-clifford-central-element}
\mathcal C_3\equiv \ii\Gamma^{\hat1}\Gamma^{\hat2}\Gamma^{\hat3},
\qquad
\mathcal C_3^2=\mathbf1_4,
\qquad
[\mathcal C_3,\Gamma^{\hat i}]=0,
\qquad
\{\mathcal C_3,\Gamma^{\hat0}\}=0.
\end{equation}
We fix the labels by
\begin{equation}
\label{eq:explicit-clifford-block-projectors}
\Pi_{\rm A}=\frac{1}{2}(\mathbf1_4+\mathcal C_3),
\qquad
\Pi_{\rm B}=\frac{1}{2}(\mathbf1_4-\mathcal C_3).
\end{equation}
The projectors therefore obey
\begin{equation}
\Pi_I\Pi_J=\delta_{IJ}\Pi_I,
\qquad
\Pi_{\rm A}+\Pi_{\rm B}=\mathbf1_4,
\qquad
[\Pi_I,\Gamma^{\hat i}]=0,
\qquad
\Gamma^{\hat0}\Pi_I=\Pi_{\bar I}\Gamma^{\hat0},
\end{equation}
where $\hat i=\hat1,\hat2,\hat3$ and $\bar I$ denotes the block complementary to $I$.
The physical adjoint exchanges these spatial blocks:
\begin{equation}
B^{-1}\mathcal C_3^\dagger B=-\mathcal C_3,
\qquad B^{-1}\Pi_I^\dagger B=\Pi_{\bar I},
\qquad \overline{\Pi_I\phi}=\bar\phi\Pi_{\bar I}.
\label{eq:physical-adjoint-block-exchange}
\end{equation}
Formal barred block labels remain independent in the complexified
variational problem. On the Lorentzian adjoint slice both the Milne phase
and the spatial block are exchanged.
Let $I_0$ denote the sector of the displayed raw modes: $I_0={\rm A}$ for the planar construction and $I_0={\rm B}$ for the explicit global Pauli realization. Convenient positive-eigenvalue representatives are
\begin{equation}
\chi_{I_0}\equiv\chi^{(+)},
\qquad
\chi_{\bar I_0}\equiv\Gamma^{\hat0}\chi^{(-)},
\qquad
\mathcal D_{\mathbb H^3}\chi_I=\ii\nu\chi_I,
\quad I\in\{{\rm A},{\rm B}\}.
\end{equation}
The nonlocal asymptotic functional for this pairing is therefore a block sum
\begin{equation}
\label{eq:full-sector-sum}
\left.\mathcal F_{\bar s r}^{\rm asy}\right|_{\rm nonlocal}
=
\mathcal F_{\bar s r,{\rm A}}^{\rm asy}[\mathbf J_{\rm A},\bar{\mathbf J}_{\rm A}]
+
\mathcal F_{\bar s r,{\rm B}}^{\rm asy}[\mathbf J_{\rm B},\bar{\mathbf J}_{\rm B}],
\qquad
\mathbf J_I=(J_I,\widetilde J_I),
\quad I\in\{{\rm A},{\rm B}\}.
\end{equation}
In the complexified variational problem, define $J_I=\Pi_IJ$ and $\bar J_I=\bar J\Pi_I$. Block diagonality then follows directly from $[A_\epsilon,\Pi_I]=0$ and $\Pi_I\Pi_J=\delta_{IJ}\Pi_I$:
\begin{equation}
\bar J_I A_\epsilon J_J
=
\bar J\,\Pi_I A_\epsilon\Pi_J J
=
\delta_{IJ}\bar J_I A_\epsilon J_I.
\end{equation}
The same block-diagonality holds for the asymptotic source--response kernel,
so no $A$--$B$ cross term occurs in that intrinsic coefficient. The complete finite-cutoff branch map also contains local cross-block terms proportional to $\Gamma^{\hat0}$; the block sum does not represent the full pairing. Relative to the raw $s=+1$ and $s=-1$
representatives, the use of $\chi_{\bar I_0}=\Gamma^{\hat0}\chi^{(-)}$
interchanges the two Milne phase assignments. Below
$\mathcal F_{\bar s r}^{\rm asy}$ denotes one block, the block label $I$ is
suppressed when no ambiguity can arise, and the phase labels are
denoted by $p,q\in\{-,+\}$, with $p=-$ for $\tau^{-\ii\nu}$ and $p=+$ for
$\tau^{+\ii\nu}$. Choose $\mathbb H^3$ eigenspinors
\begin{equation}
\label{eq:H3-positive-eigenvalue-ch4}
\mathcal D_{\mathbb H^3}\chi_{\nu,\mathfrak a}
=
\ii\nu\,\chi_{\nu,\mathfrak a},
\qquad
\nu>0,
\end{equation}
where $\mathfrak a$ collects the remaining Poincar\'e or global quantum numbers within the chosen irreducible spatial Clifford sector. Because $\{\Gamma^{\hat0},\mathcal D_{\mathbb H^3}\}=0$, the spinor $\Gamma^{\hat0}\chi_{\nu,\mathfrak a}$ supplies its complementary-sector negative-eigenvalue partner. Accordingly, the two massless solutions in \eqref{eq:eplusminus-def} are
\begin{equation}
\label{eq:massless-branch-spinors}
e_\pm[\chi_{\nu,\mathfrak a}]
=
\chi_{\nu,\mathfrak a}
\pm
\Gamma^{\hat0}\chi_{\nu,\mathfrak a},
\end{equation}
with Milne phases $\tau^{\pm\ii\nu}$. The bookkeeping used from this point onward is
\begin{equation}
\begin{array}{c|cc}
I & I_0 & \bar I_0 \\
\hline
\chi_I & \chi^{(+)} & \Gamma^{\hat0}\chi^{(-)}
\end{array}
\qquad
\begin{array}{c|cc}
p & - & + \\
\hline
\text{factor} & e_-[\chi_I]\tau^{-\ii\nu} & e_+[\chi_I]\tau^{+\ii\nu} \\
\text{source} & J_-=J & J_+=\widetilde J \\
\text{operator} & \mathcal O_-=\mathcal O & \mathcal O_+=\widetilde{\mathcal O}
\end{array}.
\label{eq:spectral-bookkeeping}
\end{equation}
For both blocks $\mathcal D_{\mathbb H^3}\chi_I=\ii\nu\chi_I$; the raw spectral sign $s$ only specifies which representative is used before the two blocks are put in this common positive-eigenvalue convention. We therefore write
\begin{equation}
\label{eq:branch-expansion}
\phi(\tau,x)
=
\int_0^\infty\frac{\dd\nu}{2\pi}
\left[
\tau^{-\ii\nu}\,
\phi_\nu^{(-)}(x)
+
\tau^{+\ii\nu}\,
\phi_\nu^{(+)}(x)
\right],
\end{equation}
where $\phi_\nu^{(-)}$ is built from $e_-[\chi_\nu]$ and $\phi_\nu^{(+)}$ from $e_+[\chi_\nu]$. These superscripts denote only the displayed Milne phases and carry no intrinsic incoming/outgoing interpretation. This is the massless specialization of the mode decomposition in section~\ref{sec:solution}; the explicit momentum or angular sums are restored in the corresponding subsections.

In the complexified variational problem, the barred field is expanded independently in the same phase basis,
\begin{equation}
\label{eq:bar-branch-expansion}
\bar\phi(\tau,x)
=
\int_0^\infty\frac{\dd\nu}{2\pi}
\left[
\tau^{-\ii\nu}\,
\bar\phi_\nu^{(-)}(x)
+
\tau^{+\ii\nu}\,
\bar\phi_\nu^{(+)}(x)
\right].
\end{equation}
The labels on the barred coefficients refer only to their displayed powers of $\tau$. At fixed $\nu>0$, the complexified variational problem treats the barred and unbarred coefficients as independent Grassmann data. Complex conjugation reverses the Milne phase and conjugates the principal-series radial weight, so a reality condition for the asymptotic coefficients depends on how the full spectral basis is extended to $\nu\in\mathbb R$. We impose no branch-by-branch relation at fixed $\nu$; the Lorentzian condition \eqref{eq:physical-adjoint-slice} is imposed only on the reconstructed full field after functional differentiation.
With $u=\log\tau$, the bilinear Mellin pairing is
\begin{align}
\int_0^\infty\frac{\dd\tau}{\tau}\,
\tau^{-\ii\nu}\tau^{+\ii\nu'}
&=
2\pi\delta(\nu-\nu'),
\\
\int_0^\infty\frac{\dd\tau}{\tau}\,
\tau^{\mp\ii\nu}\tau^{\mp\ii\nu'}
&=
2\pi\delta(\nu+\nu').
\end{align}
For $\nu,\nu'>0$, the second distribution has no support in the open integration domain. Hence, for every $\tau$-independent spatial operator
$\mathcal A$,
\begin{align}
\label{eq:independent-field-mellin-pairing}
\int_0^\infty\frac{\dd\tau}{\tau}\,
\bar\phi(\tau,x)\mathcal A\phi(\tau,x')=
\int_0^\infty\frac{\dd\nu}{2\pi}
\left[
\bar\phi_\nu^{(-)}(x)\mathcal A\phi_\nu^{(+)}(x')
+
\bar\phi_\nu^{(+)}(x)\mathcal A\phi_\nu^{(-)}(x')
\right].
\end{align}
We initially use smooth Mellin wave packets of compact support, so the temporal surface terms vanish. Pure-mode formulas are then recovered distributionally.

We use the symmetrized action and boundary polarization derived in appendix~\ref{app:dirac-boundary}. Restoring the overall normalization of \eqref{eq:action}, integration by parts gives
\begin{equation}
\label{eq:unsym-sym-relation}
I_{\rm bulk}^{\rm sym}
=
I_{\rm bulk}^{\rm unsym}
-
\frac{\mathcal N_S}{2}
\int_{\partial M}\dd^3x\,\sqrt{|h|}\,
\bar\psi\Gamma^n\psi.
\end{equation}
Thus the unsymmetrized and symmetrized actions give the same bulk Dirac equation but correspond to different boundary polarizations. It is important to distinguish the complete symmetrized on-shell action from each of its two boundary pairings.

Let $\Phi_{\rm src}$ denote the fixed radial component and $\Phi_{\rm resp}$ the dependent component determined by regularity. We label a pairing by the role of its barred and unbarred factors: $\bar s r$ means ``barred source times unbarred response,'' while $\bar r s$ means ``barred response times unbarred source.'' At the regulated boundary, define
\begin{align}
\mathcal B_{\epsilon;\bar s r}
&=
\kappa
\int_{\partial M_\epsilon}
\dd^3x\,\sqrt{|h|}\,
\bar\Phi_{\rm src}\Phi_{\rm resp},
\label{eq:pairing-barsr-regulated}
\\
\mathcal B_{\epsilon;\bar r s}
&=
\kappa
\int_{\partial M_\epsilon}
\dd^3x\,\sqrt{|h|}\,
\bar\Phi_{\rm resp}\Phi_{\rm src},
\label{eq:pairing-barrs-regulated}
\end{align}
where the sign associated with the outward normal and the chosen polarization is included in $\kappa$. The complete bare symmetrized on-shell functional is
\begin{equation}
I_{\epsilon,{\rm os}}^{\rm sym}
=
\frac{1}{2}
\left(
\mathcal B_{\epsilon;\bar s r}
+
\mathcal B_{\epsilon;\bar r s}
\right).
\label{eq:sym-action-two-orderings}
\end{equation}
The corresponding renormalized cutoff functional is
\begin{equation}
I_{\epsilon,{\rm ren}}^{\rm sym}
=
I_{\epsilon,{\rm os}}^{\rm sym}
+
I_{\epsilon,{\rm ct}}^{\rm div},
\label{eq:sym-action-renormalized}
\end{equation}
where $I_{\epsilon,{\rm ct}}^{\rm div}$ denotes the divergent counterterm
functional. The intrinsic recursion alone does not determine it; the
complete-source analysis in subsection~\ref{subsec:ct-full-pairing}
gives the weight-zero cross-block subtraction explicitly.

In the calculations below we extract the intrinsic source and response coefficients and form their bilinear, denoted by $\mathcal F_{\bar s r}^{\rm asy}$. The complete cutoff contraction $\mathcal B_{\epsilon;\bar s r}$ contains additional products and is not equated with this coefficient functional. The left-acting barred response, including the required momentum reversal and matrix ordering, is derived in subsection~\ref{subsec:ct-full-pairing}. There the two complete orderings are shown to be governed by the same full-source response operator after the actual cutoff sources are used. To keep the formulas compact, $\mathcal F_{\epsilon;\bar s r}$ below denotes the regulated intrinsic coefficient pairing and $\mathcal F_{\bar s r}^{\rm asy}$ its separated-point coefficient limit. The latter is finite, with optional finite contact terms, while the complete symmetrized functional requires the additional cross-block counterterm.

The two independent phase-labeled sources are denoted by $J,\widetilde J$ and their independent barred partners by $\bar J,\bar{\widetilde J}$. We use left functional derivatives with respect to barred sources and right functional derivatives with respect to unbarred sources:
\begin{equation}
\delta\mathcal F_{\bar s r}^{\rm asy}
=
\int\delta\bar J_\alpha\,
\frac{\delta_L\mathcal F_{\bar s r}^{\rm asy}}{\delta\bar J_\alpha}
+
\int
\left(
\frac{\delta_R\mathcal F_{\bar s r}^{\rm asy}}{\delta J_\beta}
\right)\delta J_\beta.
\end{equation}
Here $\delta_L$ and $\delta_R$ denote left and right Grassmann derivatives. Thus, for $\mathcal F_{\bar s r}^{\rm asy}=\int\bar J_\alpha\mathcal M_{\alpha\beta}J_\beta$, the mixed second derivative is $\mathcal M_{\alpha\beta}$. More explicitly, define
\begin{equation}
(\bar J_-,J_-)=(\bar J,J),
\qquad
(\bar J_+,J_+)=(\bar{\widetilde J},\widetilde J),
\end{equation}
and similarly $\mathcal O_-=\mathcal O$ and $\mathcal O_+=\widetilde{\mathcal O}$. Our asymptotic source--response convention for this pairing is
\begin{equation}
\label{eq:grassmann-derivative-convention}
\begin{aligned}
\left\langle
\mathcal O_{p,\alpha}(x,\nu)
\bar{\mathcal O}_{q,\beta}(y,\nu')
\right\rangle_{\rm asy}
&\equiv
\left.
\frac{1}{\ii}
\frac{\delta_L}{\delta\bar J_{p,\alpha}(x,\nu)}
\frac{\delta_R\mathcal F_{\bar s r}^{\rm asy}}{\delta J_{q,\beta}(y,\nu')}
\right|_{\substack{J=\bar J=0\\
\widetilde J=\bar{\widetilde J}=0}},
\\[-1mm]
&\hspace{35mm}p,q\in\{-,+\}.
\end{aligned}
\end{equation}
Consequently, the kernel appearing in the asymptotic source--response function
is $\mathscr M\equiv\mathcal M/\ii$. The complementary asymptotic
function must instead be obtained from the $\bar r s$ pairing; it is not
obtained merely by interchanging the functional derivatives acting on
$\mathcal F_{\bar s r}^{\rm asy}$.

Under the prospective CCFT interpretation, the formal operator labels $\mathcal O$ and $\widetilde{\mathcal O}$ would be coupled to the sources according to
\begin{equation}
\label{eq:prospective-source-coupling}
\mathcal I_{\rm source}
=
\int_0^\infty\frac{\dd\nu}{2\pi}
\int_{S^2}\dd\Omega_2\,
\Big(
\bar J\,\mathcal O
+\bar{\mathcal O}\,J
+\bar{\widetilde J}\,\widetilde{\mathcal O}
+\bar{\widetilde{\mathcal O}}\,\widetilde J
\Big).
\end{equation}
These symbols are only formal operator labels associated with the prospective CCFT interpretation; the present construction establishes the asymptotic source--response kernel and the fixed-Mellin radial renormalization rather than a CCFT path integral. On a Lorentzian slice the barred sources are obtained by imposing \eqref{eq:physical-adjoint-slice} on the reconstructed full field. The complementary barred radial response is derived in subsection~\ref{subsec:ct-full-pairing}. The Mellin pairing alone does not establish a CCFT interpretation. The global map to data at $\mathcal I^\pm$ requires the independent scattering energy-sign label and the continuation and inner-product prescriptions described in section~\ref{sec:cpw}.

\subsection{Poincar\'e coordinates on \texorpdfstring{$\mathbb H^3$}{H3}}

We use the radial projectors
\begin{equation}
\label{eq:poincare-radial-projectors}
\Pi_\pm=\frac{1}{2}(1\pm\Gamma^{\hat1}),
\qquad
\phi_\pm=\Pi_\pm\phi,
\qquad
\bar\phi_\pm=\bar\phi\Pi_\mp.
\end{equation}
The regulated region is $x_0\geq\epsilon$, so $\Gamma^{n_{\rm out}}=-\Gamma^{\hat1}$. Thus, the coordinate component $\phi_-$ is the outward $+$ component. We fix $\phi_-$ and $\bar\phi_-$. Using the $+$ polarization of appendix~\ref{app:dirac-boundary}, the selected pairing is
\begin{equation}
\label{eq:Sos-poincare-start}
\mathcal B_{\epsilon;\bar s r,{\rm P}}
=
\kappa_{\rm P}
\int_0^\infty\frac{\dd\tau}{\tau}
\int\dd^2\vec x\,\epsilon^{-2}\,
\bar\phi_-(\tau,\epsilon,\vec x)
\phi_+(\tau,\epsilon,\vec x)
\end{equation}
Here $\kappa_{\rm P}$ denotes the overall boundary-action normalization inherited from $\mathcal N_S$, including the chosen orientation and boundary polarization. Its phase is fixed only after a real-time convention is specified.
Let $\chi_{\nu,\vec k}$ denote the delta-normalized mode \eqref{eq:chi-mode-app-normalized} in the displayed fixed-$s$ sector. Its overall normalization cancels from the source--response ratio. Set
\begin{equation}
\chi_{\nu,\vec k}^{\pm}
=
\Pi_\pm\chi_{\nu,\vec k}.
\end{equation}
Equations \eqref{eq:chi-mode-app}--\eqref{eq:chi-mode-app-normalized} and \eqref{eq:a-minus} give, wherever $K_{\frac{1}{2}+\ii\nu}(k\epsilon)\neq0$,
\begin{equation}
\label{eq:poincare-response-map}
\chi_{\nu,\vec k}^{+}(\epsilon)
=
A_\epsilon(\vec k,\nu)\,
\chi_{\nu,\vec k}^{-}(\epsilon),
\end{equation}
where
\begin{equation}
\label{eq:Aeps-poincare}
A_\epsilon(\vec k,\nu)
=
\frac{\ii\slashed k}{k}
\frac{
K_{-\frac{1}2+\ii\nu}(k\epsilon)
}{
K_{\frac{1}2+\ii\nu}(k\epsilon)
},
\qquad
k=|\vec k|,
\qquad
\slashed k\equiv k_1\Gamma^{\hat2}+k_2\Gamma^{\hat3}.
\end{equation}
The four-dimensional branch spinors are $e_\pm[\chi_{\nu,\vec k}]$. They are not the spatial eigenspinors $\chi_{\nu,\vec k}$ alone. Since $\{\Gamma^{\hat0},\Gamma^{\hat1}\}=0$,
\begin{align}
\label{eq:branch-radial-projections-poincare}
\Pi_-\!e_\pm
&=
\chi^-_{\nu,\vec k}
\pm
\Gamma^{\hat0}\chi^+_{\nu,\vec k},
\\
\Pi_+\!e_\pm
&=
\chi^+_{\nu,\vec k}
\pm
\Gamma^{\hat0}\chi^-_{\nu,\vec k}.
\end{align}
To distinguish the finite-cutoff fields from the asymptotic coefficients, define $\sigma_-=-1$ and $\sigma_+=+1$. For a fixed Clifford block $I\in\{{\rm A},{\rm B}\}$, define
\begin{equation}
J_{\epsilon,I}^{p}(\vec k,\nu)
\in
\operatorname{im}(\Pi_I\Pi_-),
\qquad
\mathscr R_{\epsilon,I}^{p}(\vec k,\nu)
\equiv
A_\epsilon(\vec k,\nu)J_{\epsilon,I}^{p}(\vec k,\nu)
\in
\operatorname{im}(\Pi_I\Pi_+).
\end{equation}
Thus $J_{\epsilon,I}^{p}$ includes its cutoff radial dependence; it is not an amplitude multiplying a second object that already contains the same falloff. Since $A_\epsilon$ preserves the Clifford block while $\Gamma^{\hat0}$ exchanges $I$ and $\bar I$, the complete projected four-dimensional branch fields at the cutoff are
\begin{align}
\label{eq:full-finite-cutoff-branch-map}
\Phi_{-,\epsilon}^{p}
&=
J_{\epsilon,I}^{p}
+
\sigma_p\Gamma^{\hat0}A_\epsilon J_{\epsilon,I}^{p},
\\
\Phi_{+,\epsilon}^{p}
&=
A_\epsilon J_{\epsilon,I}^{p}
+
\sigma_p\Gamma^{\hat0}J_{\epsilon,I}^{p}.
\end{align}
Projecting the first equation with $\Pi_I$ recovers the source without an inverse,
\begin{equation}
J_{\epsilon,I}^{p}
=
\Pi_I\Phi_{-,\epsilon}^{p}.
\end{equation}
Consequently, on the graph subspace
\begin{equation}
\Pi_{\bar I}\Phi_{-,\epsilon}^{p}
=
\sigma_p\Gamma^{\hat0}A_\epsilon
\Pi_I\Phi_{-,\epsilon}^{p},
\end{equation}
the exact finite-cutoff map within the chosen Clifford block is
\begin{equation}
\label{eq:exact-finite-cutoff-map}
\Phi_{+,\epsilon}^{p}
=
\bigl(A_\epsilon+\sigma_p\Gamma^{\hat0}\bigr)
\Pi_I\Phi_{-,\epsilon}^{p}.
\end{equation}
No operator inverse is required in the one-block construction, but the graph representation through $A_\epsilon$ is valid only where the denominator in \eqref{eq:Aeps-poincare} is nonzero. At a zero of that denominator the regular mode is represented instead by the homogeneous source--response relation
\begin{equation}
K_{\frac{1}{2}+\ii\nu}(k\epsilon)\,
\mathscr R_{\epsilon,I}^{p}
=
\frac{\ii\slashed k}{k}
K_{-\frac{1}{2}+\ii\nu}(k\epsilon)\,
J_{\epsilon,I}^{p},
\end{equation}
which remains meaningful when the quotient defining $A_\epsilon$ does not.

For completeness, suppose that the two block sources are combined into a full source $J_\epsilon^p\in\operatorname{im}\Pi_-$. Then $1+\sigma_p\Gamma^{\hat0}A_\epsilon$ is an endomorphism of the full projected source space. Define
\begin{equation}
r_\epsilon(k,\nu)
=
\frac{K_{-\frac{1}2+\ii\nu}(k\epsilon)}{K_{\frac{1}2+\ii\nu}(k\epsilon)}.
\end{equation}
When $K_{\frac{1}{2}+\ii\nu}(k\epsilon)\neq0$ and $1+r_\epsilon(k,\nu)^2\neq0$, its inverse is
\begin{equation}
\bigl(1+\sigma_p\Gamma^{\hat0}A_\epsilon\bigr)^{-1}
=
\frac{1-\sigma_p\Gamma^{\hat0}A_\epsilon}{1+r_\epsilon^2}.
\end{equation}
For real $\nu$ and $k\epsilon>0$, $r_\epsilon$ is a phase where defined, so the full-block inverse can fail at the discrete locus $r_\epsilon=\pm\ii$. Thus the full-block formula is meromorphic in the cutoff data, while the one-block construction is most safely stated by the graph relation away from quotient poles and by the homogeneous relation at those poles.

For separated-point data we use the asymptotic coefficient prescription. Within the intrinsic fixed-block map, the regulated product differs from its finite nonlocal coefficient by local source-branch terms and subleading response corrections. Appendix~\ref{app:counterterm-audit} shows that every term of this intrinsic local series vanishes with a positive cutoff power. The additional algebraic term in the complete branch map is outside this series and must be retained in the full action. Finite local functionals of the asymptotic sources may nevertheless be added as contact-term scheme choices. In momentum space these have the schematic form
\begin{equation}
\sum_{n=0}^{N}c_n(\mu)\,
\ii\slashed k\,(k^2)^n.
\end{equation}
The nonanalytic kernel retained below is insensitive to such finite contact terms and determines the separated-point response.

Here and below, ``scheme-independent'' means unchanged by finite local momentum-polynomial additions at separated points.

To define source and response coefficients without relying on oscillatory principal-series powers, temporarily replace $\ii\nu$ by a complex parameter $\mu$ in the strip
\begin{equation}
\label{eq:mu-continuation-strip}
0<\Re\mu<\frac{1}{2}.
\end{equation}
The two leading radial powers are then $x_0^{1-\mu}$ and $x_0^{1+\mu}$, with strictly separated real parts. The coefficient of $x_0^{1-\mu}$ is defined as the source and that of $x_0^{1+\mu}$ as the response. In this strip,
\begin{equation}
\label{eq:Kratio-smallz-mu}
\frac{K_{\frac{1}2-\mu}(z)}{K_{\frac{1}2+\mu}(z)}
=
\frac{\Gamma(\frac{1}2-\mu)}{\Gamma(\frac{1}2+\mu)}
\left(\frac{z}{2}\right)^{2\mu}
+
\frac{z/2}{\mu-\frac{1}2}
+
\cdots.
\end{equation}
The first omitted contributions have powers $z^{1+4\mu}$, $z^{2+2\mu}$, and $z^3$; their ordering is understood by real part in the strip \eqref{eq:mu-continuation-strip}. For a block source, the regular mode expansions have the form
\begin{align}
J_{\epsilon,I}^{p}
&=
\epsilon^{1-\mu}J_I^p
+
\epsilon^{2+\mu}J_{{\rm sub},I}^p
+\cdots,
\\
A_\epsilon J_{\epsilon,I}^{p}
&=
\epsilon^{1+\mu}R_I^p
+
\epsilon^{2-\mu}R_{{\rm sub},I}^p
+\cdots.
\end{align}
Substitution in \eqref{eq:full-finite-cutoff-branch-map}, using the bracket
notation of \eqref{eq:asymptotic-component-notation}, shows precisely that
\begin{equation}
\bigl[\Phi_{-,\epsilon}^{p}\bigr]_{\epsilon^{1-\mu}}=J_I^p,
\qquad
\bigl[\Phi_{+,\epsilon}^{p}\bigr]_{\epsilon^{1+\mu}}=R_I^p;
\end{equation}
the remaining terms do not contribute to these specified coefficients.

We write $J_{\epsilon}^{(-)}=J_\epsilon$ and $J_{\epsilon}^{(+)}=\widetilde J_\epsilon$, with analogous notation for the independent barred source-type spinors. Using \eqref{eq:independent-field-mellin-pairing} and $\mathscr R_{\epsilon}^{p}=A_\epsilon J_{\epsilon}^{p}$ gives
\begin{align}
\label{eq:Sos-poincare-source-only}
\mathcal F_{\epsilon;\bar s r,{\rm P}}
=
\int_0^\infty\frac{\dd\nu}{2\pi}
\int\frac{\dd^2k}{(2\pi)^2}
\Big[
&
\bar{\widetilde J}_\epsilon(-\vec k,\nu)
\mathcal M_\epsilon(\vec k,\nu)
J_\epsilon(\vec k,\nu)
\nonumber\\
&+
\bar J_\epsilon(-\vec k,\nu)
\mathcal M_\epsilon(\vec k,\nu)
\widetilde J_\epsilon(\vec k,\nu)
\Big]
\end{align}
with
\begin{equation}
\label{eq:M-eps}
\mathcal M_\epsilon(\vec k,\nu)
=
\kappa_{\rm P}\epsilon^{-2}A_\epsilon(\vec k,\nu).
\end{equation}
Equation \eqref{eq:M-eps} is the regulated source-to-response kernel in the chosen asymptotic polarization. It is distinct from the exact finite-cutoff map \eqref{eq:exact-finite-cutoff-map} between the complete projected fields.
After inserting the source factors into the regulated quadratic pairing, the effective kernel in the strip \eqref{eq:mu-continuation-strip} is $\kappa_{\rm P}\epsilon^{-2\mu}A_\epsilon(\vec k,\mu)$. The first term in \eqref{eq:Kratio-smallz-mu} has a finite limit, whereas the analytic term is proportional to $\epsilon^{1-2\mu}$ and the displayed subleading nonanalytic term is proportional to $\epsilon^{1+2\mu}$; both vanish in this strip. The surviving nonlocal asymptotic kernel is therefore
\begin{equation}
\label{eq:M-ren-mu}
\mathcal M(\vec k,\mu)
=
\kappa_{\rm P}
\frac{\ii}{k}
\left(k_1\Gamma^{\hat2}+k_2\Gamma^{\hat3}\right)
\frac{\Gamma(\frac{1}2-\mu)}{\Gamma(\frac{1}2+\mu)}
\left(\frac{k}{2}\right)^{2\mu}.
\end{equation}
We then continue to the principal series by $\mu\to\ii\nu$, $\nu>0$, using
\begin{equation}
\left(\frac{k}{2}\right)^{2\mu}
=
\exp\!\left[2\mu\log\!\left(\frac{k}{2}\right)\right],
\qquad k>0,
\end{equation}
with the real logarithm. The principal-series kernel is the boundary value from $\Re\mu>0$ after the cutoff limit has been taken in the continuation strip, where the displayed momentum-polynomial term vanishes; the threshold $\nu=0$ is approached as $\nu\to0^+$. For real principal-series $\nu$, the source and response powers have the same real part, so ``source'' and ``response'' refer to these analytically continued coefficients rather than to a hierarchy based on pointwise decay.

Define the finite coefficient sources as the coefficients of the $x_0^{\,1-\ii\nu}$ falloff,
\begin{align}
\label{eq:chi-ren-poincare}
J(\vec k,\nu)
&=
\lim_{\epsilon\to0}
\epsilon^{\ii\nu-1}J_\epsilon(\vec k,\nu),
&
\widetilde J(\vec k,\nu)
&=
\lim_{\epsilon\to0}
\epsilon^{\ii\nu-1}\widetilde J_\epsilon(\vec k,\nu),
\\
\bar J(\vec k,\nu)
&=
\lim_{\epsilon\to0}
\epsilon^{\ii\nu-1}\bar J_\epsilon(\vec k,\nu),
&
\bar{\widetilde J}(\vec k,\nu)
&=
\lim_{\epsilon\to0}
\epsilon^{\ii\nu-1}\bar{\widetilde J}_\epsilon(\vec k,\nu).
\end{align}
Taking the cutoff limit, the local cutoff terms vanish and the finite coefficient is
\begin{equation}
\label{eq:M-ren}
\mathcal M(\vec k,\nu)
=
\kappa_{\rm P}
\frac{\ii}{k}
\left(
k_1\Gamma^{\hat2}
+
k_2\Gamma^{\hat3}
\right)
\frac{\Gamma(\frac{1}2-\ii\nu)}{\Gamma(\frac{1}2+\ii\nu)}
\left(\frac{k}{2}\right)^{2\ii\nu}.
\end{equation}
Hence
\begin{align}
\label{eq:Sren-poincare-momentum}
\mathcal F_{\bar s r,{\rm P}}^{\rm asy}
=
\int_0^\infty\frac{\dd\nu}{2\pi}
\int\frac{\dd^2k}{(2\pi)^2}
\Big[
&
\bar{\widetilde J}(-\vec k,\nu)
\mathcal M(\vec k,\nu)
J(\vec k,\nu)
\nonumber\\
&+
\bar J(-\vec k,\nu)
\mathcal M(\vec k,\nu)
\widetilde J(\vec k,\nu)
\Big].
\end{align}
In accordance with \eqref{eq:grassmann-derivative-convention}, define the asymptotic kernel for this pairing in momentum space by
\begin{equation}
\label{eq:Mhat-ren}
\mathscr M(\vec k,\nu)
\equiv
\frac{1}{\ii}\mathcal M(\vec k,\nu).
\end{equation}
The Fourier transform of $\mathscr M$ has the standard form
\begin{equation}
\label{eq:planar-position-kernel}
\mathcal K_\nu(\vec r)
=
c_\nu
\frac{
\Gamma^{\hat\alpha}r_{\hat\alpha}
}{
|\vec r|^{3+2\ii\nu}
},
\qquad
\hat\alpha=\hat2,\hat3,
\end{equation}
where, for the Fourier convention in \eqref{eq:Sren-poincare-momentum},
\begin{equation}
\label{eq:cnu-planar}
c_\nu=-\frac{\kappa_{\rm P}}{\ii}\frac{1+2\ii\nu}{2\pi}.
\end{equation}
Local contact terms are omitted. The mixed phase-labeled asymptotic kernels are
\begin{align}
\label{eq:spinor-2pt-mixed}
\left\langle
\mathcal O(\vec x,\nu)
\bar{\widetilde{\mathcal O}}(\vec y,\nu')
\right\rangle_{\rm asy}
&=
2\pi\delta(\nu-\nu')\,
\mathcal K_\nu(\vec x-\vec y),
\\
\left\langle
\widetilde{\mathcal O}(\vec x,\nu)
\bar{\mathcal O}(\vec y,\nu')
\right\rangle_{\rm asy}
&=
2\pi\delta(\nu-\nu')\,
\mathcal K_\nu(\vec x-\vec y).
\end{align}
The kernel corresponds to
\begin{equation}
\label{eq:Delta-nu}
\Delta_\nu=1+\ii\nu,
\qquad
\mathfrak s_{\rm A}=+\frac{1}{2},
\qquad
\mathfrak s_{\rm B}=-\frac{1}{2},
\end{equation}
where we fix the signed-spin convention by assigning $+\tfrac12$ to block $A$ and $-\tfrac12$ to block $B$ in the adopted boundary orientation and spin frame. Reversing the boundary orientation interchanges these assignments. The associated two-dimensional weights are
\begin{equation}
\label{eq:weights-nu}
h_I=\frac{\Delta_\nu+\mathfrak s_I}{2},
\qquad
\bar h_I=\frac{\Delta_\nu-\mathfrak s_I}{2},
\qquad I\in\{{\rm A},{\rm B}\}.
\end{equation}
On the physical slice, the adjoint of a reconstructed unbarred branch carries the opposite displayed phase and the complementary spatial Clifford block. The resulting relation between intrinsic coefficient sources also conjugates their principal-series weights; it is not a same-block, fixed-$\nu$ reality condition.

\subsection{Global coordinates on \texorpdfstring{$\mathbb H^3$}{H3}}

Regulate the conformal boundary at
\begin{equation}
y=Y,
\qquad
Y=\log\frac{2}{\epsilon},
\qquad
\epsilon\to0^+.
\end{equation}
The measure is
\begin{equation}
\sinh^2Y
=
\epsilon^{-2}
-
\frac{1}{2}
+
\frac{\epsilon^2}{16}.
\end{equation}
The outward normal is $\Gamma^n=\Gamma^{\hat y}$, and we use
\begin{equation}
\label{eq:Ppm-global}
P_\pm=\frac{1}{2}(1\pm\Gamma^{\hat y}),
\qquad
\phi_\pm=P_\pm\phi,
\qquad
\bar\phi_\pm=\bar\phi P_\mp.
\end{equation}
The explicit large-$y$ asymptotics of \eqref{eq:H3-spinor-mode-basis} show that, in the intrinsic representative, $P_+\chi$ contains the independent $\epsilon^{1-\ii\nu}$ coefficient, while $P_-\chi$ contains the dependent $\epsilon^{1+\ii\nu}$ coefficient. The complete projections also contain the cross-block terms described in subsection~\ref{subsec:ct-full-pairing}. We therefore fix
$\phi_+$ and $\bar\phi_+$. The selected pairing is
\begin{equation}
\label{eq:Sonshell-taufactor}
\mathcal B_{\epsilon;\bar s r,{\rm G}}
=
\kappa_{\rm G}
\int_0^\infty\frac{\dd\tau}{\tau}
\int_{S^2}\dd\Omega_2\,
\sinh^2Y\,
\bar\phi_+(\tau,Y,\Omega)
\phi_-(\tau,Y,\Omega)
\end{equation}
Here $\kappa_{\rm G}$ is the corresponding boundary-action normalization inherited from $\mathcal N_S$. We match the boundary orientation, spin frame, and source normalization conventions between the two coordinate descriptions and, in that matched convention, set $\kappa_{\rm G}=\kappa_{\rm P}$. Any remaining common complex factor is conventional.

For the positive eigenvalue representative, use
\begin{equation}
\label{eq:H3-global-positive-mode}
\chi_{\nu l m_j}(y,\Omega)
=
c_3(\nu,l)
\left[
\phi_{\nu l}(y)\hat\chi^{(-)}_{l m_j}(\Omega)
+
\ii\psi_{\nu l}(y)\hat\chi^{(+)}_{l m_j}(\Omega)
\right],
\end{equation}
which is \eqref{eq:H3-spinor-mode-basis} with $s=+1$. Since
$\hat\chi^{(+)}_{l m_j}=\Gamma^{\hat y}\hat\chi^{(-)}_{l m_j}$,
\begin{align}
\label{eq:global-projected-mode}
P_+\chi_{\nu l m_j}
&=
c_3(\nu,l)\bigl(\phi_{\nu l}+\ii\psi_{\nu l}\bigr)
P_+\hat\chi^{(-)}_{l m_j},
\\
P_-\chi_{\nu l m_j}
&=
c_3(\nu,l)\bigl(\phi_{\nu l}-\ii\psi_{\nu l}\bigr)
P_-\hat\chi^{(-)}_{l m_j}.
\end{align}
The large-$Y$ continuation of the hypergeometric functions gives
\begin{align}
\phi_{\nu l}+\ii\psi_{\nu l}
&\sim
\mathcal A_+(l,\nu)
\left(\frac{\epsilon}{2}\right)^{1-\ii\nu},
\\
\phi_{\nu l}-\ii\psi_{\nu l}
&\sim
\mathcal A_-(l,\nu)
\left(\frac{\epsilon}{2}\right)^{1+\ii\nu},
\end{align}
where
\begin{align}
\label{eq:Apm-global}
\mathcal A_+(l,\nu)
&=
2^{3-2\ii\nu}
\frac{\Gamma(l+\frac{3}2)\Gamma(2\ii\nu)}{\Gamma(l+\frac{3}2+\ii\nu)\Gamma(\ii\nu)},
\\
\mathcal A_-(l,\nu)
&=
2^{3+2\ii\nu}
\frac{\Gamma(l+\frac{3}2)\Gamma(-2\ii\nu)}{\Gamma(l+\frac{3}2-\ii\nu)\Gamma(-\ii\nu)}.
\end{align}
When $\phi_{\nu l}(Y)+\ii\psi_{\nu l}(Y)\neq0$, the exact finite-$Y$ spatial source--response ratio for a fixed harmonic is
\begin{equation}
\label{eq:AY-global-exact}
\mathscr A_Y(l,\nu)
\equiv
\frac{\phi_{\nu l}(Y)-\ii\psi_{\nu l}(Y)}{\phi_{\nu l}(Y)+\ii\psi_{\nu l}(Y)}.
\end{equation}
At a zero of the denominator, the regular mode remains well defined and is represented by the homogeneous source--response relation
\begin{equation}
\bigl[\phi_{\nu l}(Y)+\ii\psi_{\nu l}(Y)\bigr]R_{\epsilon;l m_j}
=
\bigl[\phi_{\nu l}(Y)-\ii\psi_{\nu l}(Y)\bigr]J_{\epsilon;l m_j},
\end{equation}
rather than by the quotient \eqref{eq:AY-global-exact}.
The nonlocal ratio of the two asymptotic coefficients is instead
\begin{equation}
\label{eq:Reps-global}
\mathcal R_\epsilon(l,\nu)
=
\frac{\mathcal A_-(l,\nu)}{\mathcal A_+(l,\nu)}
\left(\frac{\epsilon}{2}\right)^{2\ii\nu}
=
\frac{\Gamma(\frac{1}2-\ii\nu)}{\Gamma(\frac{1}2+\ii\nu)}
\frac{\Gamma(l+\frac{3}2+\ii\nu)}{\Gamma(l+\frac{3}2-\ii\nu)}
\left(\frac{\epsilon}{2}\right)^{2\ii\nu}.
\end{equation}
This equality is exact at the level of the displayed asymptotic coefficients; no $O(\epsilon)$ correction is implied by the exact definition $\epsilon=2e^{-Y}$.

Define normalized source and response angular harmonics
\begin{equation}
\label{eq:global-boundary-harmonics}
u_{l m_j}(\Omega)
=
\sqrt2\,P_+\hat\chi^{(-)}_{l m_j}(\Omega),
\qquad
v_{l m_j}(\Omega)
=
\sqrt2\,P_-\hat\chi^{(-)}_{l m_j}(\Omega).
\end{equation}
The quantities $\mathscr A_Y(l,\nu)$ and $\mathcal R_\epsilon(l,\nu)$ are used only within the displayed irreducible Clifford sector. We do not promote them to an operator inverse on the full four-component spinor space. As in the Poincar\'e patch, the $\Gamma^{\hat0}$ partner exchanges the radial projectors: in the complete branch spinor, it contributes no $\epsilon^{1-\mu}$ coefficient to the fixed $P_+$ component and no $\epsilon^{1+\mu}$ coefficient to the dependent $P_-$ component. Thus the designated intrinsic source and response coefficients acquire no extra numerical factor from that partner. This statement concerns extraction of the individual coefficients; the complete-source renormalized functional has the factor of two derived in \eqref{eq:audit-complete-functional-relation}.

For separated-point data we use the same continuation \eqref{eq:mu-continuation-strip}. Replacing $\ii\nu$ by $\mu$ gives
\begin{equation}
\mathcal R_\epsilon(l,\mu)
=
\frac{\Gamma(\frac{1}2-\mu)}{\Gamma(\frac{1}2+\mu)}
\frac{\Gamma(l+\frac{3}2+\mu)}{\Gamma(l+\frac{3}2-\mu)}
\left(\frac{\epsilon}{2}\right)^{2\mu}.
\end{equation}
After including the two source factors, the regulated harmonic kernel is proportional to $\epsilon^{2-2\mu}\sinh^2Y\,\mathcal R_\epsilon(l,\mu)$. Since $\epsilon^2\sinh^2Y\to1$, its finite nonlocal limit in the strip is
\begin{equation}
\label{eq:Mren-harm-mu}
\mathcal M(l,\mu)
=
\kappa_{\rm G}2^{-2\mu}
\frac{\Gamma(\frac{1}2-\mu)}{\Gamma(\frac{1}2+\mu)}
\frac{\Gamma(l+\frac{3}2+\mu)}{\Gamma(l+\frac{3}2-\mu)}.
\end{equation}
Only after this extraction do we continue $\mu\to\ii\nu$. As in the Poincar\'e patch, the limit is already finite; the coefficient prescription isolates its separated-point nonlocal part, modulo finite local contact terms.

Expand the source-type cutoff boundary spinors as
\begin{align}
J_\epsilon^p(\Omega,\nu)
&=
\sum_{l,m_j}J_{\epsilon;l m_j}^p(\nu)u_{l m_j}(\Omega),
\\
\bar J_\epsilon^p(\Omega,\nu)
&=
\sum_{l,m_j}\bar J_{\epsilon;l m_j}^p(\nu)v_{l m_j}^\dagger(\Omega),
\qquad
p\in\{-,+\}.
\end{align}
The response-type asymptotic coefficients satisfy
\begin{equation}
R_{\epsilon;l m_j}^p(\nu)
=
\mathcal R_\epsilon(l,\nu)
J_{\epsilon;l m_j}^p(\nu).
\end{equation}
After the Mellin pairing, the regulated asymptotic functional is
\begin{align}
\label{eq:Sonshell-harmonic-branches}
\mathcal F_{\epsilon;\bar s r,{\rm G}}
={}&
\int_0^\infty\frac{\dd\nu}{2\pi}
\sum_{l=0}^\infty\sum_{m_j}
\Big[
\bar J_{\epsilon;l m_j}^{(-)}(\nu)
\mathcal M_\epsilon(l,\nu)
J_{\epsilon;l m_j}^{(+)}(\nu)
\nonumber\\[-1mm]
&\hspace{35mm}
+
\bar J_{\epsilon;l m_j}^{(+)}(\nu)
\mathcal M_\epsilon(l,\nu)
J_{\epsilon;l m_j}^{(-)}(\nu)
\Big]
\end{align}
where
\begin{equation}
\mathcal M_\epsilon(l,\nu)
=
\kappa_{\rm G}\sinh^2Y\,
\mathcal R_\epsilon(l,\nu).
\end{equation}
We define global finite coefficient sources in the boundary Weyl frame associated with the defining function $\epsilon=2e^{-Y}$,
\begin{align}
J_{l m_j}^p(\nu)
&=\lim_{\epsilon\to0}\epsilon^{\ii\nu-1}J_{\epsilon;l m_j}^p(\nu),
\nonumber\\
\bar J_{l m_j}^p(\nu)
&=\lim_{\epsilon\to0}\epsilon^{\ii\nu-1}\bar J_{\epsilon;l m_j}^p(\nu),
\qquad p\in\{-,+\}.
\end{align}
Since $\sinh^2Y\,\epsilon^2\to1$, the finite nonlocal harmonic coefficient is
\begin{equation}
\label{eq:Mren-harm}
\mathcal M(l,\nu)
=
\kappa_{\rm G}2^{-2\ii\nu}
\frac{\Gamma(\frac{1}2-\ii\nu)}{\Gamma(\frac{1}2+\ii\nu)}
\frac{\Gamma(l+\frac{3}2+\ii\nu)}{\Gamma(l+\frac{3}2-\ii\nu)}.
\end{equation}
Define the harmonic asymptotic kernel for this pairing by
\begin{equation}
\label{eq:Mhat-harm}
\mathscr M(l,\nu)
\equiv
\frac{1}{\ii}\mathcal M(l,\nu).
\end{equation}

Appendix~\ref{app:modesum} gives the fixed-$l$ addition theorem and the
Abel-regulated Jacobi sum. The position-space transform of $\mathscr M$ is
\begin{equation}
\label{eq:sphere-spinor-kernel}
\mathcal K_\nu^{S^2}(\Omega,\Omega')
=
c_\nu^{S^2}
\frac{
P_-(\Omega)\,
\Gamma^{\hat\alpha}\Xi_{\hat\alpha}(\Omega,\Omega')\,
\mathcal P(\Omega,\Omega')\,
P_+(\Omega')
}{
\bigl[2(1-\cos\gamma)\bigr]^{1+\ii\nu}
},
\qquad 0<\gamma<\pi,
\end{equation}
where $\mathcal P(\Omega,\Omega')$ is the spin parallel propagator along the unique minimizing geodesic and $\Xi_{\hat\alpha}=\nabla_{\hat\alpha}\gamma$. The closed form is therefore understood for $0<\gamma<\pi$; the harmonic expansion defines the global distribution, including the antipodal point. The normalization is
\begin{equation}
\label{eq:cnu-sphere}
c_\nu^{S^2}=-\frac{\kappa_{\rm G}}{\ii}\frac{1+2\ii\nu}{2\pi}.
\end{equation}
Upon imposing the matched normalization convention $\kappa_{\rm G}=\kappa_{\rm P}$, this agrees with \eqref{eq:cnu-planar}. The outer projectors may be
suppressed when the kernel is understood as a map from the $P_+$ source
subspace to the $P_-$ response subspace. The mixed asymptotic kernels are
\begin{align}
\label{eq:2pt-sphere-mixed}
\left\langle
\mathcal O(\Omega,\nu)
\bar{\widetilde{\mathcal O}}(\Omega',\nu')
\right\rangle_{\rm asy}
&=
2\pi\delta(\nu-\nu')\,
\mathcal K_\nu^{S^2}(\Omega,\Omega'),
\\
\left\langle
\widetilde{\mathcal O}(\Omega,\nu)
\bar{\mathcal O}(\Omega',\nu')
\right\rangle_{\rm asy}
&=
2\pi\delta(\nu-\nu')\,
\mathcal K_\nu^{S^2}(\Omega,\Omega').
\end{align}
For $\gamma\to0$,
\begin{equation}
\mathcal K_\nu^{S^2}
\sim
\slashed{\Xi}\,
\gamma^{-2-2\ii\nu},
\end{equation}
in agreement with \eqref{eq:planar-position-kernel}. This agreement is a local short-distance check. A full finite-separation identification additionally requires the stereographic Weyl factors, endpoint spin rotations, and the transformation of the spin parallel propagator; those frame-matching steps are not derived in the present paper.

\bibliographystyle{JHEP}
\bibliography{bibliography}

\end{document}